\documentclass{cernrep}
\usepackage{amssymb,graphicx}
\usepackage{epstopdf}
\usepackage{amsmath,amsfonts}
\usepackage{epsfig}
\usepackage[dvipsnames]{xcolor}
\def\const{\mbox{const}}
\def\e{{\rm e}}

\def\d{\partial}
\def\l{\left(}
\def\r{\right)}
\def\la{\langle }
\def\ra{\rangle }
\newcommand{\be}{\begin{equation}}
\newcommand{\ee}{\end{equation}}
\newcommand{\bea}{\begin{eqnarray}}
\newcommand{\eea}{\end{eqnarray}}

\renewcommand{\ln}{\mathop{\rm ln}\nolimits}

\begin{document}

\title{Cosmology and dark matter}


\author{V.A.Rubakov}

\institute{Institute for Nuclear Research 
of the Russian Academy of Sciences, \\
60th October Anniversary Prospect, 7a, 
Moscow, 117312, Russia\\
and\\
Department of Particle Physics and Cosmology,
Physics Faculty, Moscow State University,\\ Vorobjevy Gory,
119991, Moscow, Russia}

\maketitle

\abstract{Cosmology and astroparticle physics give strongest
  possible evidence for the incompleteness of the Standard
  Model of particle physics. Leaving aside misterious dark energy,
  which may or may not be just the cosmological constant,
  two properties of the Universe cannot be explained by
  the Standard Model: 
  dark matter and matter-antimatter asymmtery.
  Dark matter particles may well be discovered in foreseeable future;
  this issue is under intense experimental investigation.
  Theoretical hypotheses on the nature of the dark matter particles
  are numerous, so we concentrate on
  several well motivated candidates, such as WIMPs,
  axions and sterile neutrinos, and also give examples of less
  motivated and more elusive candidates such as fuzzy dark matter.
 This gives an idea of the spectrum of conceivable
  dark matter candidates, while certainly not exhausting it.
We then consider the matter-antimatter asymmetry and discuss whether
it may result from physics at 100~GeV -- TeV scale.
Finally, we 
turn to the earliest epoch of the cosmological evolution.
Although the latter topic does not appear immediately related to
contemporary particle physics, it is of great interest due to its
fundamental nature.
We emphasize that the cosmological data, notably, on
CMB anisotropies, unequivocally show that 
the well understood hot stage was not the earliest one.
The best guess for the earlier stage is inflation, which is consistent with 
everything known to date; however,
there are alternative scenarios. We discuss the ways to study the earliest
epoch, with emphasis on  future cosmological observations.
}

\tableofcontents

\section{Introduction}
\label{Intro}

It is a commonplace by now that
cosmology and astroparticle physics, on the one side, and particle physics,
on the other, are
deeply interrelated. Indeed, the gross properties of the Universe --
the existence of dark matter and the very presence of conventional, baryonic
matter -- call for the extension of the
Standard Model of particle physics. A fascinating possibility is that
the physics behind these phenomena is within reach of current or future
terrestrial experiments.  The experimental programs in these directions
are currently intensely pursued.

Another aspect of cosmology, which currently does not apppear directly
related to terrestrial particle physics experiments, is the
earliest epoch of the evolution of the Universe. On the one hand, there is
no doubt that the usual hot epoch was preceded by another, much
less conventional stage. This knowledge comes from the study of
inhomogeneities in the Universe through the measurements of CMB
anisotropies, as well as matter distribution (galaxies, clusters of galaxies,
voids) in the present and recent Universe. On the the other hand,
we know only rather general properties of the cosmological perturbations,
which, we are convinced, were generated before the hot epoch.
For this reason, we cannot be sure about the
earliest epoch; the best guess is inflation, but alternatives to
inflation have not  yet been ruled out. It is conceivable that future
cosmological observations will be able to disentangle between different
hypotheses; it is amazing that the study of the Universe at large will
possibly reveal the properties of the very early epoch characterized by
enormous energy density and evolution rate.

Cosmology and astroparticle physics is a large area of research,
so we will be unable to cover it to any level of completeness.
On the dark matter side, the number of proposals for dark
matter objects invented by theorists in more than 30 years is enormous,
so we do not attempt even to list them.
Instead, we concentrate on a few hypotheses which may or may not
have to do with reality. Namely, we study reasonably well
motivated candidates -- WIMPs, axions, sterile neutrinos -- and also
discuss more exotic possibilities.
On the baryon asymmetry side, we focus on scenarios for its generation
which employ physics accessible by terrestrial experiments.
A particular mechanism of this sort is
the electroweak baryogenesis.
The last part of these lectures deals with the earliest cosmology --
inflation and its alternatives.

To end up this Introduction, we point out that
most of the topics we discuss are studied, in one or another way,
in books~\cite{books}. There are of course numerous reviews, some of which
will be referred to in appropriate places.

\section{Homogeneous and isotropic Universe}

\subsection{FLRW metric}
\label{sub:FLRW}

When talking about the Universe, we will always mean 
its visible part. The visible part is, almost for sure,
a small, and maybe even tiny patch of a huge space; for the time being
(at least) we cannot tell what is outside the part we observe.
At large scales the (visible part of the) Universe is 
{\it homogeneous and isotropic}: all regions of the Universe are
the same, and no direction is preferred.
Homogeneous and isotropic three-dimensional 
spaces can be of three types.
These are
three-sphere, flat 
(Euclidean) space 
and three-hyperboloid.

A basic 
property of our Universe is that 
it  expands: the space stretches out. 
This is encoded in the space-time metric
(Friedmann--Lema\^itre--Robertson--Walker, FLRW)
\begin{equation}
 ds^2 = dt^2 - a^2(t) d{\bf x}^2 \; ,
\label{FRW}
\end{equation}
where $d{\bf x}^2$ is the distance on unit 
three-sphere or Euclidean space or hyperboloid,
$a(t)$ is the scale factor. Observationally, the three-dimensional space is
Euclidean (flat) to good approximation
(see, however, Ref.\cite{DiValentino:2019qzk} where it is claimed that
Planck lensing data prefer closed Universe), so we will
treat  $d{\bf x}^2=\delta_{ij} dx^i dx^j$, $i,j =1,2,3$,
as line interval in three-dimensional Euclidean space.

The coordinates ${\bf x}$ are comoving. This means that they label
positions of free, static particles in space (one has to check that
world lines of free static particles obey
${\bf x}=\mbox{const}$; this is indeed the case). 
As an example, distant galaxies stay at fixed  ${\bf x}$ (modulo peculiar
motions, if any). In our expanding Universe,
the scale factor $a(t)$ increases in time, so
the distance between free masses of fixed spatial
coordinates
${\bf x}$ grows,
$  dl^2 = a^{2}(t) d{\bf x}^2$. The
galaxies run away from each other.

Since the space stretches out, so does the wavelength of a photon;
photon experiences redshift.
If the wavelength at emission (say, by distant star)
is   $\lambda_e$, then the wavelength
we measure is
\[
\lambda_0 = (1 + z) \lambda_e \; , \;\;\;\;
\mbox{where} \;\;\;\;\; z = \frac{a(t_0)}{a(t_e)} - 1 \; .
\]
Here $t_e$ is the time at emission, and $z$ is redshift.
Hereafter we denote by subscript $0$ the quantities measured at
the present time. We sometimes set $a_0\equiv a(t_0)=1$
and put ourselves at the
origin of coordinate frame, then $|{\bf x}|$ is the {\it present} distance
to a point with coordinates ${\bf x}$. We also call this {\it
  comoving distance}. 

Clearly, the further from us is the source, the longer it takes for
light, seen by us today, to travel, i.e., the larger $t_0 - t_e$. High redshift
sources are far away from us both in space and in time. For not so
distant sources, we have $t_0 - t_e =r$, where $r$
is the physical distance to the source\footnote{Hereafter we use
  the natural units,
with the speed of light, Planck and Boltzmann constants equal to 1,\\
$c=\hbar = k_B =1$.
Then Newton's gravity constant is
$G=M_{Pl}^{-2}$, where $M_{Pl} = 1.2\cdot 10^{19}$~GeV is the Planck mass.}.
For $z \ll 1$ we thus have the Hubble law,
\be
z = H_0 r \; .
\label{mar22-15-1}
\ee
$H_0 \equiv H(t_0)$ is the Hubble constant, i.e., the
present value of the Hubble parameter
\[
H (t) = \frac{\dot{a}(t)}{a(t)} \; .
\]
The value of the Hubble constant is a subject of some controversy.
While
the redshift of an object can be measured with high precision
($\lambda_e$ is the wavelength of a photon
emitted by an excited atom; 
one identifies a series of emission 
lines, thus determining $\lambda_e$, 
and measures their actual wavelengths $\lambda_0$, both with
spectroscopic precision; absorption lines are used as well), 
absolute distances to astrophysical sources have considerable systematic
uncertainty. The precise value of $H_0$ will not be important for our
semi-quantitative discussions; we quote here the value found by
the Planck collaboration~\cite{Aghanim:2018eyx},
\be
     H_0 = (67.7 \pm 0.4)~\frac{\mbox{km/s}}{\mbox{Mpc}} \approx
(14.4 \cdot 10^9~\mbox{yrs})^{-1} \; .
     \label{H00}
     \ee
     Here Mpc is the length unit used in cosmology and astrophysics,
\[
1~\mbox{Mpc} \approx 3\cdot 10^6~\mbox{light~years} \approx 
3\cdot 10^{24}~\mbox{cm} \; .
\]
The funny unit used in the first expression in
\eqref{H00} has to do with (somewhat misleading)
interpretation of redshift as Doppler effect: galaxies run away from us
at velocity $v=z$. 
To account for uncertainties in $H_0$ one writes for
the present value of the Hubble parameter 
\begin{equation}
   H_0 = h \cdot 100 ~\frac{\mbox{km}/\mbox{s}}{ \mbox{Mpc}} \; .
\label{H0}
\end{equation}
Thus $   h \approx 0.7$.
We will use this value in estimates.

Concerning length scales characterstic of various objects,
we quote the following:
\begin{itemize}
  \item sizes of visible parts of dwarf galaxies are of order
  1~kpc and even smaller;
  \item  sizes of  visible
    parts of galaxies like ours are of order 10~kpc;
  \item
    dark halos of galaxies
    extend to distances of order 100~kpc and larger;
  \item
    clusters of galaxies have
    sizes of order $1-3$~Mpc;
  \item  homogeneity scale\footnote{Regions of this size
    and larger look all the same, while smaller regions differ from each
    other; they contain different numbers of galaxies.}
  today is of order 200~Mpc;
\item
  the size of the visible Universe is 14~Gpc. 
  \end{itemize}


\subsection{CMB}

One of the fundamental discoveries of 1960's was 
cosmic microwave background (CMB).
These are
photons with black-body spectrum of temperature
\be
   T_0 = 2.7255 \pm 0.0006~\mbox{K} \; .
\label{temperature}
\ee
Measurements of this spectrum are quite precise and show
no deviation from the Planck 
spectrum (although some deviations are predicted,
see Ref.~\cite{black-body} for review).
The 
energy density of CMB photons is given by the  Stefan--Boltzmann formula
\be 
  \rho_{\gamma, 0} = \frac{\pi^2}{15} T_0^4 =
2.7 \cdot 10^{-10} ~\frac{\mbox{GeV}}{\mbox{cm}^3} \; .
\label{rhogamma} \; 
\ee
while the number density of CMB photons is
$n_{\gamma, 0} = 410~~\mbox{cm}^{-3}$.

The discovery of CMB has shown that
the Universe was hot at early times, and cooled down due to expansion.
As we pointed out,
the wavength of a photon increases  in time as $a(t)$, so the
energies and hence temperature of photons scale as 
\[
\omega(t) \propto a^{-1}(t) \; , \;\;\;\;\;\;\;
T(t) = \frac{a_0}{a(t)} T_0 = (1+z) T_0 \; .
\]
Importantly, the energy density of CMB photons scales as
\[
\rho_\gamma \propto T^4 \propto a^{-4} \;.
\]
This is in contrast with the scaling of energy density (mass density)
of non-relativistic particles (baryons, dark matter)
\[
\rho_M \propto a^{-3} \; ,
\]
which is obtained by simply noting that the mass in comoving volume
remains constant.

\subsection{Friedmann edquation}
\label{subsec:Friedeq}

The expansion of the spatially flat Universe
is governed by the Friedmann equation, 
\begin{equation}
    H^2 \equiv \left( \frac{\dot{a}}{a} \right)^2 
= \frac{8\pi}{3M_{Pl}^2}  \rho\; ,
\label{Friedmann}
\end{equation}
where  $\rho$ is the
{\it total} energy density in the Universe.
This is nothing but the $(00)$-component
of the Einstein equations of General Relativity,
$R_{\mu \nu} - \frac{1}{2}g_{\mu\nu}R = 8\pi T_{\mu \nu}$, 
specified to spatially flat FLRW metric and homogeneous and
isotropic matter.

One conventionally defines the parameter (critical density),
\be
\rho_c = \frac{3}{8\pi} M_{Pl}^2 H_0^2 \approx 5\cdot 10^{-6} ~
\frac{\mbox{GeV}}{\mbox{cm}^3} \; .
\label{rhoc-new}
\ee
It is equal to the sum of all forms
of energy density
in the {\it present} Universe.

\subsection{Present composition of the Universe}

The {\it present} composition of the Universe
is characterized by the parameters
\be
   \Omega_\lambda = \frac{\rho_{\lambda,0}}{\rho_{c}} \; .
\nonumber
\ee
where $\lambda$ labels various forms of energy:
relativistic matter  ($\lambda=rad$), non-relativistic matter
  ($\lambda=M$),
dark energy ($\lambda=\Lambda$).
Clearly,  eq.~(\ref{Friedmann}) gives
\[
\sum_\lambda \Omega_\lambda = 1\; . 
\]
Let us quote the numerical values:
\begin{subequations}
\begin{align}
\Omega_{rad} &= 8.6 \cdot 10^{-5} \; ,
\label{mar23-15-1}
\\
\Omega_M &= 0.31 \; ,
\\
\Omega_\Lambda &= 0.69 \; .
\label{mar13-15-11}
\end{align}
\end{subequations}
A point concerning
$\Omega_{rad}$ is in order.
Its value in
eq.~\eqref{mar23-15-1} is calculated for unrealistic case in which
{\it all neutrinos are relativistic today}, so the radiation component
even at present consists of CMB photons and three neutrino species.
This prescription is convenient
for studying the early Universe, 
since the energy density of relativistic neutrinos scales in the
same way as that of photons,
\[
\rho_\nu \propto T^4 \propto a^{-4} \; ,
\]
and at temperatures above neutrino masses (but below 1~MeV) we have
\[
\rho_\nu = 
=  \Omega_\nu \rho_c \left(\frac{a_0}{a}\right)^4 \; .
\]


Non-relativistic matter consists of baryons and dark matter.
Their contributions are~\cite{Aghanim:2018eyx}
\begin{subequations}
\label{nov12-19-14}
  \begin{align}
\Omega_B &= 0.049 \; ,
\\
\Omega_{DM} &=0.26 \; .
  \end{align}
  \end{subequations}

As we pointed out above, energy densities of various species
evolve as follows:
\begin{itemize}
\item radiation (photons and neutrinos at temperatures above
  neutrino mass):
\be
\rho_{rad} (t) = \left(\frac{a_0}{a(t)}\right)^4 \rho_{rad, 0}
= (1+z)^4 \, \Omega_{rad} \rho_c \; .
\label{mar23-15-3}
\ee
\item Non-relativistic matter:
  \be
\rho_M (t) = \left(\frac{a_0}{a (t)}\right)^3 \rho_{M, 0}
= (1+z)^3 \, \Omega_{M} \rho_c \; .
\label{mar23-15-4}
\ee
\item The dark energy density  does not change in time,
  or changes very slowly. In what follows we take
  it constant in time,
\be
\rho_\Lambda = \Omega_\Lambda \rho_c = \mbox{const} \; .
\label{mar23-15-5}
\ee
This assumption is not at all innocent. It means that dark energy
is assumed to be a cosmological constant. However,
even slight dependence of $\rho_\Lambda$ on time would mean that we
are dealing with something different from the cosmological constant.
In that case
the dark energy density would be associated with some field;
there are various theoretical  proposals concerning the properties of
this field. Present data are consistent with time-independent
$\rho_\Lambda$, but the precision of this statement is not yet very high.
It is extremely important to study the time-(in)dependence of
$\rho_\Lambda$ with high precision; several experiments are aimed at that.
\end{itemize}

\subsection{Cosmological epochs}

The Friedmann equation \eqref{Friedmann} is now written as
\begin{align*}
H^2(t) &= \frac{8\pi}{3 M_{Pl}^2} [\rho_\Lambda + \rho_M (t)
+ \rho_{rad}(t)]\\
&= H_0^2 \left[ \Omega_\Lambda +  \Omega_M
 \left(\frac{a_0}{a(t)}\right)^3
+ \Omega_{rad} \left(\frac{a_0}{a(t)}\right)^4 \right]
\end{align*}
This shows that the dominant term in the right hand side at early times
(small $a(t)$)
was $\rho_{rad}$, i.e., the expansion was dominated by ultrarelativistic
particles (radiation). This is radiation domination epoch. Then the term
$\rho_M$ took over, and matter dominated epoch began.
The redshift at radiation--matter equality, when the energy densities of 
radiation and matter were equal, is
\[
1+ z_{eq} = \frac{a_0}{a(t_{eq})} = \frac{\Omega_M}{\Omega_{rad}}
\approx 3500 \; ,
\]
and using the Friedmann equation one finds the age of the Universe at
equality
\[
t_{eq} \approx 50~000~\mbox{years} \; .
\]
The present Universe is at the end of the
transition from matter domination to $\Lambda$-domination:
the dark energy density $\rho_\Lambda$
will  completely dominate over non-relativistic
matter in future.

So, we have the following sequence
of the regimes of evolution:
\be
\dots \Longrightarrow \mbox{Radiation~domination}
\Longrightarrow \mbox{Matter~domination}\Longrightarrow 
\Lambda\mbox{--domination} \; .
\label{nov8-19-1}
\ee
Dots here denote some cosmological
epoch preceding the hot stage. We discuss this point
later on.

\subsection{Radiation domination}
\label{sub:RD}

\subsubsection{Expansion law}

The evolution of the scale factor at radiation
domination is obtained 
by using $\rho_{rad} \propto a^{-4}$ in the Friedmann equation
\eqref{Friedmann}:
\[
\frac{\dot{a}}{a} = \frac{\mbox{const}}{a^2} \; .
\]
This gives 
\be
a (t) = \mbox{const} \cdot \sqrt{t} \; .
\label{sep13-11-5}
\ee
The constant here does not have physical significance,
as one can rescale the coordinates ${\bf x}$ at one moment
of time, thus changing the normailzation of $a$.

There are several properties that immediately follow from the result
\eqref{sep13-11-5}. First, the expansion {\it decelerates}:
\[
\ddot{a} < 0 \; .
\]
Second, time $t=0$ is the Big Bang singularity
(assuming, for the sake of argument,
that the Universe starts right from radiation domination epoch).
The expansion rate
\[
H(t) = \frac{1}{2t}
\]
diverges as $t \to 0$, and so does the energy
density $\rho(t) \propto H^2 (t)$ and temperature
$T \propto \rho^{1/4}$. This is ``classical'' singularity
(singularity in classical General Relativity) which,
one expects, is resolved in one or another way in complete
quantum gravity theory. One usually assumes (although this is not
necessarily correct) that the classical expansion begins just after the
Planck epoch,
when $\rho \sim M_{Pl}^4$, $H \sim M_{Pl}$, etc.

\subsubsection{Particle horizon}

The third observation has to do with the causal structure of space-time
in the Hot Big Bang Theory (theory that assumes that the evolution starts from
the singularity directly into radiation domination --- no dots is
\eqref{nov8-19-1}).
Consider signals emitted right after the Big Bang singularity
and travelling at the speed of light. The
light cone obeys
$ds=0$, and hence $a(t) dx = dt$. So, the coordinate distance
that a signal travels from the Big Bang to time $t$ is
\be
x = \int_0^{t} \frac{dt}{a(t)} \equiv \eta \; .
\label{sep13-11-6}
\ee
In the radiation dominated Universe 
\[
\eta = \mbox{\const} \cdot \sqrt{t} \; .
\]
The physical distance from the emission point to
the position of the signal is
\be
l_{H}( t) = a(t) x = a(t) \int_0^{t} \frac{dt}{a(t)} \; .
\label{nov8-19-2}
\ee
This physical distance is finite; it 
is the size of a causally connected region at time $t$.
It is called the horizon size (more precisely, the
size of particle horizon).  In other words, 
an observer at time $t$ can have information only on the part
of the Universe whose physical size at that time is $l_{H} (t)$.
At radiation domination, one has
\[
l_H (t) = 2t \; .
\]
Note that this horizon size is of order of the Hubble size,
\be
l_{H}(t) \sim H^{-1}(t) \; .
\label{mar23-15-7}
\ee
The notion of horizon is straightforwardly extended to matter
dominated epoch and to the present time: relation \eqref{nov8-19-2}
is of general nature, while the scale factor $a(t)$ has to be
calculated anew.
To give an idea of numbers, the horizon size at 
the present epoch is
\[
l_{H}(t_0) \approx 14~\mbox{Gpc} \simeq 4 \cdot 10^{28}~\mbox{cm}
\; .
\]

\subsubsection{Energy density}

At radiation domination, cosmic plasma is almost always in thermal
equilibrium, and interactions between particles are almost always weak.
So, the plasma properties are determined by thermodynamics of
a gas of free relativistic particles.
At different times, the number of relativistic species that
contribute into energy density, is different. As an example,
at temperatures above 1~MeV but below 100~MeV, relativistic are
photons, three types of neutrinos, electrons and positrons,
while at temperatures of about 200~GeV all Srtandard Model
particles are relativistic. In most cases, one can neglect
chemical potentials, i.e., consider cosmic plasma symmetric
under interchange of
particles with antiparicles (chemical potentaial of photons is zero,
since photons can be created in processes like $e^- e^- \to
e^- e^- \gamma$; since particle and its antiparticle can annihilate into
photons, e.g., $e^+ e^- \to \gamma \gamma$,
chemical potentials of particles and antiparticles
are equal in modulus and opposite in sign, e.g., $\mu_{e^+}
= - \mu_{e^-}$; in symmetric plasma  $\mu_{e^+}
= - \mu_{e^-} = 0$). Then the
Stefan--Boltzmann law gives 
for the energy density 
\be
\rho_{rad} = \frac{\pi^2}{30} g_* T^4 \; ,
\label{sep13-11-2}
\ee
where $g_*$ is the effective number of degrees of freedom,
\[
g_* = \sum_{bosons} g_i + \frac{7}{8}\sum_{fermions} g_i \; ,
\]
$g_i$ is the number of spin states of a particle $i$,
the factor 7/8 is due
to Fermi-statistics. The parameter $g_*$ depends on temperature, and
hence on time: as the temperature decreases below mass of a particle,
this particle drops out from the sum here.
The formula \eqref{sep13-11-2}
enables one to write
the Friedmann equation \eqref{Friedmann}
as 
\be
H = \frac{T^2}{M_{Pl}^*} \; , \;\;\;\;\; M_{Pl}^* = \frac{M_{Pl}}{1.66 \sqrt{g_*}}
\; .
\label{mar25-15-10}
\ee
We use this simple result in what follows.

\subsubsection{Entropy}

The cosmological expansion 
is slow, which implies conservation of entropy
(modulo quite exotic scenarios with
large entropy generation). 
The entropy density of free relativistic
gas in thermal equilibrium is given by
\[
s = \frac{2\pi^2}{45} g_* T^3 \; .
\]
The conservation of entropy means that
the entropy density scales {\it exactly} as $a^{-3}$,
\be
sa^3 = \mbox{const} \; ,
\label{sep13-11-1}
\ee
while 
temperature scales {\it approximately} as $a^{-1}$
(this is because $g_*$ depends on time). We note for future reference that
the effective number of degrees of freedom
in the Standard Model at $T \gtrsim 100$~GeV
is 
\[
g_* (100~\mbox{GeV}) \approx 100 \; .
\]
The present  entropy density in the Universe, still with the prescription that
neutrinos are relativistic, is
\be
s_0 \approx 3000~\mbox{cm}^{-3} \; .
\label{e-density}
\ee
The precise meaning of this number is that at high temperatures
(when there is thermal equilibrium), the entropy density is
$s (t) = (a_0/a(t))^3 s_0$.

Notion of entropy is convenient, in particular, for characterizing
asymmetries which can exist 
if there are conserved quantum numbers, such as the baryon number
after baryogenesis. The density of a conserved number
also scales as $a^{-3}$, so the time independent
characteristic of, say, the baryon abundance is the baryon-to-entropy ratio
\[
\Delta_B = \frac{n_B}{s} \; .
\]
At late times, one can use another parameter, 
baryon-to-photon ratio
\be
\eta_B= \frac{n_B}{n_\gamma}
\; ,
\label{nov10-19-1}
\ee
where $n_\gamma$ is photon
number density.
It
is related to $\Delta_B$ by a numerical factor, but this factor
depends on time through $g_*$ and stays constant only after 
$e^+ e^-$-annihilation, i.e., at $T \lesssim 0.5$~MeV. Numerically,
\be
\Delta_B = 0.14 \eta_{B, 0} = 0.86 \cdot 10^{-10} \; .
\label{mar28-15-1}
\ee
In what follows we  discuss the ways to obtain this number
from observations.

\subsection{Matter domination}

At matter domination, we have $\rho \propto a^{-3}$, and the Friedmann equation
\eqref{Friedmann} gives
\[
a(t) = \mbox{const} \cdot t^{2/3}
\]
Qualitatively, matter domination is similar to radiation domination:
expansion is decelerated, the size of particle horizon is of order
of the Hubble size, $l_{H} (t) \sim H^{-1} (t) \sim t$. An important
difference between radiation and matter dominated epochs is that
inhomogeneities in energy desity (``scalar perturbations'') grow rapidly
at matter domination and slowly at radiation domination. Thus, matter
domination is the epoch of structure formation in the Universe.

\subsection{Dark energy domination}
\label{sec:de}

The expansion of the Universe is accelerated today.
Within General Relativity this is attributed to
dark energy.
 We know very little
 about this ``substance'': we know its energy density,
 eq.~\eqref{mar13-15-11},
 and also know that this energy density changes in time very
 slowly, if at all. The latter fact is quantified
 in the following way. Let us denote by $p$ the effective pressure, i.e.,
 spatial component of the energy-momentum tensor in locally-Lorentz frame
 $T_{\mu \nu} =\mbox{diag} (\rho, p, p, p)$. Then covariant conservation of
 the energy-momentum in expanding Universe gives for any fraction
 that does not interact with other fractions
 \[
\dot{\rho} = -3 \frac{\dot{a}}{a} (\rho + p) 
\]
(note that relativistic and non-relativistic  matter have
$p=\rho/3$ and $p=0$, respectively, so this equation gives
for them $\rho \propto a^{-4}$ and $\rho \propto a^{-3}$, as it should).
A simple parametrization of time-dependent dark energy
is $p_\Lambda = w_\Lambda \rho_\Lambda$ with time-independent $w_\Lambda$.
The combination of comsological data gives~\cite{Aghanim:2018eyx}
\be
w_\Lambda \approx -1.03 \pm 0.03 \; .
\label{mar23-15-12}
\ee
Thus, with reasonable precision one has  $p_\Lambda = - \rho_\Lambda$,
which corresponds to time-independent dark energy density.

\begin{figure}[!htb]
\hskip 0.2\textwidth
\includegraphics[width=0.6\linewidth, angle=90]{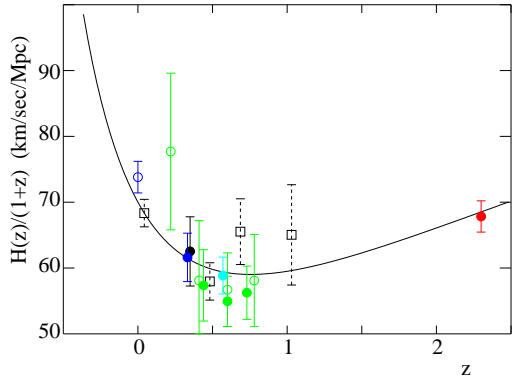}
\caption{Observational data on the time derivative of the scale factor
  as function of redshift $z$~\cite{Busca:2012bu}. The change
  of the behavior from
decreasing to increasing, as
$z$ decreases, means the change from decelerated to
accelerated expansion. The theoretical curve corresponds to
spatially flat Universe with $h=0.7$ and $\Omega_\Lambda = 0.73$. 
\label{adot}
}
\end{figure}

The solution
to the Friedmann equation \eqref{Friedmann} with constant
$\rho = \rho_\Lambda$ is
\[
a (t) = \e^{H_\Lambda t} \; ,
\]
where $H_\Lambda = (8\pi \rho_\Lambda/3M_{Pl}^2)^{1/2} = \mbox{const}$.
This gives accelerated expansion,
$\ddot{a} > 0$, unlike at radiation or matter domination.
The transition from
decelerated (matter dominated)
to accelerated expansion (dark energy dominated)
has been confirmed quite some time ago by combined observational
data, see Fig.~\ref{adot},
which shows the dependence on redshift of the quantity
$H(z)/(1+z)=\dot a(t)/a_0$. 

\vspace{-4cm}

In the case of the cosmological constant,
energy-momentum tensor is proportional to metric, and
in locally-Lorentz frame it reads
\[
T_{\mu \nu} = \rho_\Lambda \eta_{\mu \nu} \; ,
\]
where $\eta_{\mu \nu}$ is the Minkowski tensor. Hence $w_\Lambda = -1$.
One can view this as the 
characteristic of vacuum, whose energy-momentum tensor must
be Lorentz-covariant. As we pointed out above,
any deviation from $w=-1$ would mean that we are dealing
with something other than vacuum energy density.

The problem with dark energy is that 
its present value is extremely small by particle physics standards, 
\begin{eqnarray*}
\rho_{DE}\approx 4~\mbox{GeV/m$^3$}=(2\times10^{-3}\mbox{eV})^4\,.
\end{eqnarray*}
In fact, there are two hard problems.
One is that the dark energy density is zero to an excellent approximation. 
Another is that
it is non-zero nevertheless, and one has to understand its energy scale.
We are not going to discuss these points anymore,
and only emphasize that we are not aware of a compelling mechanism
that solves any of the two cosmological constant problems
(with possible exception of anthropic argument  due to Weinberg 
and Linde \cite{Weinberg:1987dv,Linde:1986dq}).

\section{Cornerstones of thermal history}

\subsection{Recombination~=~photon last scattering}

Going back in time, we reach so high temperatures that 
the usual matter (electrons and protons 
with rather small admixture of light nulei, mainly $^4$He) 
is in the plasma phase. In plasma,
photons interact with electrons 
due to the Thomson scattering
and protons have Coulomb interaction with electrons. 
These interactions are strong enough to keep photons, electrons and protons
in thermal equilibrium. 
When the temperature drops to
\[
T_{rec} \approx 3000~{\mbox K} \; , \;\;\;\;\;\; z_{rec} \approx 1090 \; ,
\]
almost all electrons ``recombine'' with protons into
neutral hydrogen atoms (helium recombined earlier).
The number density of atoms at that time is  quite small,
$250~\mbox{cm}^{-3}$, so
from that time on, 
the Universe is transparent to photons\footnote{Modulo effects of
  re-ionization that occured much later and affected a small
  fraction of CMB photons.}.
Thus, $T_{rec}$ is
{\it photon last scattering temperature}.
At that time
the age of the Universe is $t_{rec} \approx 380$~thousand years
(for comparison, its present age is about 13.8 billion years).

CMB photons give us (literally!)
the photographic picture of the Universe at photon last scattering
epoch. 
The last scattering epoch lasted considerably shorter
than the then Hubble time
$H^{-1}(t_{rec}) \sim t_{rec}$; to a meaningful (although rather
crude)
approximation,
recombination occured instantaneously. This is important, since in the
opposite case of long recombination, the photographic picture
would be strongly washed out.

\begin{figure}[htb!]
\begin{center}
\includegraphics[width=0.7\textwidth]{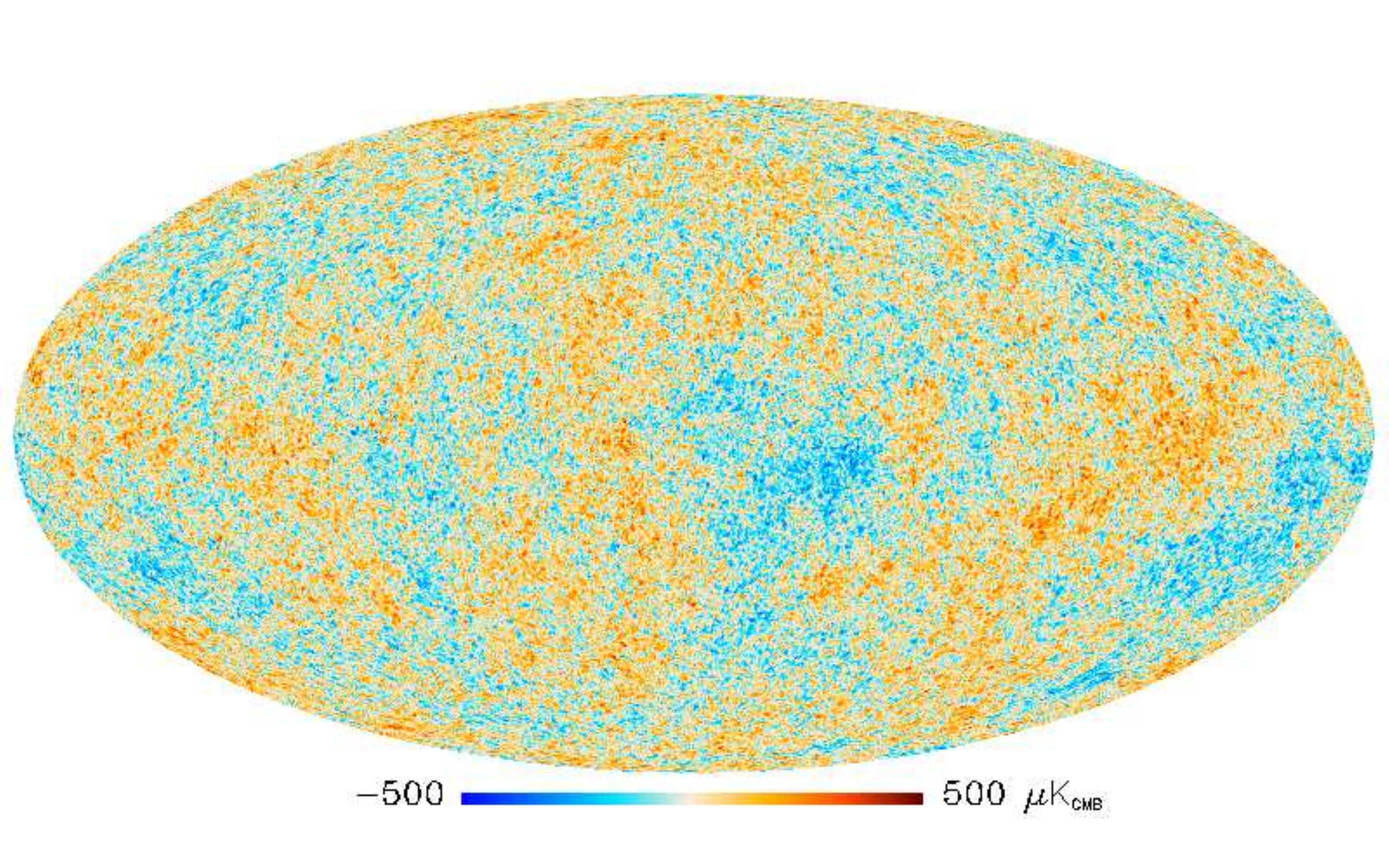}
\end{center}
\caption{\label{Planck-sky}  CMB sky as seen by Planck.
 }
\end{figure}

This photographic picture is shown in Fig.~\ref{Planck-sky}.
Here brighter (darker) regions correspond to higher (lower)
temperatures. The relative temperature fluctuation is of order
$\delta T/T = 10^{-4} - 10^{-5}$, so the 380 thousand year old
 Universe was much more homogeneous than today.
 
One performs Fourier decomposition of the temperatue fluctuations,
i.e., decomposition in spherical harmonics:
\[
    \frac{\delta T}{T}        (\theta, \varphi)
    = \sum_{l,m} a_{lm} Y_{lm}(\theta, \varphi) \; .
    \]
    Here
    $a_{lm}$ are independent Gaussian random variables (no
    non-Gaussianities have been found so far) with
 $\langle a_{lm} a^*_{l'm'} \rangle \propto \delta_{ll'} \delta_{mm'}$
and
$ \langle a^*_{lm} a_{lm} \rangle =  C_{l}$. The multipoles
$C_l$, or, equivalently,
\[
D_l = \frac{l(l+1)}{2\pi} C_l 
\]are the main quantities of interest.
The larger $l$, the smaller angular scales, hence the shorter
wavelengths of density perturbations producing the temperature
anisotropy.

It is worth noting that averaging here is understood in terms of
an ensemble of Universes, while we have just one Universe. So, there is
inevitable uncertainty in $C_l$, called
cosmic variance. For given $l$, one has $(2l+1)$ quantities
$a_{lm}$, $m=0, \pm 1, \dots , \pm l$, so the uncertainty is
$\Delta C_l/C_l \simeq 1/{\sqrt{2l}}$.

\begin{figure}[htb!]
\begin{center}
\includegraphics[width=0.7\textwidth]{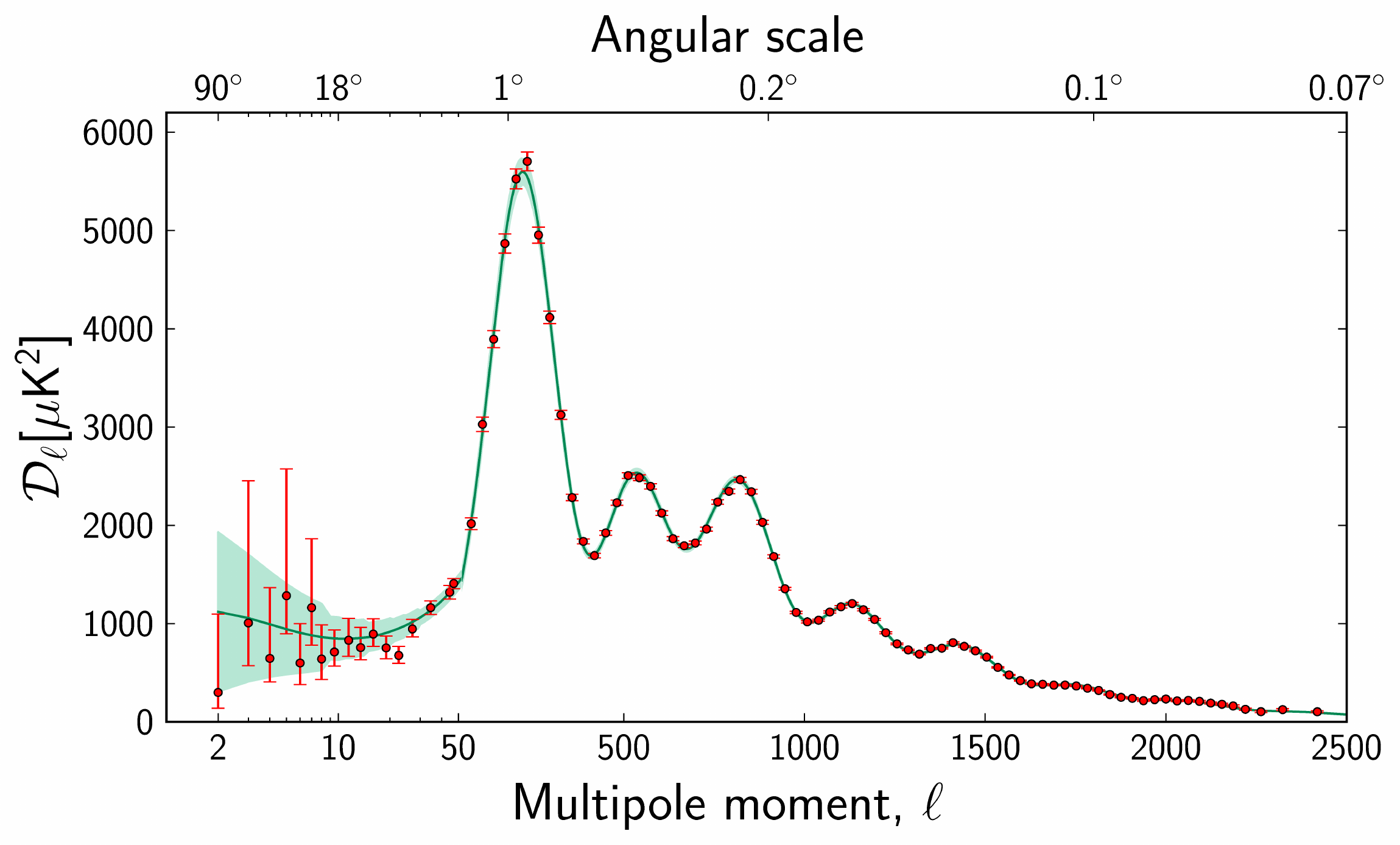}
\end{center}
\caption{\label{Planck-Dl}  Multipoles $D_l$ as measured by Planck.
 }
\end{figure}

CMB temperature multipoles are shown in Fig.~\ref{Planck-Dl}
(error bars there are due to cosmic variance, not the measurement errors).
Also measured are CMB polarization multipoles and
temperature-polarization cross-correlation multipoles.
There is a lot of physics behind these quantities,
which has to do with
\begin{itemize}
\item primordial perturbations: the perturbations that
  are built in already at the beginning of the hot cosmological epoch,
  see Sec.~\ref{sec:pertu};
\item development of sound waves in cosmic plasma from the 
 early hot stage to recombination; 
  gravitational potentials due to dark matter 
  at recombination (which are sensitive to the
  composition of cosmic medium);
\item propagation of photons after 
recombination (which is
sensitive to
expansion history of the Universe and structure formation).
\end{itemize}
Clearly, CMB measurements are a major source of the cosmological
information. We come back to CMB  in due course.

\subsection{Big Bang Nucleosynthesis}

As we go back in time further, we get to 
the temerature in
the Universe in MeV range. The epoch characterized by temperatures
1~MeV --- 30 keV is the epoch of
Big Bang Nucleosynthesis. That epoch starts 
at temperature 1~MeV, when the age of the Universe
is 1~s. At temperatures above 1~MeV, there are rapid
weak processes like
\be
e^- + p \longleftrightarrow n + \nu_e \; .
\label{mar22-15-2}
\ee
These processes keep neutrons and protons in chemical equilibrium;
the ratio of their number densities is determined by
the Boltzmann factor,
$
n_n/n_p = \mbox{exp}\left(-\frac{m_n - m_p}{T}\right)$.
At 
$T_{n} \approx 1$~MeV neutron-proton transitions \eqref{mar22-15-2} 
switch off,  and neutron-proton ratio is frozen out at the value
\[
\frac{n_e}{n_p} = \e^{-\frac{m_n - m_p}{T_{n}}} \; .
\]

Interestingly,  
$m_n - m_p \sim T_n$, so the neutron-proton ratio
at neutron freeze-out and later was neither equal to 1, nor very small.
Were it equal to 1, protons would in the end
combine with neutrons into 
$^4$He, and there would remain no hydrogen in the
Universe. On the other hand, for very small $n_n/n_p$, too few light nuclei
would be formed, and we would not have any observable
remnants of the BBN epoch. In either case the Universe would be quite
different from what it actually is. It is worth noting that the approximate
relation $m_n - m_p \sim T_n$ is a coincidence: 
$m_n - m_p$ is determined by light quark masses and electromagnetic
coupling, while $T_n$ is determined by the strength of
weak interactions (the rates of the processes \eqref{mar22-15-2})
and gravity (the expansion of the Universe).
This is one of numerous coincidences we encounter in cosmology.

\begin{figure}[htb!]
\begin{center}
\includegraphics[width=0.8\textwidth,angle=0]{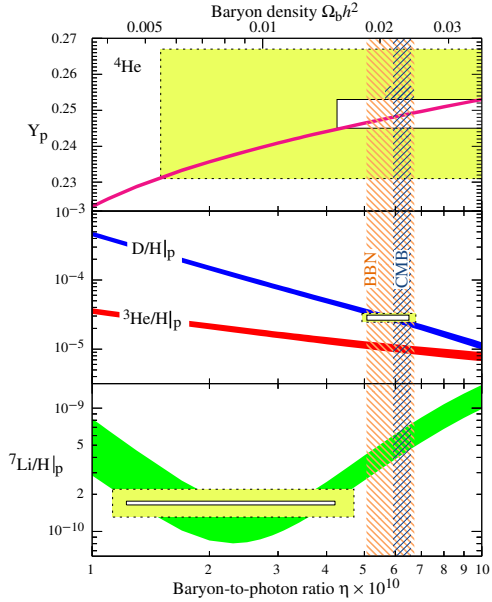}
\end{center}
\vspace{-4cm}
\caption{Abundances of light elements, measured (boxes;
larger boxes include systematic uncertainties) and calculated
as functions of baryon-to-photon ratio $\eta$~\cite{PDG-11}.
The determination of $\eta \equiv \eta_B$ from BBN (vertical range marked BBN)
is in excellent agreement with the determination from the analysis
of CMB temperature fluctiations  (vertical range marked CMB).
 \label{bbn09}
 }
\end{figure}

At temperatures 100 -- 30~keV,  
neutrons combined with protons into
light elements in thermonuclear reactions 
\begin{eqnarray}
  p + n &\to& ^2\mbox{H} + \gamma\; ,
\nonumber\\
  ^2\mbox{ H} + p &\to& ^3\mbox{ He} + \gamma \; ,
\nonumber\\
   ^3\mbox{He} + ^2\mbox{H} &\to&  ^4\mbox{ He} + p  \; ,
\end{eqnarray}
etc., up to $^7$Li. The abundances of light elements
have been measured, see Fig.~\ref{bbn09}.
The only parameter relevant for calculating
these abundances (assuming negligible neutrino-antineutrino asymmetry)
is the baryon-to-photon ratio 
$\eta_B \equiv \eta$, see eq.~\eqref{nov10-19-1}, which
determines the number
density of baryons. 
Comparison of the Big Bang Nucleosynthesis
theory with the observational determination of the composition
of cosmic medium enables one to determine $\eta_B$ and check the
overall consistency of the BBN picture. It is even more reassuring
that a completely
independent measurement of $\eta_B$ that makes use of the CMB
temperature fluctuations is in excellent agreement with BBN.
Thus, BBN 
gives us confidence that we understand the
Universe at $T\sim 1$~MeV, $t \sim 1$~s. 
In particular, we are convinced that
the cosmological expansion was
governed by General Relativity.

\vspace{-4cm}

\subsection{Neutrino decoupling}

Another class of processes of interest at temperatures in the
MeV range is neutrino production, annihilation and scattering,
\[
\nu_\alpha + \bar{\nu}_\alpha \longleftrightarrow e^+ + e^-
\]
and crossing processes. Here the subscript $\alpha$ labels neutrino flavors.
These processes switch off at $T \sim 2 -3$~MeV, depending on
neutrino flavor. Since then neutrinos do not interact with cosmic medium
other than gravitationally,
but they do affect the properties of CMB and distribution of galaxies
through their gravitational interactions. Thus, observational 
data can be used to establish, albeit somewhat indirectly,
the existence of relic neutrinos and set limits on neutrino masses.
We quote here the limit reported by Planck
collaboration~\cite{Aghanim:2018eyx}
\[
\sum m_\nu <0.12~\mbox{eV} \; ,
\]
where the sum runs over the three neutrino species.
Other analyses give somewhat weaker limits. Also, the data can be used to
determine the effective number of neutrino species that counts the 
number of relativistic degrees of freedom~\cite{Aghanim:2018eyx}:
\[
N_{\nu, \, eff} = 2.99 \pm 0.17 \; ,
\]
which is consistent with the Standard Model value $N_\nu = 3$.
 We see that cosmology {\it requires}
 relic neutrinos.

\section{Dark matter: evidence}
\label{sec:dm}

Unlike dark energy,
dark matter experiences the same gravitational force as
the baryonic matter. Dark matter is discussed in numerous reviews,
see, e.g.,
Refs.~\cite{dark-rev,Roszkowski:2017nbc,Arcadi:2017kky,Bullock:2017xww}.
It
consists presumably of new stable massive
particles. These make clumps of mass 
which constitute 
most of the mass of galaxies and  
clusters of galaxies.
Dark matter  is characterized by the mass-to-entropy ratio,
\be
   \left( \frac{\rho_{DM}}{s} \right)_0 = \frac{\Omega_{DM} \rho_c}{s_0}
\approx 
\frac{0.26 \cdot 5 \cdot 10^{-6}~ \mbox{GeV} \cdot \mbox{cm}^{-3}}{3000 
~\mbox{cm}^{-3}} = 4 \cdot 10^{-10}~ \mbox{GeV} \; .
\label{10p*}
\ee
This ratio is constant in time since the freeze out of
dark matter density: both number density of dark matter particles
$n_{DM}$ (and hence their mass density 
$\rho_{DM}=m_{DM} n_{DM}$) and entropy density
decrease exactly as $a^{-3}$. 

There are various ways of measuring the contribution of
non-baryonic dark matter into the total energy density of various objects and
the
Universe as a whole.

\subsection{Dark matter in galaxies}

\begin{figure}[b!]
\begin{center}
\hskip 0.05\textwidth
\includegraphics[width=0.5\textwidth]{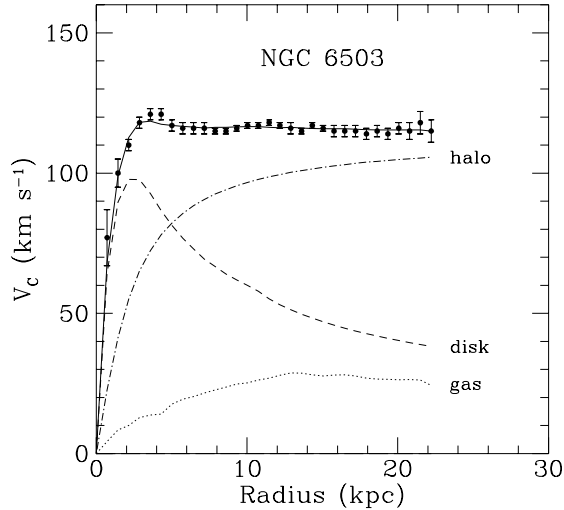}
\end{center}
\vspace{-2cm}
\caption{\label{rotation-curve} 
Rotation velocities of hydrogen gas clouds around a galaxy
NGC 6503~\cite{Extended-curves}. Lines show the
contributions of the three main components that produce the
gravitational potential. The main contribution at large distances
is due to dark matter, labeled ``halo''.
}
\end{figure}

Dark matter exis in galaxies. Its distribution is
measured by the observations of rotation velocities of distant stars
and gas clouds around a galaxy, Fig.~\ref{rotation-curve}.
If the mass was concentrated in a luminous central part of a galaxy,
the velocities of objects away from the central part would decrease with
the distance $r$ to the center as
$v \propto r^{-1/2}$ -- this immediately follows from the second Newton's
law
\[
\frac{v^2}{r} = G \frac{M(r)}{r^2} \; .
\]
In reality, 
rotation curves are typically flat up to distances
exceeding the size of the bright part by a factor of 10 or so. 
The fact that dark matter halos are so large is explained by
the defining property of dark matter particles: they do not lose their
energies by emitting photons, and, in general, interact with conventional 
matter very weakly.

\subsection{Dark matter in clusters of galaxies}

\begin{figure}[htb]
  \vspace{-2.5cm}
\centerline{\includegraphics[width=0.39\textwidth]{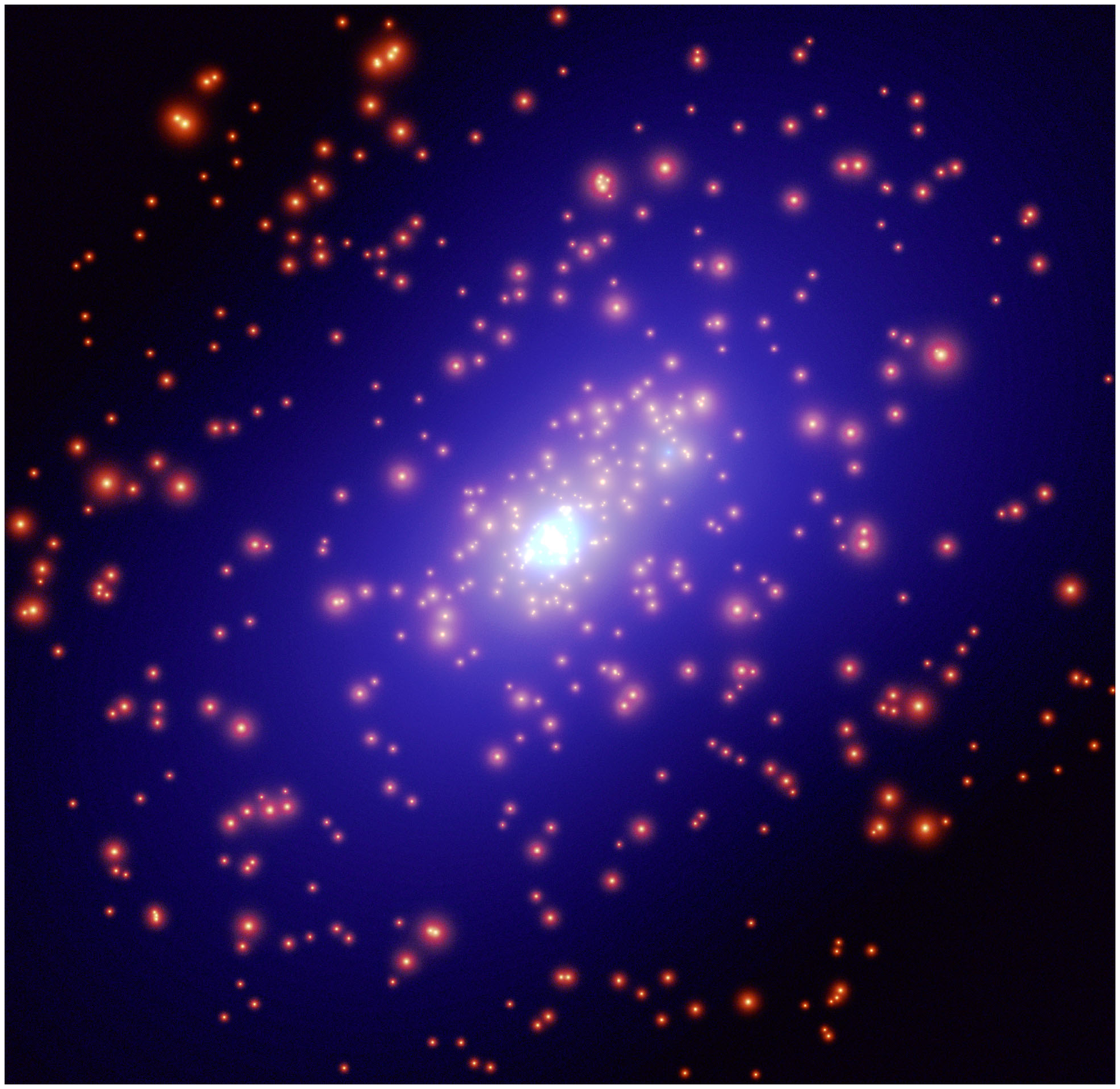}
\includegraphics[width=0.329\textwidth]{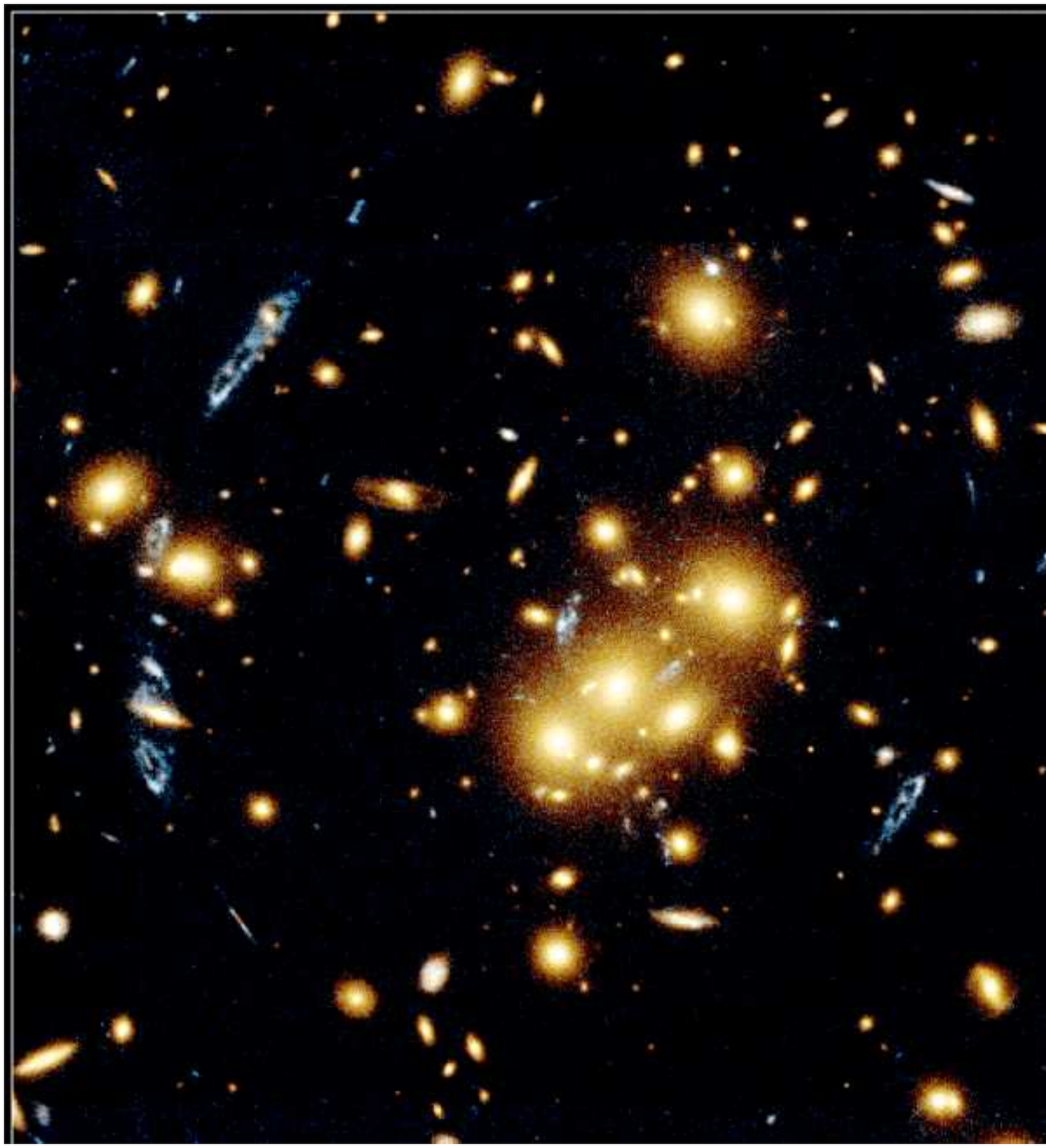}}
\caption{\label{grav-lens-dark} 
Cluster of galaxies
CL$0024+1654$~\cite{CL0024+1654}, acting as gravitational lens. 
Right panel: cluster in visible light.
Round yellow spots are galaxies in the cluster.
Elongated blue strips are images of one and the same galaxy behind the
cluster. Left panel: reconstructed distribution of gravitating mass in
the cluster; brighter regions have larger mass density.}
\end{figure}

Dark matter makes most of the mass of  the largest 
gravitationally bound objects -- clusters of
galaxies.  There are various methods to determine the gravitating mass
of a cluster, and mass distribution in a cluster, which give
consistent results. These include measurements of rotational
velocities of galaxies in a cluster (original Zwicky argument that
goes back to 1930's), measurements of temperature of hot gas
(which actually makes most of baryonic matter in clusters),
observations of 
gravitational lensing of
extended light sources (galaxies) behind the cluster,
see  Fig.~\ref{grav-lens-dark}. All these determinations show that 
baryons (independently measured through their X-ray emission)
make less than 1/4 of total mass in clusters. The rest is dark matter.


\begin{figure}[htb!]
  \vspace{-3cm}
\centerline{
\includegraphics[width=0.4\textwidth]{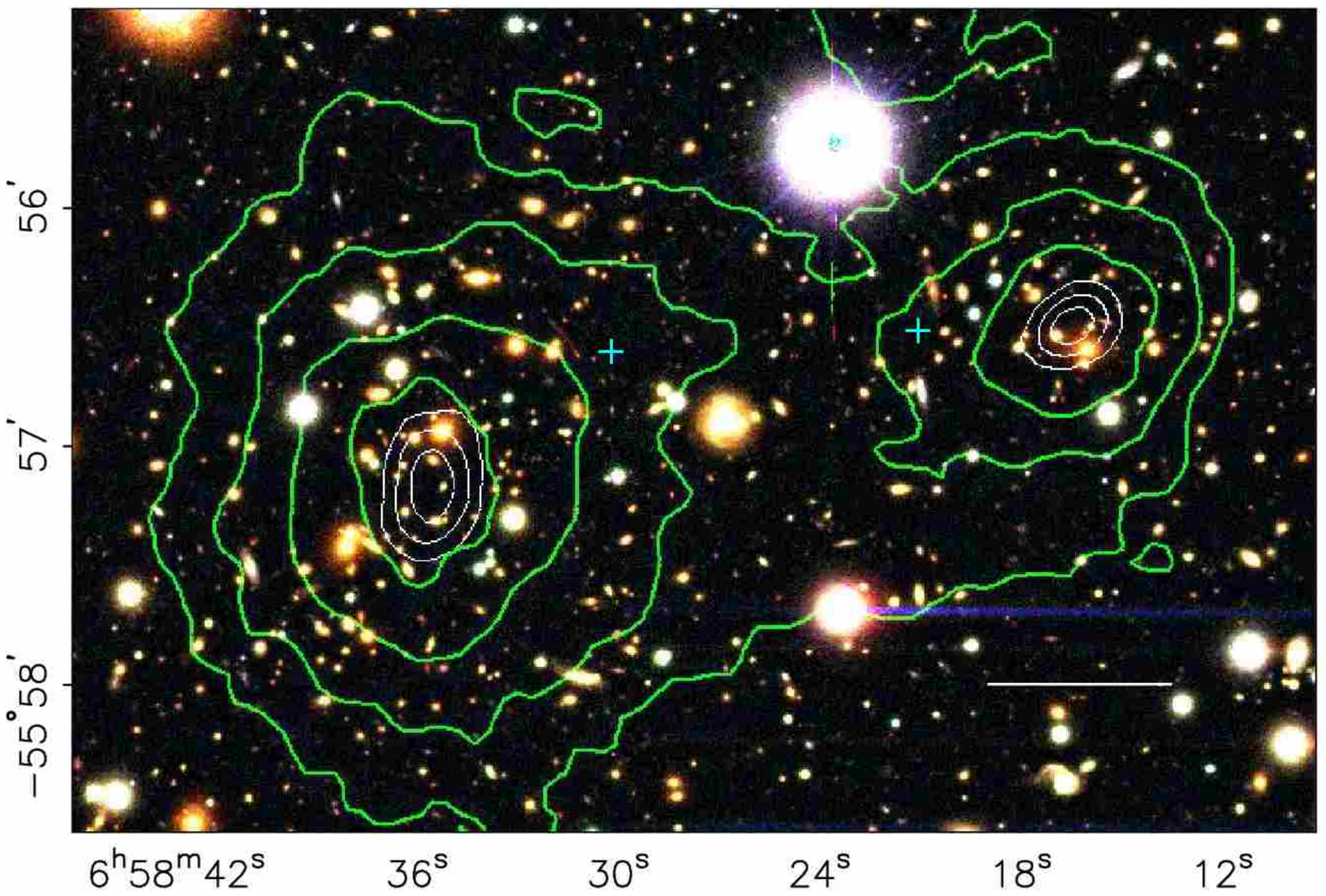}
\includegraphics[width=0.4\textwidth]{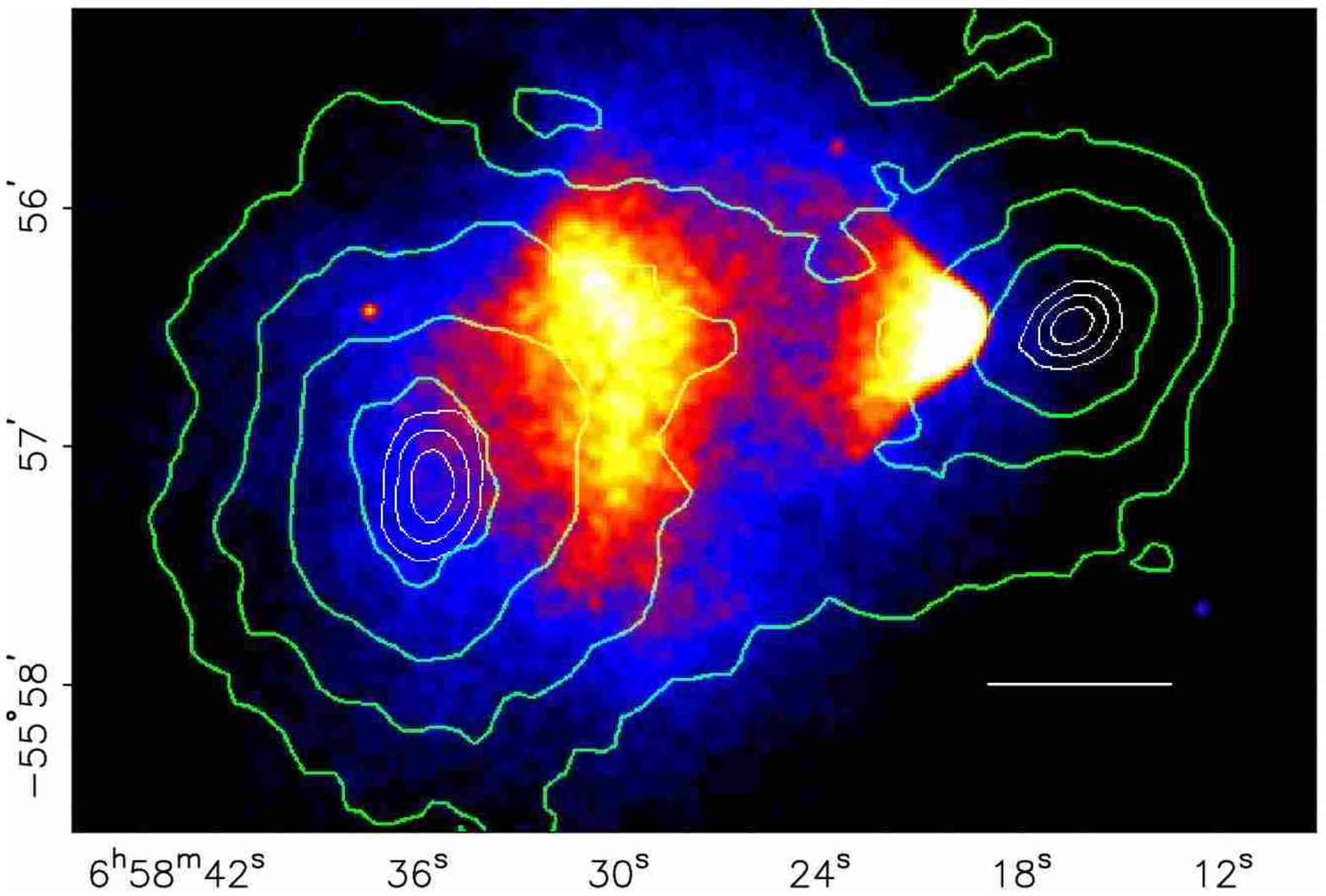}
}
\caption{ Observation~\cite{Clowe:2006eq} of the Bulet Cluster
1E0657\hbox{--}558 at $z=0.296$. Closed lines show the gravitational potential
produced mainly by dark matter and measured through gravitational
lensing. Bright regions in the right panel
show X-ray emission of hot baryon gas,
which makes most of the baryonic matter in the clusters. The length
of white
interval is 200~kpc in comoving frame.
\label{colliding-clusters}
}
\end{figure}

Concerning galaxies and
clusters of galaxies, we note that there are attempts to
attribute the properties of rotation curves and other phenomena,
which are usually considered as evidence for dark matter, to
modification of gravity, and in this way
get rid of dark matter altogether. There are several strong arguments that
rule out this idea. One argument has to do with 
the Bullet Cluster, 
Fig.~\ref{colliding-clusters}. Shown are two galaxy clusters that
passed through each other. The dark matter and galaxies do not 
experience friction and thus do not lose their velocities.
On the contrary baryons in hot, X-ray emitting gas do experience
friction and hence get slowed down and lag behind dark matter and galaxies.
In this way the baryons (which are mainly in hot gas) and dark matter
are separated in space. Since the baryonic mass
and gravitational potentials are not concentric, one cannot
attribute gravitational potentials solely to baryons, even assuming the
modification of Newton's gravity law. As a remark, the fact that dark matter
moves after cluster collision considerably faster than baryonic gas means that
elastic scattering between dark matter particles is weak. Quantitatively, the
limit on the dark matter elastic scattering cross section is
\be
\sigma^{(el)}_{DM-DM} < 1 \cdot 10^{-24}~\mbox{cm}^2 \; .
\label{nov15-19-3}
\ee
This limit is not particularly strong, but it does rule out part of the
parameter space of strongly interacting massive particle (SIMP) dark matter
models, see Sec.~\ref{subs:hints}.

\subsection{Dark matter imprint in CMB}
\label{sec:imprint}

Composition of the Universe strongly affects the CMB angular anisotropy 
and polarization. Before recombination,
the energy  density perturbation is a sum of perturbation in
baryon-electron-photon component and dark matter component,
\[
\delta \rho = \delta \rho_B + \delta \rho_{DM}
\]
(we simplify things here, as there is also perturbation in
gravitational potentials induced by density perturbation).
The
tightly coupled
baryon-electron-photon plasma has high pressure (due to photon component
with $p_\gamma = \rho_\gamma/3$), so density perturbations  in this
  fraction undergo acoustic oscillations: every Fourier mode
oscillates in time as
\be
\delta \rho_B ({\bf k}, t) = A({\bf k}) \mbox{cos} \left(
\int_0^t v_s  \frac{k}{a(t)} dt \right) \; ,
\label{nov12-19-10}
\ee
where ${\bf k}$ is comoving momentum (and ${\bf k}/a(t)$ is physical
momentum which gets redshifted), $v_s \approx 1/\sqrt{3}$ is sound speed,
and $A({\bf k})$ is the amplitude that varies slowly with $k$ (in
statistical sense: $\delta \rho ({\bf k})$ is Gaussian random field).
We comment in Sec.~\ref{sec:pertu}
on the fact that the phase of cosine in \eqref{nov12-19-10}
is well defined. 
On the contrary, dark matter is pressureless, so its perturbation is
almost independent of time,
\[
\delta \rho_{DM} \approx \delta \rho_{DM} ({\bf k}) \; ,
\]
where $ \delta \rho_{DM} ({\bf k})$ slowly varies with $k$.
At recombination time $t_r$, the energy density perturbation is a sum
\be
\delta \rho ({\bf k}, t_r) =  A({\bf k}) \mbox{cos} \left(
\int_0^{t_r} v_s  \frac{k}{a(t)} dt \right) +  \delta \rho_{DM} ({\bf k}) \; .
\label{nov12-19-11}
\ee
The first term here oscillates {\it as function of $k$}, while the second
term is a smooth, non-oscillating function of $k$.

Now, behavior of $\delta \rho (t_r)$ as function of
spatial momentum $k$ translates into behavior of CMB temperature
fluctuation $\delta T$ as function of multipole number $l$.
$\delta T$ at a given point in space at recombination epoch
is proportional to $\delta \rho$ (here we again simplify things,
this time quite considerably). We see CMB coming from
a {\it photon last scattering sphere}; smaller angular scale in
this photographic picture corresponds to smaller spatial scale
at recombination epoch, hence larger multipole $l$ corresponds
to higher three-momentum $k$. Thus, oscillatory formula \eqref{nov12-19-11}
translates into oscillatory behavior in Fig.~\ref{Planck-Dl}.
Both oscillatory part of temperature angular spectrum
(which is due to the first, baryonic term in \eqref{nov12-19-11})
and smooth part (due to the second, dark matter  term in \eqref{nov12-19-11})
are clearly visible in  Fig.~\ref{Planck-Dl}. The detailed  analysis of this
angular spectrum 
enables one to determine both baryon content
and dark matter content in the Universe, $\Omega_B$ and $\Omega_{DM}$
quoted in \eqref{nov12-19-14}.

\subsection{Dark matter and structure formation}

Dark matter is crucial for our existence, for the following reason.
As we discussed above, density perturbations in
baryon-electron-photon plasma before recombination
do not grow because of high pressure; 
instead,  they oscillate 
with time-independent
amplitudes.
Hence, in a Universe without dark matter, density perturbations
in baryonic component would start to grow only after baryons decouple from
photons, i.e., after recombination. The mechanism of the growth is
qualitatively simple: an overdense region gravitationally attracts
surrounding matter; this matter falls into the overdense region,
and the density contrast increases. In the expanding, matter dominated
Universe this gravitational instability results in the density contrast
growing like $(\delta \rho/\rho) (t) \propto a(t)$. Hence, in a Universe
without dark matter, the growth factor for baryon density perturbations
would be at most\footnote{Because of the presence of dark energy,
the growth factor is even somewhat smaller.} 
\be
   \frac{a(t_0)}{a(t_{rec})} = 1+z_{rec} = \frac{T_{rec}}{T_0} \approx 10^3 \; .
\label{bgrow}
\ee
The initial amplitude of density perturbations is very well known from
the CMB anisotropy measurements, $(\delta \rho /\rho)_i = 5 \cdot 10^{-5}$.
Hence, a Universe without dark matter would still be nearly homogeneous:
the density contrast would be in the range of a few per cent. No structure
would have been formed, no galaxies, no life. No structure would be formed
in future either, as the accelerated expansion
 due to dark energy will soon terminate
the growth of perturbations.

Since dark matter particles decoupled from plasma much earlier
than baryons,
perturbations in dark matter started to grow much earlier.
The corresponding growth factor is larger than (\ref{bgrow}),
so that the dark matter density contrast at galactic and 
sub-galactic scales
becomes of order one, perturbations  enter non-linear regime,
collapse and form
dense dark matter clumps at $z = 5~-~10$. Baryons fall into 
potential wells formed by dark matter, so dark matter and baryon
perturbations work together soon after recombination. Galaxies get 
formed in the regions where dark matter was overdense originally.
For this picture to hold, dark matter particles
must be non-relativistic early enough, as relativistic particles
fly through gravitational wells instead of being trapped there.
This means, in particular, that neutrinos cannot 
constitute a considerable part
of dark matter. 

\subsection{Digression. Standard ruler: BAO}

Before recombination, the sound speed in baryon-electron-photon component
is about $v_s \approx 1/\sqrt{3}$. After recombination, baryons (atoms)
decouple from photons, sound speed in baryon component is practically
zero, and baryons no longer move in space. This leads to a feature in the
spatial distribution of matter (galaxies) which is known as
Baryon Acoustic Oscillations (BAO). It is worth noting that
similar phenomenon was described by A.D.~Sakharov~\cite{Sakharov:BAO}
back in 1965, but
in the context of cold cosmological model (Sakharov's paper was written
before the discovery of CMB).

\begin{figure}[htb!]
  \vspace{-3cm}
\centerline{
\includegraphics[width=0.5\textwidth]{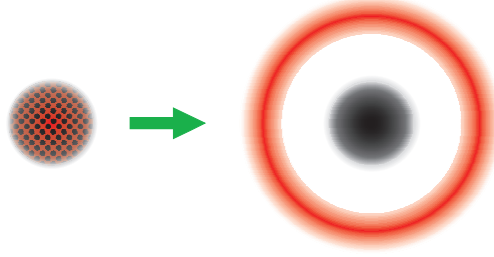}
}
\vspace{-3cm}
\caption{Schematic picture of Baryon Acoustic Oscillations.
  Dark regions show dark matter overdensity, less dark (red)
  regions are the ones with baryon overdensity. Left: initial
  condition. Right: at recombination and later.
\label{BAO-BAO}
}
\end{figure}

Phyiscs behind BAO is illustrated in Fig.~\ref{BAO-BAO}. Suppose there
is an overdense region in the very early Universe (in the beginning
of the hot epoch). Importantly, the initial conditions for
baryon-electron-photon component and dark matter are the same:
overdensity exists in both of them in the same place in space
(this is the property
of adiabatic scalar perturbations; CMB measurements ensure that
primordial perturbations are indeed adiabatic). This initial condition is
shown in the left panel of Fig.~\ref{BAO-BAO}. Before recombination,
dark matter perturbation stays in the same place, while perturbation
in baryon-electron-photon component moves away with the sound speed.
If the initial perurbarion is spherically symmetric, then the sound wave
is spherical, as shown in the right panel. At recombination, the baryon
perturbation is frozen in, and the whole picture expands merely
due to the cosmological expansion. The comoving distance between the
dark matter overdensity and baryon overdensity shell is
the comoving sound horizon at recombination
\[
r_s = 
\int_0^{t_r} v_s  \frac{k}{a(t)} dt
\]
(this is precisely the argument of cosine in \eqref{nov12-19-11});
its present value is $r_s \simeq 150$~Mpc (we set $a_0=1$ here),
and the value at
redshift $z$ is $r_s/(1+z)$. 

\begin{figure}[!htb]
  \hskip 0.15\textwidth
  \vspace{-2cm}
\includegraphics[width=0.6\textwidth]{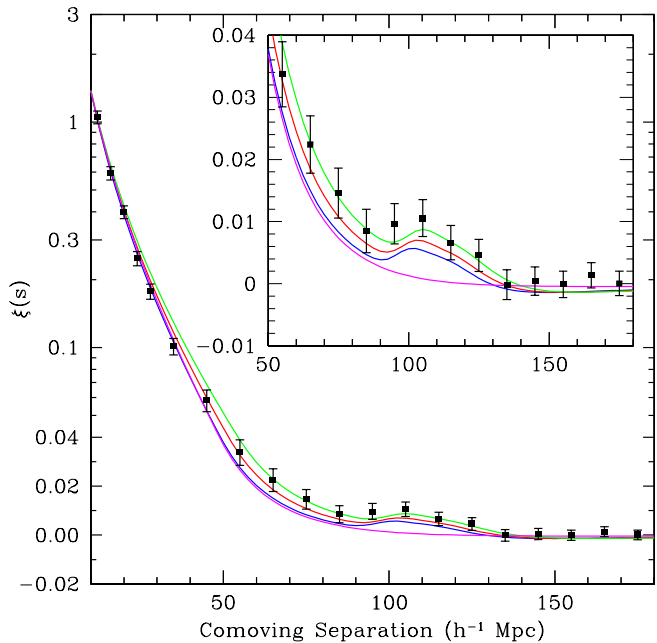}
\vspace{-0.5cm}
\caption{The first detection of BAO: the correlation function
$\xi\l s\r$ determined by the analysis of the 
SDSS data on the distribution of distant galaxies.
Solid lines show the predictions of various cosmological models.
Green, red and blue lines correspond to  $\Omega_M h^2= 0.12, 0.13, 0.14$,
respectively, with
$\Omega_Bh^2=0.024$, $n_s=0.98$ in all cases. Magenta line 
corresponds to unrealistic Universe without baryons. The parameter $h$
is defined in \eqref{H0}; numerically, $h_0 \approx 0.7$.
\label{BAO-corr}
}
\end{figure}

Due to BAO, there is correlation between the matter densities
(dark matter plus baryons) separated by comoving distance $r_s$.
It
shows up as a feature in the galaxy-galaxy correlation function
$\xi(s)$, where $s$ is comoving distance.
This bump in the correlation function was detected in
Ref.~\cite{Eisenstein:2005su}, see Fig.~\ref{BAO-corr}.
Clearly, BAO serves as a standard ruler at various redshifts,
which can be used to study the evolution of the Universe in not
so distant past.

Currently, BAO is a very powerful tool of
observational cosmology. It is used, in particular, to study
time (in)dependence of dark energy.

The bump in the spatial correlation function
translates into oscillations in momentum space, hence the name.

\section{Astrophysics: more hints
  on dark matter properties}

Important information on dark matter properties is obtained by
theoretical analysis of structure formation and its comparison
with observational data. Indeed, as we discussed above, dark mater
plays the key role in structure formation, so properties of galaxies
and their distribution in space potentially
tell us a lot about dark matter.

Currently, theoretical studies are made mostly
via numerical simulations, many of which ignore effects due to
baryons (dark-matter-only). Thus, these simulations give the
dark matter distribution. To compare it with observed structures,
one often assumes that baryons 
trace dark matter, with qualification that
baryons are capable of losing their
kinetic energy and forming more compact structures inside
dark matter halos. In other words, simulated dark matter collapsed
clump of a mass charactersitic of a galaxy is associated with
a visible galaxy, heavier dark matter clumps are interpreted as
clusters of galaxies, etc.

Currently, the most popular dark matter scenario is  cold
dark matter, CDM. It consists of particles whose velocities
are negligible at all stages of structure formation, and whose
non-gravitational interactions with themselves and with
baryons are
negligible too (from the viewpoint of structure formation).
The CDM numerical simulations (plus the above assumption concerning
baryons) are in very good agreement with observations {\it at
  relatively large spatial scales}. This is an important result
that implies interesting limits on dark matter properties, which we
discuss below.

However, there are astrophysical phenomena at shorter scales
that may or may not hint
towards something different from weakly interacting CDM. The situation
is
inconclusive yet, but it is worth keeping in mind these phenomena,
which we now discuss in turn.

\subsection{Missing satellite problem: astrophysics vs warm dark matter}
\label{subsec:missing}

\begin{figure}[htb!]
\centerline{
\includegraphics[width=0.34\textwidth]{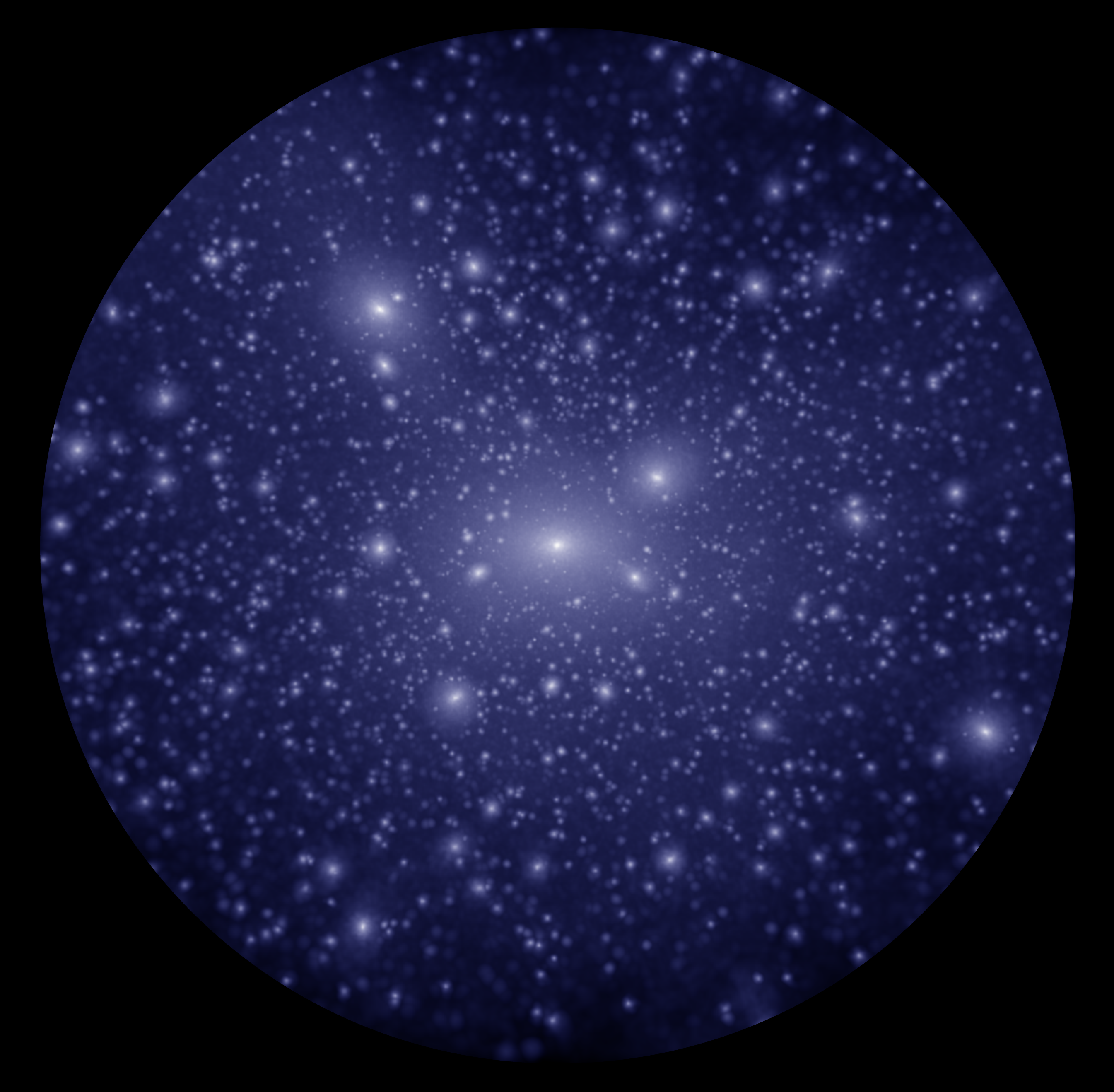}
\includegraphics[width=0.34\textwidth]{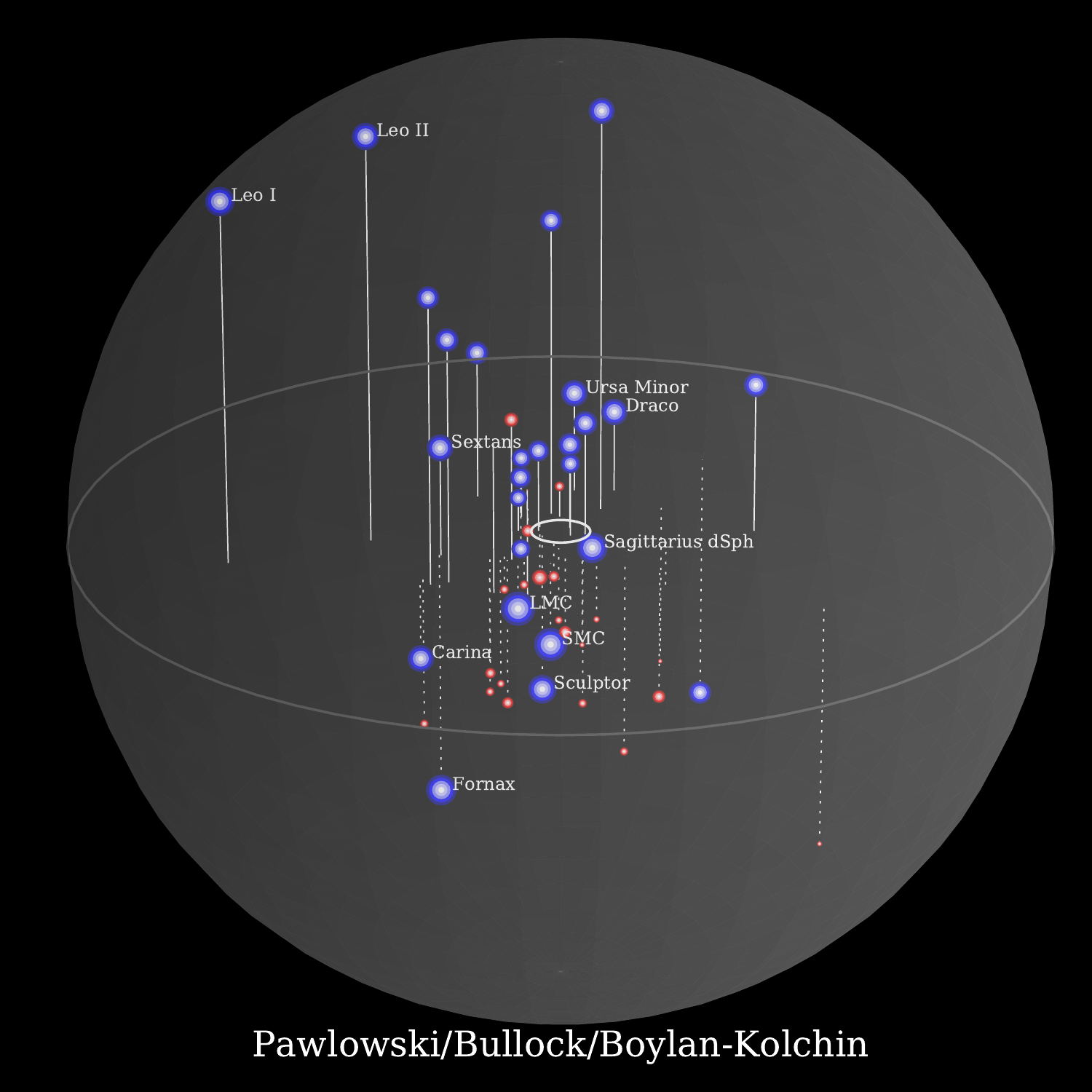}
}
\caption{Left: CDM-only simulation of
  250 kpc vicinity of a galaxy like Milky Way;
  right: actual distribution of satellite galaxies in  250 kpc
  vicinity of Milky Way~\cite{Bullock:2017xww}.
\label{subhalos}
}
\end{figure}

It has long been known that CDM-only simulations produce a lot of
small mass halos, $M \lesssim 10^9 M_\odot$ where $M_\odot$ is the Solar mass.
Galaxies like Milky Way have masses $(10^{11} - 10^{12})M_\odot$, so we
are talking about dwarf galaxies. As an example, the left panel of
Fig.~\ref{subhalos} shows the simulated
dark matter distribution in a ball of radius 250 kpc around a
galaxy similar to Milky Way. Assuming that baryons trace dark matter,
one observes that there must be hundreds of satellite galaxies there.
The actual Milky Way satellites are shown in the right panel of
Fig.~\ref{subhalos}; clearly the number of satellites is a lot smaller.
This is the missing satellite problem.

It is conceivable that this problem has astrophysical solution within
CDM model. One point is that the number of observed faint
satellite galaxies around Milky Way is not that small any longer:
while a few years ago this number was about 20,
it is currently about 60, and this is not a comlete sample
because of limited detection efficiency -- the
expectation~\cite{Kim:2017iwr}
for a
complete sample
is 150 -- 300 with masses exceeding $10^{8}M_\odot$. Another property is that
dark matter halos of mass $M<10^9M_\odot$ appear
fairly inefficient in forming luminous
component\footnote{Another effect, important for satellite galaxies close to the center of
Milky Way, is the tidal force due to
gravitational potential produced by the disk of
the host galaxy~\cite{Garrison-Kimmel:2017zes}.}
--- this has been suggested  by simulations that include numerous
effects due to baryons~\cite{Shen:2013wva,Revaz:2018bkf}.
Thus, if CDM model is correct, and missing satellite problem
has astrophysical solution, there must be a large number of
ultra-faint dwarf galaxies with masses $(10^8 - 10^9)M_\odot$ and even
larger number
of
non-luminous
dark matter
halos with $M \lesssim 10^8 M_\odot$ in the vicinity of Milky Way.
This prediction will be possible to check in near future,
notably, with Large Synoptic Survey Telescope, LSST~\cite{Bechtol:2019acd}.


An alternative, particle physics solution to the
missing satellite problem is {\it warm dark matter}, WDM.
A reasonably well motivated WDM candidate is sterile neutrino,
which we discuss in Sec.~\ref{sterilnu}. Another popular candidate is
light
gravitino.
In WDM case, dark matter particles decouple from kinteic equilibrium with
baryon-photon component when they are
relativistic. Let us assume for definiteness
that they are in {\it kinetic} equilibrium
with cosmic plasma at temperature $T_f$
when their number density freezes out (there is no {\it chemical}
equilibrium 
at $T=T_f$, otherwise the dark matter would be overabundant). After
kinetic equilibrium breaks down at temperature $T_d \leq T_f$, 
the spatial momenta decrease as $a^{-1}$,
i.e., the momenta are  of order $T$ all the time after decoupling.
When dark matter particles are relativistic, the density perturbations
do not grow: relativistic particles escape from the
gravitational potentials, so they do not
experience the gravitational instability; in fact, the density
perturbations, and hence
the gravitational wells get smeared out instead of getting deeper.
WDM particles become non-relativistic at $T\sim m$, where $m$ is their mass.
Only after that the WDM perturbations start to 
grow. 
Before becoming non-relativistic, WDM particles travel the distance of the
order of the horizon size; the WDM perturbations therefore are suppressed
at those scales.
The horizon size at the time $t_{nr}$ when $T\sim m$ is of order
\[
   l_H(t_{nr}) \simeq
H^{-1} (T\sim m) = \frac{M_{Pl}^*}{T^2} \sim  \frac{M_{Pl}^*}{m^2} 
\; .
\]
Due to the expansion of the Universe,
the corresponding length at present is
\be
  l_0 =  l_H(t_{nr}) \frac{a_0}{a(t_{nr})} \sim 
l_H(t_{nr}) \frac{T}{T_0} \sim \frac{M_{Pl}}{m T_0} \; ,
\label{l0dwarf}
\ee
where we neglected  (rather weak) dependence on $g_*$.
Hence, in WDM scenario,
 structures of comoving sizes smaller than $l_0$ are less abundant
as compared to CDM. Let us point out that $l_0$ refers to the
size of the perturbation in the linear regime; in other 
words, this is the size of the region from which matter collapses into a
compact object.

To solve the missing satellite problem, one requires that
the mass of dark matter  which was originally
distributed over
the volume of comoving size $l_0$, and
collapsed later on, is of order of the mass of satellite galaxy,
\[
\frac{4\pi}{3} l_0^3 \Omega_{DM} \rho_c \sim M_{dwarf} \; .
\]
With $M_{dwarf} \sim 10^{8}M_\odot$ we find $l_0 \sim 100$~kpc, and
eq.~\eqref{l0dwarf} gives
the estimate for the mass of a dark matter particle
\be
 \mbox{WDM}\; : \;\;\;\; m_{DM} = 3~-~ 10~\mbox{keV} \; .
\label{wdmmass}
\ee
On the other hand, this effect is absent, i.e., dark matter is cold,
for
\be
\mbox{CDM}\; : \;\;\;\; m_{DM} \gtrsim 10~\mbox{keV} \; .
\label{cdmmass} 
\ee
Let us recall that these estimates apply to particles that
are initially in kinetic equilibrium with cosmic plasma.
They do {\it not} apply in the opposite case; an example is
axion dark matter, which is cold despite of very small axion mass.

Reversing the argument, one obtains a limit on the mass of WDM particle which
decouples in kinetic equilibtium~\cite{Kim:2017iwr},
\be
m \gtrsim 4~\mbox{keV} \; .
\label{nov15-19-1}
\ee

\subsubsection{Digression: phase space bound}

\begin{figure}[htb!]
\centerline{
\includegraphics[width=0.6\textwidth]{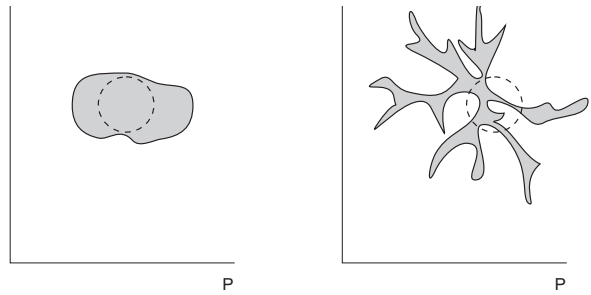}
}
\vspace{-8cm}
\caption{Sketch of the behavior of an ensemble of particles in
  phase space. As the ensemble evolves, initial compact
  distribution (left panel) becomes
  less compact.
\label{ameba}
}
\end{figure}

In fact there are other ways to obtain the limits on $m$.
One has to do with phase space density: the maximum value 
of coarse grained phase space density
\[
f (p,x)_{coarse~grained}
= \left( \frac{dN}{d^3 p\, d^3 x} \right)_{coarse~grained}
\]
does not decrease in the course of the evolution (here $N$ is the number
of particles). Indeed, Liouville theorem
tells that the microscopic phase space density is time-independent.
What happens in the course of evolution is that particles penetrate
initially unoccupied regions of phase space, see Fig.~\ref{ameba}.
While the maximum value of the microscopic phase space density
remains constant in time, the maximum value of coarse grained
phase space density (average over phase space volume shown by
dashed line in Fig.~\ref{ameba}) decreases.

The initial phase space density of particles in kinetic equilibrium
is
\[
f_i =\frac{A}{(2\pi)^3} \frac{1}{\mbox{e}^{p/T} + 1} \; ,
\]
where we consider fermions for definiteness. The parameter $A$
is determined by requiring that the number density $n$ takes
prescribed value, so that
\[
n_0= \frac{\Omega_{DM} \rho_c}{m} \; .
\]
We find
\[
n= \int f_i d^3p = A \cdot \frac{3\zeta(3)}{4\pi^2}T^3 \; ,
\]
where $\zeta(3) \approx 1.2$.
So, the maximum of the initial phase space density
is
\[
f_{i , \, max} = \frac{n}{12 \pi \zeta(3) T^3}
= \frac{\Omega_{DM} \rho_c}{12 \pi \zeta(3) m T_{0 \, eff}^3} \; ,
\]
where $T_{0 \, eff}$ depends on the decoupling temperature and is somewhat
lower than the present photon temperature.

On the other hand, one can measure a quantity
\[
Q = \frac{\rho_{DM, \, gal}}{\langle v_{gal}^2/3 \rangle^{3/2}}
\]
where $\rho_{DM, \, gal}$ is mass density (say, in a central part of
dwarf galaxy), $\langle v_{gal}^2 \rangle$ is average velocity squared,
and hence
$\langle v_{gal}^2/3 \rangle$ is the average velocity
squared
along the line of sight (of stars, and hence dark matter particles,
in a virialized galaxy). Since $v_{gal}=p_{gal}/m$ and
$\rho_{DM, \, gal} = m n_{gal}$,
one obtains an estimate for the phase space density of dark matter
particles in a dwarf galaxy,
\[
f \simeq \frac{n_{gal}}{\langle p_{gal}^2
    \rangle^{3/2}} = \frac{Q}{3^{3/2} m^4} \; .
  \]
  One requires that
  \[
  f < f_{i \, max}
  \]
  and obtains the bound on the mass of the
  dark matter particle
  \[
  m \gtrsim 3\cdot \left(\frac{Q}{\Omega_{DM} \rho_c}\right)^{1/3} T_{0 \, eff} \; .
  \]
  The values of $Q$ measured in compact dwarf galaxies are in the range
  \[
  Q \sim (5\cdot 10^{-3} - 2\cdot 10^{-2})\cdot
  \frac{M_\odot/\mbox{pc}^3}{(km/s)^3}
  \]
  while for relic that decouples at $T= (1 - 100)~\mbox{MeV}$
  one has $T_{0 \, eff} = 2.0~\mbox{K}$. This gives~\cite{Boyarsky,Gorby}
  \[
  m \gtrsim 6~\mbox{keV} \; .
  \]
Accidentally, this bound is similar to   
\eqref{nov15-19-1}. We note that bounds coming from phase
space density considerations are called bounds of
Tremain--Gunn type.

We also note that similar (in fact, slightly stronger but less
robust) bounds are obtained by the study of Lyman-$\alpha$ forest,
see, e.g., Ref~\cite{Irsic:2017ixq}.

\subsection{Other hints, SIMP and fuzzy DM}
\label{subs:hints}

There are two other issues that may or may not be problematic for CDM.
One is the ``core-cusp problem'': CDM-only simulations show singular
mass density profiles (cusps)
in the centers of galaxies, $\rho_{DM} (r) \propto r^{-1}$, while
observations imply enhanced but smooth profiles (cores).
Another is ``too-big-to-fail'' problem, which currently means that
the densities in large satellite galaxies ($M \sim 10^{10}~M_\odot$),
predicted by CDM-only simulations, are systematically higher than
the observed mass densities~\cite{Bullock:2017xww}.

The astropysical solutions to these problems again have to do with
baryons (supernovae feedback, etc.), and also interactions
of satellite galaxies with large host galaxy, Milky Way, see, e.g.,
Refs.~\cite{Bullock:2017xww,Lovell:2016nkp} for discussion.
On the particle physics side, WDM may again help out.
Two other particle physics solutions are Strongly Intracting
Massive Particles (SIMP) as dark matter, and fuzzy dark matter.

The idea of SIMP~\cite{SpSt} is that dark matter is cold, but
elastic scattering of dark matter
particles smoothes out the cuspy mass distribution in galactic
centers. Elastic scattering can also lead to decrease of the
dark matter density and thus alleviate the too-big-to-fail
problem. To give an idea of the elastic scattering cross section,
we take mass density of dark matter of order
$\rho_{DM} \sim 1~\mbox{GeV}/\mbox{cm}^3$
and require that the mean free path of dark matter
particle is of order $l \sim 1$~kpc (typical values, by order of magnitude,
both for centers of large galaxies and for dwarf galaxies),
\[
1 \sim l \sigma^{(el)} n_{DM} = l \sigma^{(el)} \frac{\rho_{DM}}{m} \; ,
  \]
  and obtain
  \[
  \frac{\sigma^{(el)}}{m} \sim  \frac{1}{l\rho_{DM}} \sim 10^{-24}
  \frac{\mbox{cm}^2}{\mbox{GeV}} \; .
  \]
  This is a very large cross section by particle physics standards,
  and, in view of \eqref{nov15-19-3}, dark matter particle
  must be fairly light, $m \lesssim 1$~GeV. The large
  elastic cross section
  may be due to $t$-channel exchange of light mediator with
  $m_{med} \sim 10 - 100$~MeV. This mediator must decay into
  $e^+e^-$, $\gamma \gamma$, etc., otherwise it would be
  dark matter itself. All these features make SIMP scenario interesting
  from the viewpoint of collider (search in $Z$-decays) and
  ``beyond collider'' experiments, such as SHiP. 

  Yet another proposal is fuzzy dark matter consisting of very light
  bosons,
  \[
  m \sim (10^{-21} - 10^{-22})~\mbox{eV} \; .
  \]
  The mechanism of their production must ensure that all of them are born
  with
  zero momenta, i.e., these particles form scalar condensate.
  An oversimplified picture is  that the de~Broglie wavelength
  of these particles at velocities typical for galactic centers
  and dwarf galaxies, $v \sim 10~\mbox{km}/\mbox{s}$, is about
  1~kpc:
 \[
 \frac{2\pi}{mv} \sim 1~\mbox{kpc} \; .
 \]
 Detailed discussion of advantages of fuzzy dark matter
 is given, e.g., in Ref.~\cite{Hui:2016ltb}.
 A way to constrain this scenatio is again to study Lyman-$\alpha$
 forest; current constraints  \cite{Irsic:2017yje} are at the level
 $2\cdot 10^{-21}$~eV. Interestingly, effects of fuzzy dark matter
 may in future be detected by pulsar timing method~\cite{Khmelnitsky:2013lxt}.

 From particle physics viewpoint, fuzzy dark matter particles
 may emerge as pseudo-Nambu--Goldstone bosons, similar to axions.
 We discuss axions later, and here we borrow the main ideas.
 The axion-like Lagrangian for the  pseudo-Nambu--Goldstone
 scalar field
 $\theta$ reads
  \[
        L= \frac{F^2}{2}  (\partial \theta)^2 -
          {\mu^4} (1-\cos^2 \theta) \approx
  \frac{ F^2}{2} (\partial \theta)^2 -
  \frac{ \mu^4}{2} \theta^2 \; ,
 \]
 where $F$ is the expectation value of a field that spontaneously
 breaks
 approximate $U(1)$ symmetry, and $\mu$ is the parameter of the explicit
 violation of this
 symmetry. Then the mass of the axion-like particle is
 \[
 m = \frac{\mu^2}{F} \; .
 \]
 The mechanism that creates the scalar condensate is misalignment.
 The initial value of $\theta$ is an arbitrary number between $-\pi$
 and $\pi$, so that $\theta_i \sim 1$. The field starts to oscillate
 when the expansion rate becomes small enough, $H \sim m$.
 The calculation of the present mass density is a simplified version of
 the axion calculation that we give in Sec.~\ref{subs:axions}; one finds that
 $\Omega_{DM} \sim 0.25$ is obtained for
 $m=10^{-22}$~eV if
 \[
 F \sim 10^{17}~\mbox{GeV} \; .
 \]
 This is in the ballpark of GUT/string scales, which is intriguing.

 \subsection{Summary of DM astrophysics}

 Let us summarize the astrophysics of dark matter.
 \begin{itemize}

 \item Cold dark matter descibes remarkably well the distrubution and
   properties of
   structures in the Universe at relatively large scales, from galaxies
   like Milky Way or somewhat smaller ($M \gtrsim 10^{11} M_\odot$) to larger
   structures like clusters of galaxies, filaments, etc.; also, CDM is
   remarkably consistent with CMB data which probe even larger scales.
   
 \item Currently, data and simulations at shorter scales are inconclusive:
   they may or may not show that there are ``anomalies'', the features that
   contradict CDM model.

 \item It will become clear fairly soon whether these ``anomalies''
   are real or not. The progress will come from refined simulations with
   all effects of baryons included, and from  new instruments,
   notably LSST.

 \item If the ``anomalies'' are real, we will have to give up CDM, and,
   responding to the data, will narrow down the set of dark matter
   models (WDM, or SIMP, or fuzzy dark matter, or something else). This
   will have a profound effect on the strategy of search for dark matter
   particles.

 \item  If the ``anomalies'' are not there, astrophysics will have
   to deliver the confirmation of CDM model by the discoveries of
   relatively light ultra-faint dwarf galaxies ($M = (10^{8} - 10^9) M_\odot$)
   and dark objects of even smaller mass.

 \end{itemize}

 All this makes astrophysics a powerful tool of studying dark matter and 
   directing particle physics in its search for dark matter particles.

   \section{Thermal WIMP}

   \subsection{WIMP abundance: annihilation cross section}

Thermal WIMP  (weakly interacting massive
particle) is a scenario featuring
a simple mechanism of the dark matter generation in the early
Universe. WIMP is a
{\it cold} dark matter candidate. Because of its simplicity
and robustness, it has been  considered by many as the most likely one.

Let us not go into all details of (fairly straightforward)
calculation of the thermal WIMP
abundance. These details are given in several textbooks, and also
presented in proceedings of similar Schools, see, e.g., 
Ref.~\cite{Rubakov:2017zvc}. Instead, we give the main assumptions
behind this mechanism and describe the main steps of the calculation.

One assumes that there exists a heavy stable neutral particle $\chi$, and
that 
$\chi$-particles can only be destroyed or created in cosmic plasma
via their
pair-annihilation or creation, with annihilation products being
the particles of the Standard Model\footnote{The latter assumption
  can be relaxed: decay products of $\chi$-particles may be
  new particles which sufficiently strongly interact with the Standard Model
  particles and in the end disappear from cosmic
  plasma. Also, destruction and creation of $\chi$-particles may occur
  via co-annihilation with their nearly degenerate partners and inverse
  pair creation processes; this occurs in a class of supersymmetric models
  where $\chi$ is the lightest supersymmetric particle and its
  partner is  the next-to-lightest supersymmetric particle.}. We note that
there is a version of WIMP model in which
particle $\chi$ is not truly neutral, i.e., it
does not coincide with its own antiparticle. In that case one assumes that
the production and destruction occurs only
via $\chi - \bar{\chi}$ annnihilation, and there is no asymmetry between
$\chi$ and $\bar{\chi}$ in cosmic plasma, $n_\chi = n_{\bar{\chi}}$.
The calculation in the $\chi-\bar{\chi}$ model is identical to the case
of truly neutral particle, so we consider the latter case only.

One also assumes that the $\chi$-particles are not
strongly coupled, but
$\chi \chi$-annihilation cross section is
sufficiently large, so the $\chi$-particles are in complete thermal
equilibrium at high temperatures. The latter assumption is justified in
the end of the calculation.
The thermal equilibrium means, in particular, that the abundance of
$\chi$-particles is given by the standard Bose--Einstein or Fermi--Dirac
distribution formula.

The
cosmological behaviour of $\chi$-particles is as follows.
At high temperatures, $T \gg m_\chi$, the number density of
$\chi$-particles is high, $n_\chi (T) \sim T^3$.
As the temperature drops below $m_\chi$, the equilibrium number
density decreases,
\be
   n^{(eq)}_\chi 
\propto \mbox{e}^{-\frac{m_\chi}{T}}\; ,
\label{mar26-15-2}
\ee
 At some ``freeze-out'' temperature $T_f$ the 
number density becomes so small, that $\chi$-particles can no longer
meet each other during the Hubble time, and their annihilation
terminates\footnote{This is a slightly oversimplified picture, which,
  however, gives a correct estimate, modulo  factor of order 1 in the
  argument of logarithm.}. After that the number density of  survived
$\chi$-particles decreases as $a^{-3}$, and these relic particles
form CDM.
The freeze-out temperature $T_f$
is obtained by equating the
mean free time of $\chi$-particle with respect to annihilation,
\[
\tau_{ann} (T_f) = \left( \sigma_0 (T_f) n_\chi (T_f) \right)^{-1}
\]
to the Hubble time (see \eqref{mar25-15-10})
\[
H^{-1} (T_f) = \frac{M_{Pl}^*}{T_f^2} \; .
\]
Here we introduced the weighted annihilation cross section
\[
\sigma_0 (T) = \langle \sigma_{ann} v \rangle_T \; ,
\]
where $v$ is the relative velocity of $\chi$-particles
(in the non-relativistic regime relevant here we have $v \ll 1$),
and we average over the thermal ensemble.

Thus, freeze-out occurs when
\[
\sigma_0 (T_f) n_\chi (T_f) =  \frac{T_f^2}{M_{Pl}^*} \; .
\]
Because of exponential decay of $n_\chi^{(eq)}$ with temperature,
eq.~\eqref{mar26-15-2}, freeze-out temperature is smaller than the mass
by a logarithmic factor only,
\be
   T_f \approx \frac{m_\chi}{\ln (M_{Pl}^{*} m_\chi \sigma_0)} \; .
\label{may10-1}
\ee
Note that due to large logarithm,  $\chi$-particles are
indeed non-relativistic at freeze-out: their velocity squared
is of order
\[
v^2 (T_f) \simeq 0.1 \; .
\]
At freeze-out, the number density is
\be
n_\chi (T_f) = \frac{T_f^2}{M_{Pl}^{*} \sigma_0 (T_f)} \; ,
\label{mar26-15-6}
\ee
Note that this density is inversely proportional to the annihilation
cross section (modulo logarithm). The reason is that
for higher annihilation cross section, the creation-annihilation
processes are longer in equilibrium, and less $\chi$-particles survive.
Up to a numerical factor of order 1, the number-to-entropy ratio 
at freeze-out
is
\be
   \frac{n_\chi}{s} \simeq \frac{1}{g_* (T_f) M_{Pl}^{*}T_f \sigma_0(T_f)}  \; .
\label{nchis}
\ee
This ratio stays constant until the present time, so 
 the present number density of
$\chi$-particles is
$   n_{\chi, 0} = s_0 \cdot  
\left(n_\chi/s \right)_{freeze-out}$,
and the mass-to-entropy ratio is
\be
  \frac{\rho_{\chi, 0}}{s_0} = \frac{m_\chi n_{\chi,0}}{s_0} 
   \simeq   
\frac{\ln (M_{Pl}^{*} m_\chi \sigma_0)}{g_*(T_f) M_{Pl}^{*} \sigma_0(T_f)}
\simeq \frac{\ln (M_{Pl}^{*} m_\chi \sigma_0)}{\sqrt{g_*(T_f)} M_{Pl} \sigma_0 (T_f)} \; ,
\nonumber
\ee
where we made use of (\ref{may10-1}).
This formula is remarkable. The mass density depends mostly on
one parameter, the annihilation cross section $\sigma_0$. The
dependence on the mass
of $\chi$-particle is through the logarithm and through
$g_* (T_f)$; it is very
mild. 
Plugging in  
$g_* (T_f) \sim 100$, as well as numerical factor omitted
in Eq.~(\ref{nchis}), and comparing with (\ref{10p*}) we obtain the estimate
\be
   \sigma_0 (T_f) \equiv \langle \sigma v \rangle (T_f)
= 1 \cdot 10^{-36}~\mbox{cm}^2 = 1~\mbox{pb}\; .
\label{estim}
\ee
This is a weak scale cross section, 
which tells us that the relevant energy scale
is 100~GeV -- TeV. We note in passing that the estimate
(\ref{estim}) is quite precise and robust.

The annihilation cross section can be parametrized as
$   \sigma_0 = \alpha^2/ M^2$
where $\alpha$ is some coupling constant, and $M$ is
a mass scale responsible for the annihilation processes\footnote{For
  $s$-wave annihilation, $\sigma_0$ is independent
  of particle velocity, and hence temperature;
  if annihilation is in $p$-wave, 
there is an additional
suppression by $v^2(T_f) \sim 0.1$.}
(which may be higher than 
$m_\chi$).
 This parametrization is suggested by the picture of
$\chi$ pair-annihilation via the exchange by another particle of mass 
$M$. With $\alpha \sim
10^{-2}$, the estimate for the mass scale is roughly
$ M \sim 1~\mbox{TeV}$.
Thus, with mild assumptions, we find that the
WIMP dark matter may naturally originate from
the TeV-scale physics. In fact, what we have found can be understood as
an approximate equality between the cosmological parameter, mass-to-entropy
ratio of dark matter, and the particle physics parameters,
\[
\mbox{mass-to-entropy} \simeq \frac{1}{M_{Pl}}
\l \frac{\mbox{TeV}}{\alpha_W} \r^2 \; .
\]
Both are of order $10^{-10}~\mbox{GeV}$, and it is very tempting
to think that this ``WIMP miracle'' is not a mere coincidence.
For long time the above argument has been -- and still is -- a strong
motivation for  WIMP search. 

\subsection{WIMP candidates: ``minimal'' and SUSY; direct searches}

\subsubsection{``Minimal'' WIMP}

Even though the name -- {\it Weakly} Interacting Massive Particle 
-- suggests that this particle participates in the Standard Model
weak interactions, in most theoretical models this is not so.
An exception is ``minimal'' WIMP~\cite{Cirelli:2005uq}. This is a member
of electroweak multiplet with zero electric charge and zero coupling to
$Z$-boson (couplings to photon and $Z$ would yield to too strong
interactions with the Standard Model particles
which are forbidden by direct searches). This is possible
for vector-like 5-plet (weak isospin 2) with zero weak hypercharge.
Another, albeit fine-tuned option is vector-like triplet (weak isospin 1)
with zero weak hypercharge. Particles in vector-like representations
may have ``hard'' masses (not given by Englert--Brout--Higgs mechanism).
The right annihilation cross section \eqref{estim} is obtained for
masses of these particles
\[
\mbox{5-plet}:~~m_5 = 9.6~\mbox{TeV} \; , \;\;\;\;\;\;\;\;\;
\mbox{3-plet}:~~m_3 = 3~\mbox{TeV} \; .
\]
These particles are on the verge of being ruled out by direct
searches.

\subsubsection{Neutralino}

A well motivated WIMP candidate
is neutralino of  supersymmetric extensions of the Standard
Model. The situation with neutralino is rather tense, however.
One point is that the pair-annihilation of neutralinos often occurs
in $p$-wave, rather than $s$-wave. This gives the suppression factor
in $\sigma_0$, 
proportional to $v^2 \sim 0.1$.
Hence, neutralinos tend to be overproduced in large part of the parameter
space of MSSM and other SUSY models.

Another point is the null results of the direct searches for
WIMPs 
in underground
laboratories. The idea of direct search
is that WIMPs orbiting around the center of our
Galaxy with velocity of order $10^{-3}$ sometimes hit a nucleus in a
detector and deposit small energy in it. The relevant parameters for
these searches are WIMP-nucleon elastic scattering cross section
and WIMP mass. One distinguishes spin-independent and spin-dependent
scattering. In the former case, the WIMP-{\it nucleus}
cross secion is proportional to $A^2$, where $A$ is the number of nucleons
in the nucleus (this is effect of coherent scattering), while in the
latter case the cross section is proportional to $J(J+1)$ where $J$
is the spin
of the nucleus. 

\begin{figure}[htb!]
  \begin{center}
    \vspace{-2cm}
\includegraphics[angle=0,width=0.7\textwidth]{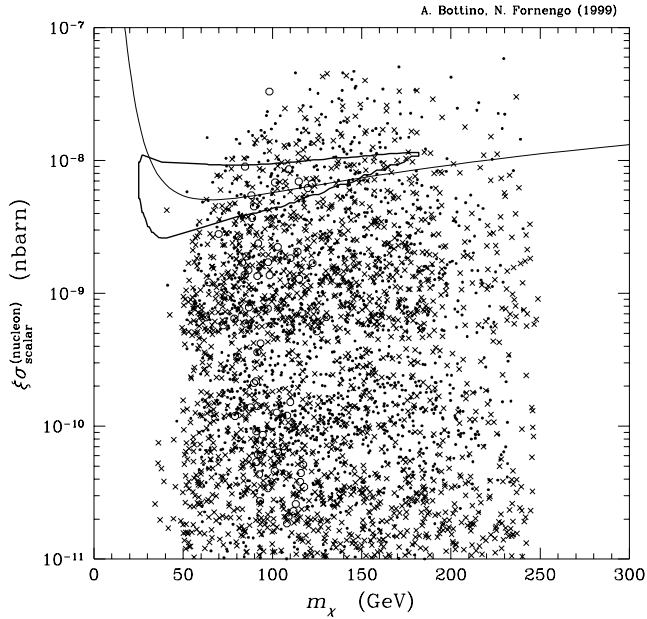}
\end{center}
\vspace{-4cm}
\caption{The situation with neutralinos and their direct searches in
  1999~\cite{Bottino:1998zd}.
  Shown are
  theoretical predictions (crosses and dots) and direct detection limits
  (open solid line; closed solid line is DAMA hint). Vertical axis:
  spin-independent cross section of elastic WIMP-nucleus scattering
  per nucleon; parameter $\xi$ takes value
  1 for spin-1/2 neutralino; note that
  $10^{-10}~\mbox{nbarn} = 10^{-43}~\mbox{cm}^2$.
\label{bottino}
}
\end{figure}

\begin{figure}[htb!]
\begin{center}
\includegraphics[angle=0,width=0.7\textwidth]{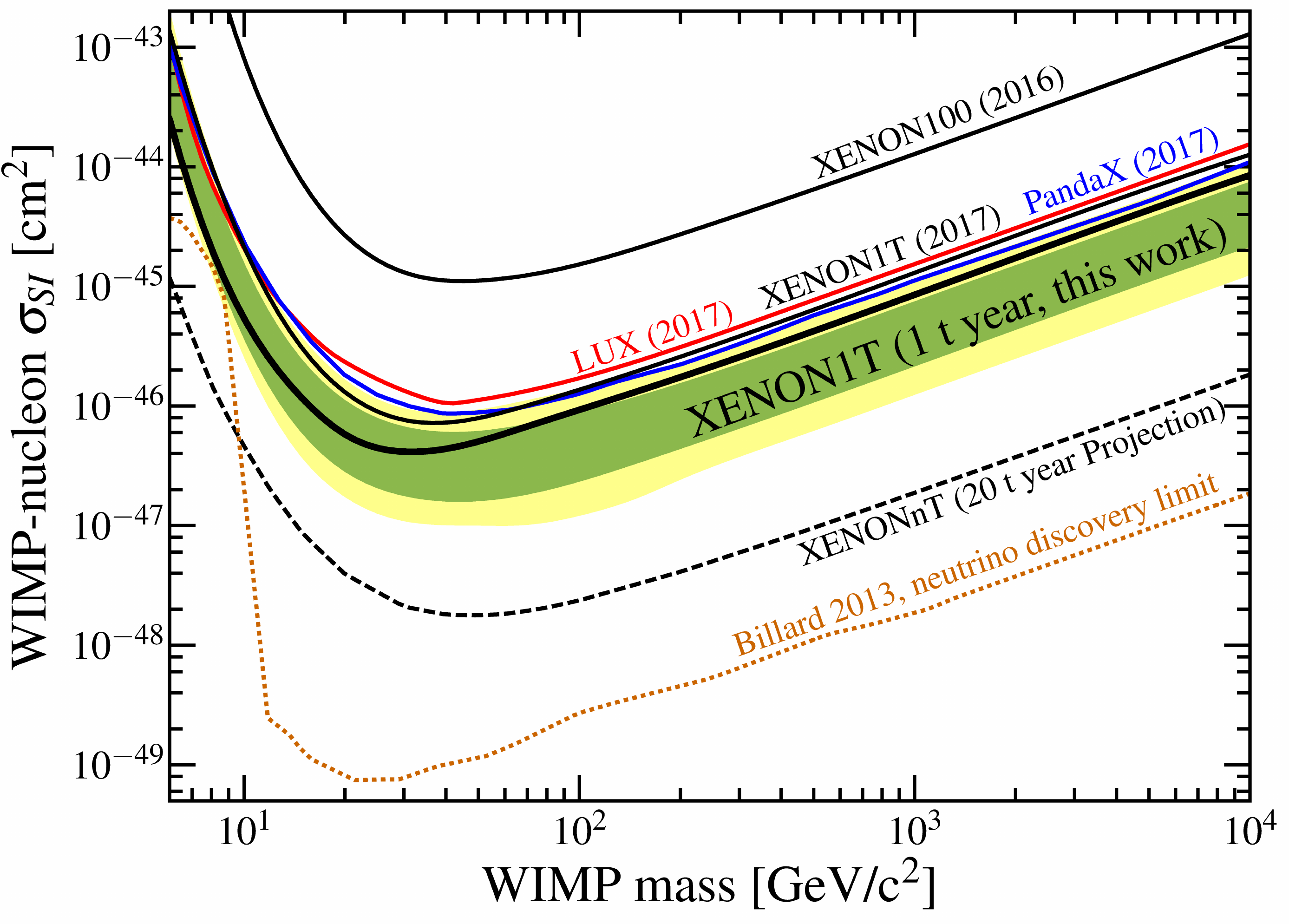}
\end{center}
\caption{Current results of
  direct searches for WIMPs: best limits on spin-independent WIMP-nucleus
  cross section per nucleon come from XENON-1T
  experiment~\cite{Aprile:2018dbl}. Note that cross section
  $10^{-10}~\mbox{nbarn} = 10^{-43}~\mbox{cm}^2$ in the lower part of
  Fig.~\ref{bottino}
  is in the upper part of this figure.
\label{xenon}
}
\end{figure}

To illustrate the progress in
direct search,
we show in Fig.~\ref{bottino} the situation with neutralinos and
their direct searches as of 1999, Ref.~\cite{Bottino:1998zd},
while Fig.~\ref{xenon} shows the best current limits on spin-independent
cross section~\cite{Aprile:2018dbl}.


\begin{figure}[htb!]
\vspace{-5.5cm}
  \begin{center}
\includegraphics[angle=0,width=0.7\textwidth]{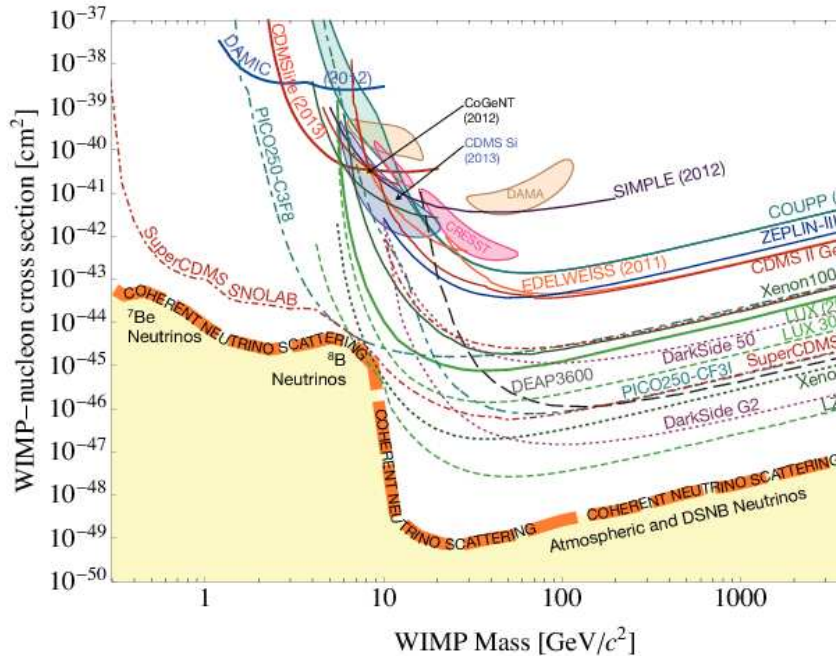}
\end{center}
\caption{Current limits and projected sensitivities of
  direct searches for WIMPs (spin-independent WIMP-nucleus 
  cross section per nucleon). Yellow band in the lower part
  is ``neutrino floor'',
  at which interactions of cosmic neutrinos become an important
  background.
\label{WIMPtoday}
}
\end{figure}

Figure~\ref{WIMPtoday} shows both current limits (solid lines) and projected
sensitivities of future dark matter detection experiments, again
for spin-independent interactions~\cite{Roszkowski:2017nbc}. We see that,
on the one hand, the progress in experimental search is truly remarkable,
and, on the other, the null results of this search are becoming alarming.
The null results of direct (and also indirect, see below)
searches are particularly worrying in view of null results of
SUSY searches at the LHC. 

\subsection{Ad hoc WIMP candidates; indirect searches and the LHC}

In view of the strong direct detection limits and null results of the
SUSY searches at the LHC, it makes sense to consider less motivated,
ad hoc WIMP candidates. The simplest assumption is that WIMP is
not nearly degenerate with any other new particle, so that the calculation
of its abundance outlined above applies, and that
there is one particle that mediates its pair-annihilation.
This mediator can be either a Standard Model particle or a new one;
we give examples of both cases.
The models of this sort are often called
simplified. We emphasize that the two examples of simplified models
which we are going to
discuss do not exhaust all possible WIMPs and mediators.
Some of the models that we leave aside are actually consistent with
both cosmology (they give the right value of $\Omega_{DM}$)
and experimental limits. The study of numerous
simplified models is given, e.g., in Ref.~\cite{Arcadi:2017kky}.

\begin{figure}[!tb]
\centerline{\includegraphics[width=0.5\textwidth,angle=0]{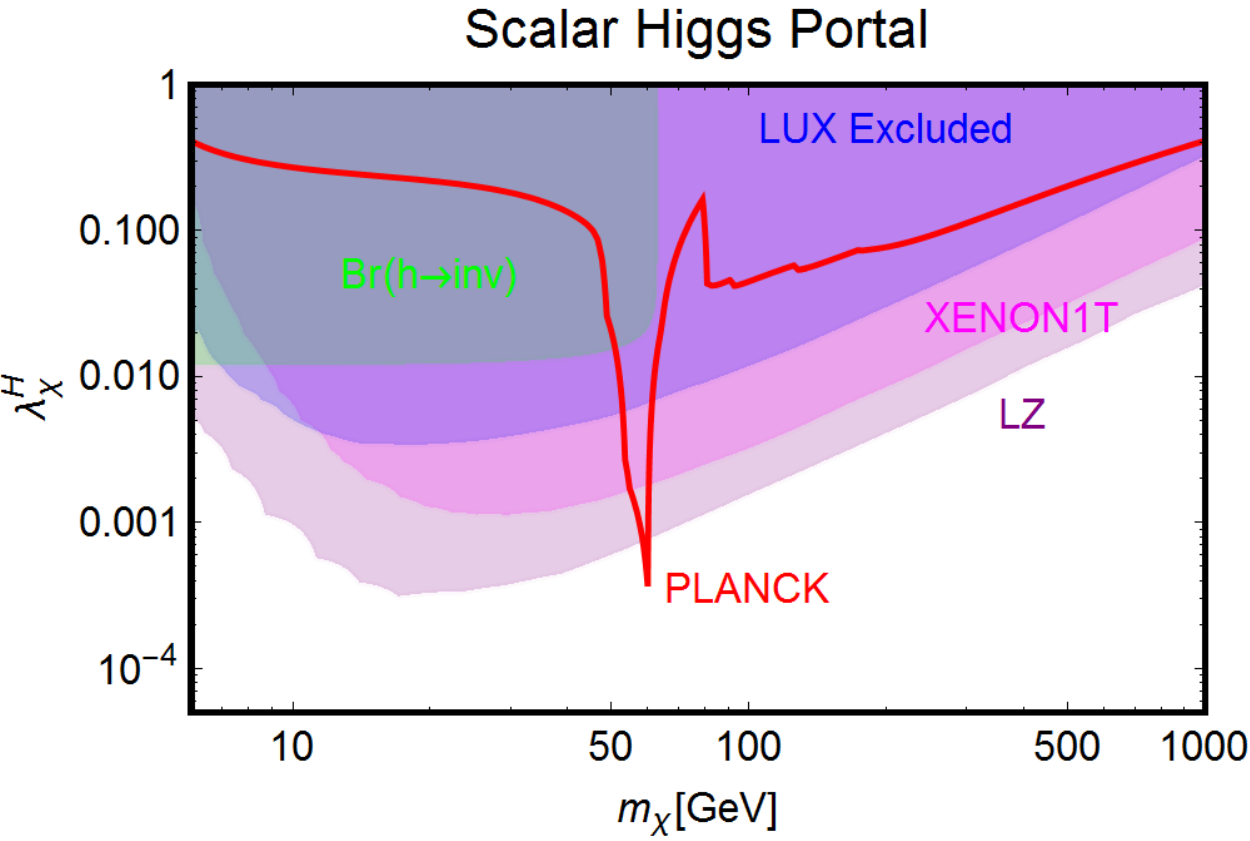}%
  ~~~~~~~~~~~\includegraphics[width=0.5\textwidth,angle=0]{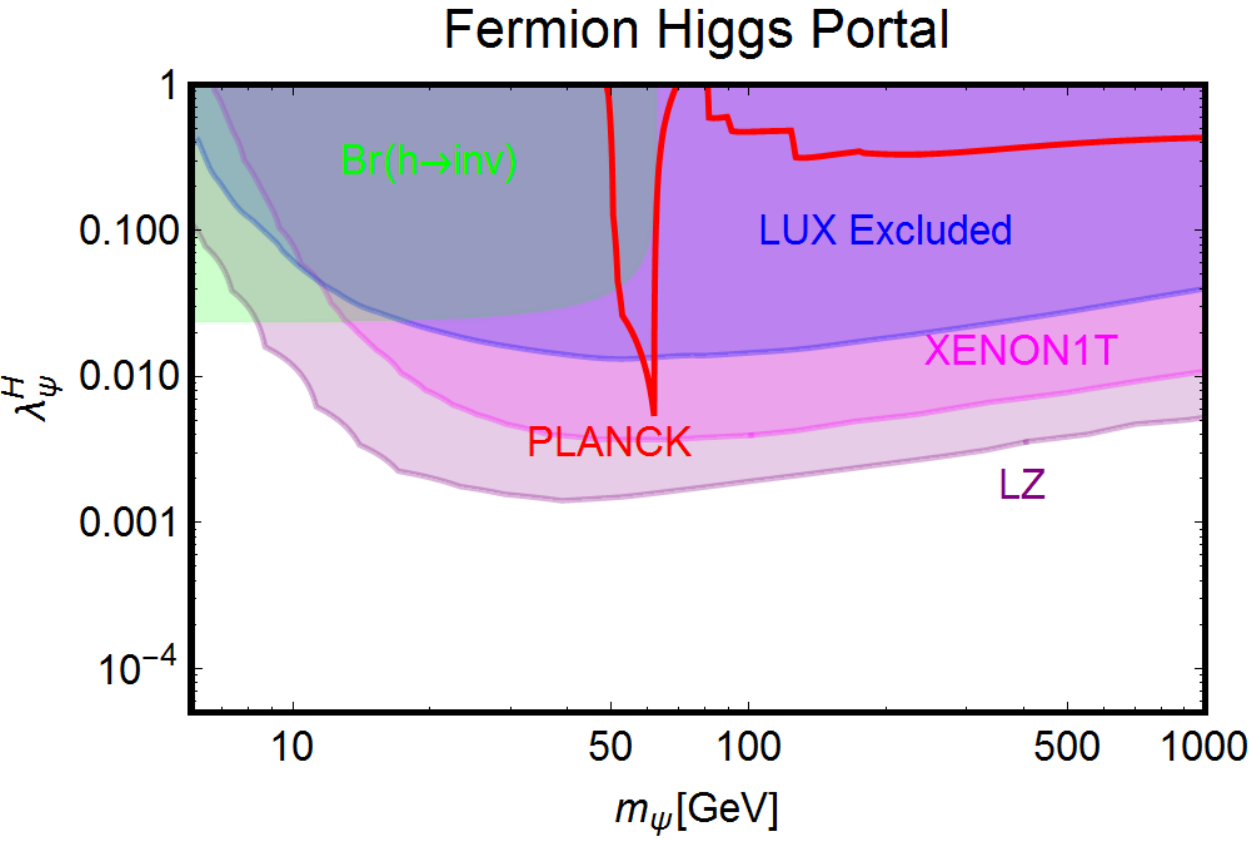}}%
\caption{Predictions from dark matter abundance (red solid lines
  labeled ``PLANCK'')
    and direct detection limits (shadows) in the Higgs
    portal models~\cite{Arcadi:2017kky}. Left panel:
    spin-0 WIMP; right panel: spin-1/2 WIMP.
\label{Hport}
}
 \end{figure}

\begin{figure}[htb!]
\begin{center}
\includegraphics[angle=0,width=0.5\textwidth]{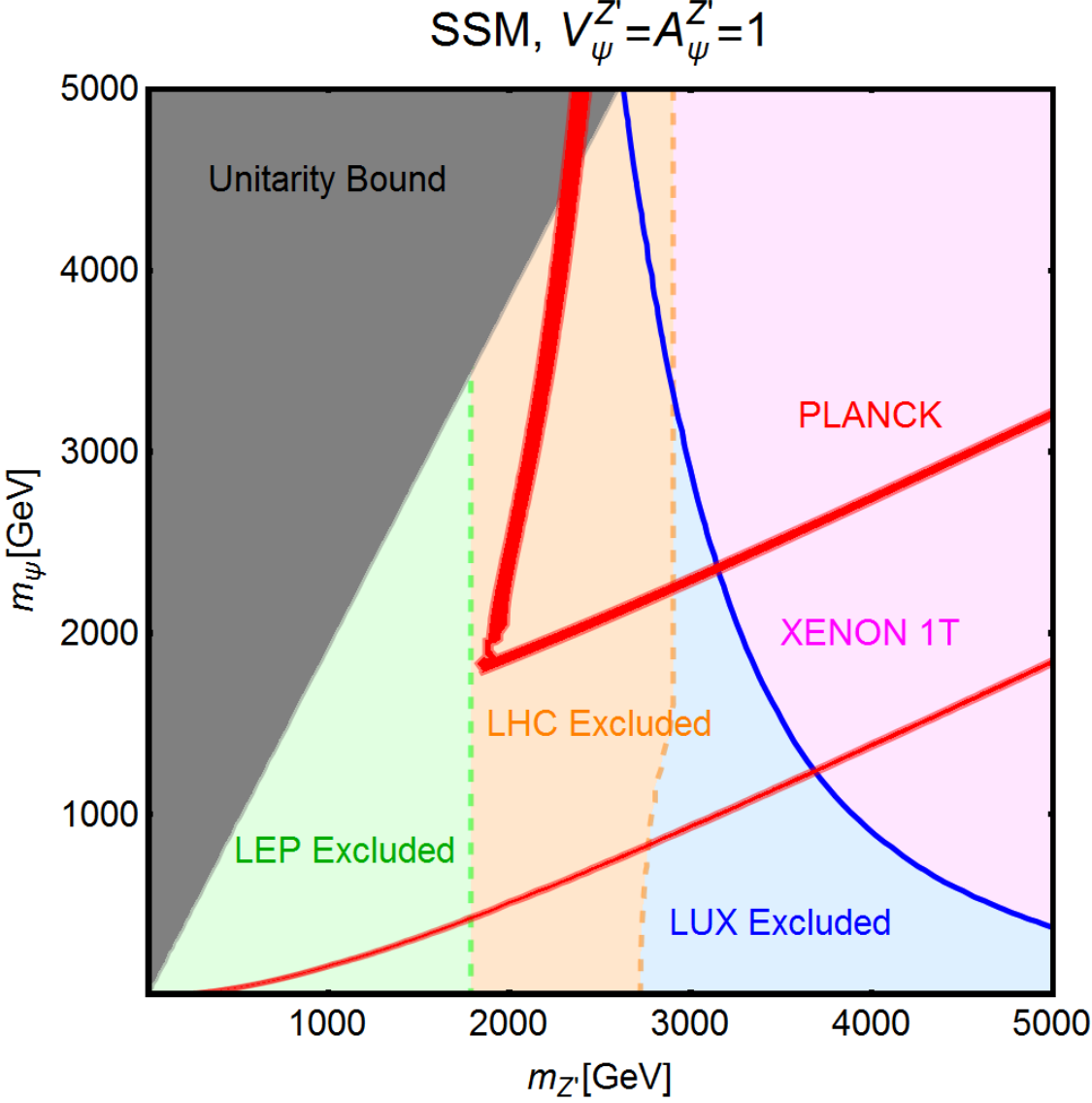}
\end{center}
\caption{Same as in Fig.~\ref{Hport} but for $Z'$-portal
  \eqref{nov18-19-1} with $g_\psi=g_f=0.65$ and $V_\psi = A_\psi=V_f=A_f =1$.
\label{Zprimeport}
}
 \end{figure}

With this reservation, it is fair to say
that many simplified models are either already ruled out
or will be ruled out soon. As one illustration, we consider
``Higgs portal'', a set of models where the only
field which interacts directly with WIMPs is Englert--Brout--Higgs field.
The lowest dimension Higgs-WIMP interaction terms in the cases of
spin-0 WIMP $\chi$ and  spin-1/2 WIMP $\psi$ are
\[
\lambda^H_{\chi} \chi^{*} \chi H^\dagger H \; ,
\;\;\;\;\;\; \frac{\lambda^H_{\psi}}{\Lambda} \bar{\psi} \psi H^{\dagger} H \; ,
\]
where $H$ is EBH field.
 Here $\lambda^H_{\chi}$, $\lambda^H_{\psi}$ are dimensionless parameters,
 while $\Lambda$ has dimension of mass. In both cases $\chi$ ($\psi$)
 is a
 Standard Model singlet with zero weak hypercharge; it has ``hard'' mass
 $m_{\chi (\psi)}$. Since the vacuum expectation value of EBH field $H$ is
 non-zero, the above interaction terms induce trilinear
 WIMP-WIMP-Higgs responsible for $s$-channel WIMP annihilation
 via the Higgs exchange. It is this annihilation that
 is relevant in the early Universe.
 The trouble is that almost entire parameter space
 of the Higgs portal is ruled out by direct searches. This is illustrated in
 Fig.~\ref{Hport}, Ref.~\cite{Arcadi:2017kky}.
 Another illustration is $Z'$-portal. One assumes that both
 WIMP (say, spin-1/2 particle $\psi$) and Standard Model fermions 
  interact with a new vector boson $Z'$:
 \be
 g_\psi \bar{\psi}(V_\psi - A_\psi \gamma^5) \psi Z'
 + \sum_f \, g_f 
 \bar{f}(V_f - A_f \gamma^5) f \, Z'\; ,
 \label{nov18-19-1}
 \ee
 where sum runs over all Standard Model fermions (important role is
 played by
 quarks). The coupling constants $g_\psi$, $g_f$ are often chosen
 to be of order 0.5, as suggested by GUTs. Almost all
 parameter space of $Z'$-portal models with $V_\psi \neq 0$
 is also ruled out by direct searches~\cite{Arcadi:2017kky},
 as shown in Fig.~\ref{Zprimeport}.

\begin{figure}[htb!]
\begin{center}
\includegraphics[angle=0,width=0.5\textwidth]{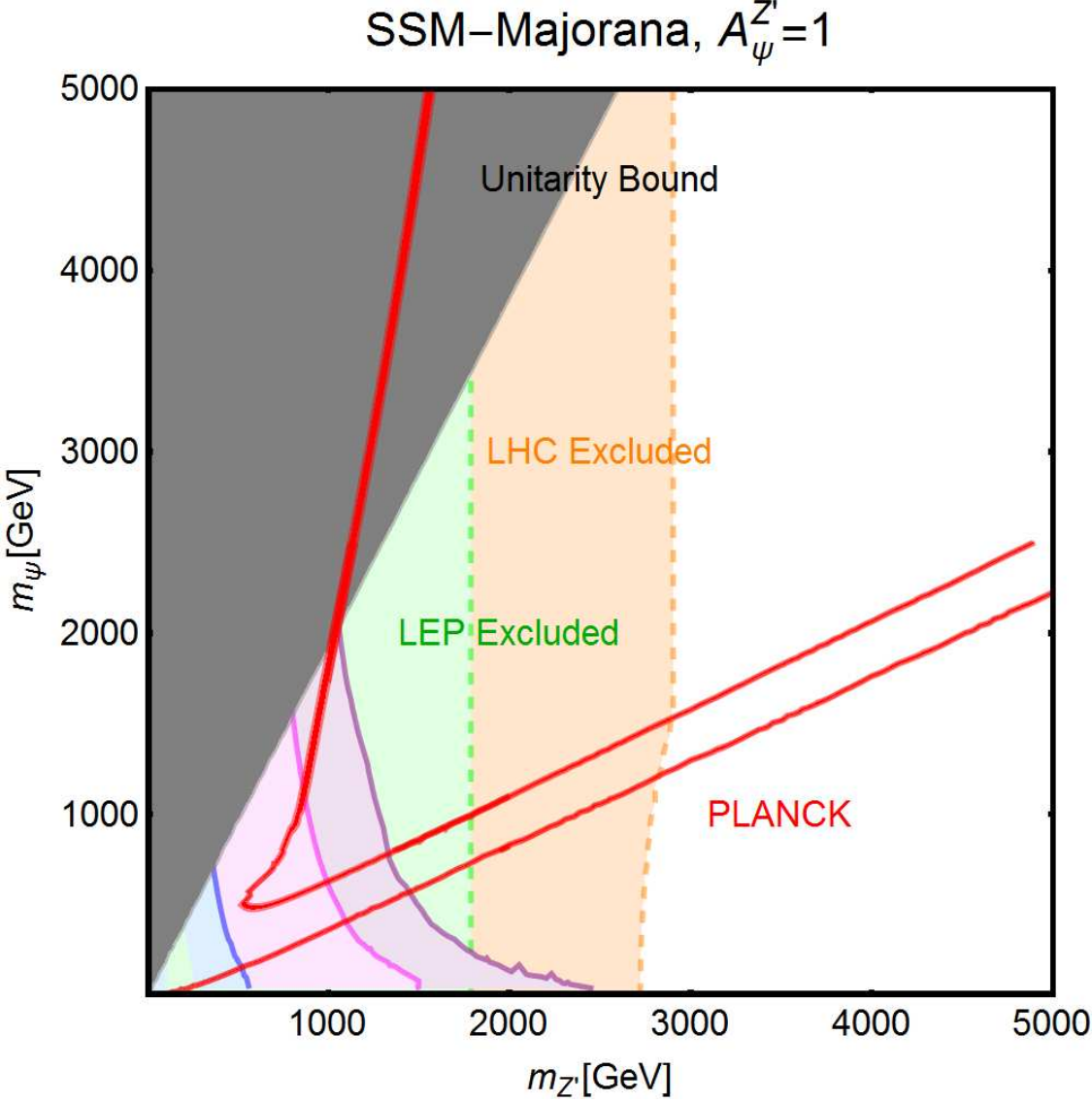}
\end{center}
\caption{Same as in Fig.~\ref{Zprimeport} but for axial-vector
  $Z'$-portal
  \eqref{nov18-19-1} with  and $V_\psi=V_f=0$, $A_\psi = A_f=1$.
\label{spin-dep}
}
\end{figure}

 The situation is better in models with axial-vector interactions of
 new vector boson (we still call it $Z'$) with both
 the Standard Model particles and WIMPs,
 \[
 V_\psi = V_f =0 \; .
 \]
 In that case, interaction of WIMPs with nucleons is spin-dependent,
 the elastic WIMP-nucleus cross section is not
 enhanced  by $A^2$, so the direct detection limits are not as
 strong as in the case of spin-independent interaction.
 An important player here is the LHC, whose limits are the most
 stringent~\cite{Arcadi:2017kky},
 see Fig.~\ref{spin-dep}. We see from Fig.~\ref{spin-dep} that models with
 $M_{Z'} \gtrsim 2.8$~TeV
 are capable of producing the correct abundance of
 dark matter and at the same time are not ruled out experimentally.

\begin{figure}[htb!]
  \vspace{-7cm}
\begin{center}
\includegraphics[angle=0,width=0.8\textwidth]{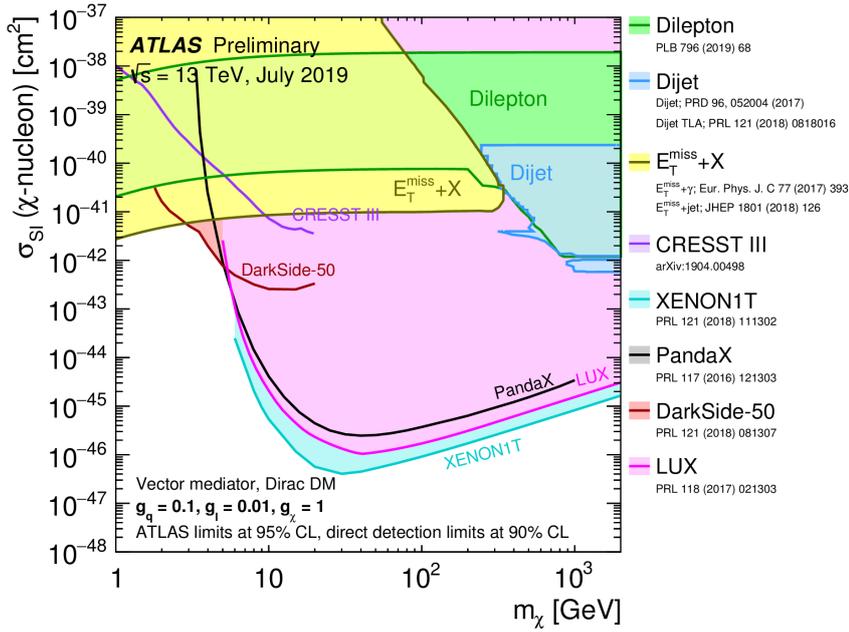}
\end{center}
\caption{LHC and direct detection limits in the case of spin-independent
  WIMP-nucleon elastic cross section.
\label{dir-vs-LHC1}
}
\end{figure}

\begin{figure}[htb!]
\vspace{-7cm}
  \begin{center}
\includegraphics[angle=0,width=0.8\textwidth]{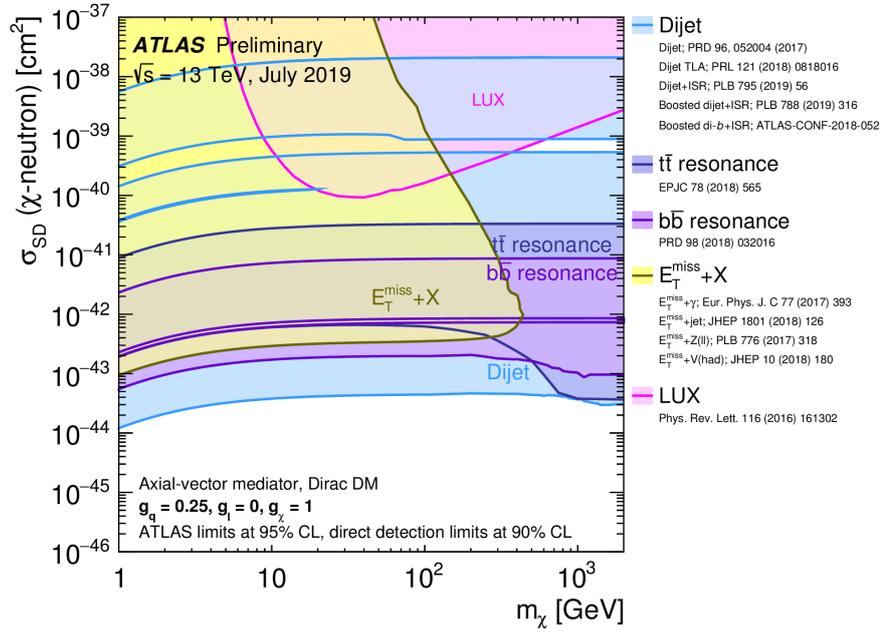}
\end{center}
\caption{The same as in Fig.~\ref{dir-vs-LHC1} but for spin-dependent
   WIMP-nucleon elastic cross section.
\label{dir-vs-LHC2}
}
\end{figure}

Another way of comparing current
sensitivities of direct and LHC searches is
given in Figs.~\ref{dir-vs-LHC1}, \ref{dir-vs-LHC2}.
The plots (compiled by ATLAS collaboraion)
refer to the model \eqref{nov18-19-1} with vector boson $Z'$ and
coupling constants with quarks $g_q$, leptons $g_l$ and WIMPs
$g_\psi \equiv g_\chi$ whose values are written in figures.
Figure~\ref{dir-vs-LHC1} 
shows the limits in the vector case, $A_\psi = A_f =0$, $V_\psi = V_f =1$,
while   Fig.~\ref{dir-vs-LHC2} refers to axial-vector case
$A_\psi = A_f =1$, $V_\psi = V_f =0$. Clearly, the direct searches are
more sensitive than the LHC in vector case (spin-independent WIMP-nucleon
elastic cross section),
while the LHC wins in the axial-vector case (spin-dependent elastic
cross section).
Overall, the LHC has become an important source of limits on WIMPs.

\begin{figure}[htb!]
\begin{center}
\includegraphics[angle=0,width=0.8\textwidth]{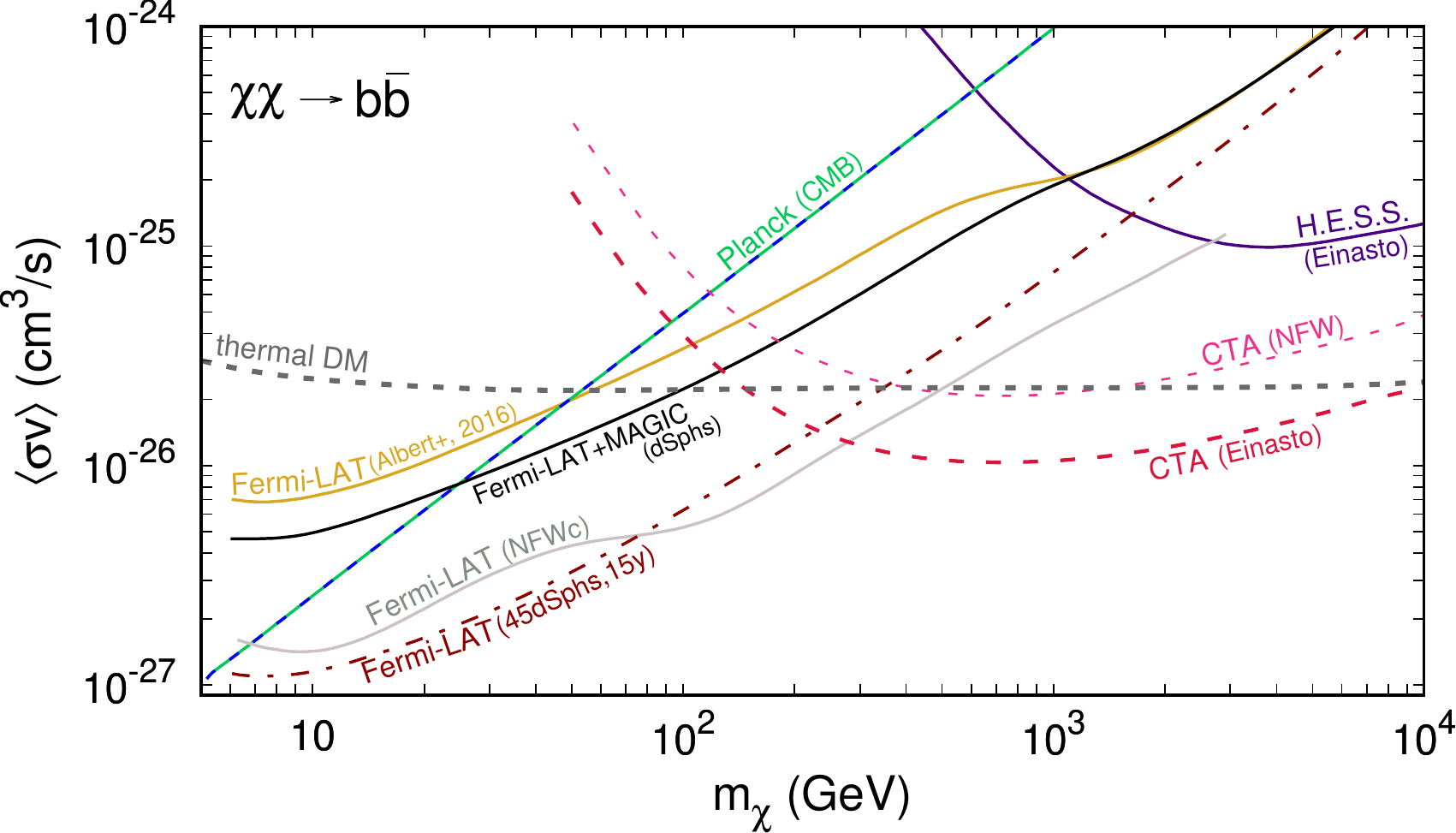}
\end{center}
\caption{Limits on WIMP annihilation cross section
  obtained by $\gamma$-ray telescopes (solid lines) and projected
  sensitivities of future $\gamma$-ray observatories (dashed lines).
  ``NFW'' and ``Einasto'' refer to different dark matter profiles in
  galaxies. Dashed line ``thermal DM'' is the predicion from cosmology
  \eqref{estim} under assumption of $s$-wave annihilation. Note the different
  units for $\langle \sigma v \rangle$ used in this figure and in
   \eqref{estim}.
\label{gamma-limits}
}
\end{figure}

Besides direct  and LHC searches for cosmic and collider-produced
WIMPs, respectively,
important ways to address WIMPs are indirect searches. One approach is to
search for high energy $\gamma$-rays which are produced in annihilations
of WIMPs in various cosmic sources, from dwarf galaxies to Galactic
center to clusters of galaxies, and also diffuse $\gamma$-ray flux
coming from the entire Universe. This approach is particularly
relevant if WIMP annihilation proceeds in $s$-wave: in that case
 the non-relativistic annihilation rate is determined  by
 \eqref{estim}, which is velocity-independent (modulo possible
 Sommerfeld enhancement, see Ref.~\cite{Lisanti:2016jxe}
 for detailed discussion).
 On the contrary, for $p$-wave annihilation
the rate $\sigma v$ is proportional to $v^2$, and since the
velocities in the sources are small ($v^2 \lesssim 10^{-6}$ as compared 
to $v^2 \simeq 0.1$ relevant to \eqref{estim}), the annihilation
cross section is strongly suppressed in the present Universe.
Thus, meaningful limits are obtained by $\gamma$-ray observatories
for WIMPs annihilating in $s$-wave.
The current situation and future prospects
are illustrated in
Fig.~\ref{gamma-limits}, Ref.~\cite{Roszkowski:2017nbc}. The assumption that
enters this compilation is that the major WIMP annihilation channel
is $b\bar{b}$. Clearly, already existing instruments, and to even larger
extent
future experiments are sensitive to a wide class of WIMP models.

Indirect searches for dark matter WIMPs include the search for
neutrinos coming from the centers of the Earth and Sun
(WIMPs may concentrate and annihilate there), see, e.g.,
Ref.~\cite{Avrorin:2014swy}, positrons and
antiprotons in cosmic rays (produced in WIMP annihilations
in our Galaxy), see, e.g., Ref.~\cite{Bergstrom:2013jra}. 
These searches have produced interesting, albeit model-dependent
limits on WIMP properties.

\subsection{WIMP summary}

\begin{itemize}

\item While WIMP hypothesis was very attractive for long time,
  and SUSY neutralino was considered the best candidate, today the WIMP
  option is highly squeezed. On the one hand,
  the parameter space of most of the
  concrete models is strongly constrained by direct, LHC and indirect
  searches. On the other hand, SUSY searches at the LHC have moved colored
  superpartner masses into TeV region, thus making SUSY less attractive
  from the viewpoint of solving the gauge chierarchy problem.

\item This does not mean too much, however: we would like to
  discover {\it one theory}
  and {\it one point in its parameter space}.

\item Hunt for WIMPs continues in numerous directions. Their potential 
  is far from being exhausted. Concerning direct searches,
  we will soon face
  the neutrino floor problem -- the situation where cosmic neutrino
  background will show up. It is time to look into ways to go beneath
  the neutrino floor.

\item With null results of WIMP searches, it makes a lot of sense to
  strengthen also searches for other dark matter candidates.

  \end{itemize}
  
\section{Axions}
\label{subs:axions}

Axion is a consequence of the Peccei--Quinn solution to the
strong CP-problem. It is a pseudo-Nambu--Goldstone boson
of an approximate Peccei--Quinn symmetry.

\subsection{Strong CP problem}
\label{sec:strongCP}

To understand the strong CP-problem, we begin with considering QCD
in the chiral limit $m_u = m_d = m_s =0$. The Lagrangian is
\begin{align*}
  L_{QCD, m=0} &
  = - \frac{1}{4} G_{\mu \nu}^a G^{a \mu \nu}
 + \sum_i \bar{q}_i i \gamma^\mu D_\mu q_i
\\
& =  - \frac{1}{4} G_{\mu \nu}^a G^{a \mu \nu}
+ \sum_i \left(\bar{q}_{L,i} i \gamma^\mu D_\mu q_{L,i}
+ \bar{q}_{R,i} i \gamma^\mu D_\mu q_{R,i}\right) \; ,
\end{align*}
where $i=u,d,s$.
As it stands, it is invariant under independent transformations
of left and right quark fields $ q_{L,i}$ and $ q_{R,i}$, each with
arbitrary unitary
matrices.  Naively, this means that the theory possesses
large symmetry
\be
 SU(3)_L \times  U(1)_L\times SU(3)_R \times U(1)_R 
= 
SU(3)_L \times SU(3)_R \times U(1)_B \times U(1)_A
\label{nov27-19-1}
\ee
  where vector $U(1)_B$ is baryon number symmetry,
 $q_i \to \e^{i\alpha} q_i$, while axial $U(1)_A$   
  act as $q_i \to \e^{i\beta \gamma^5} q_i$.

  The symmtery \eqref{nov27-19-1}
  is spontaneously broken: there exist quark condensates
  in QCD vacuum:
  \be
  \langle \bar{u}_L u_R \rangle =  \langle \bar{d}_L d_R \rangle =
 \langle \bar{s}_L s_R \rangle=
  \frac{1}{2}
 \langle \bar{q} q \rangle \sim \Lambda_{QCD}^3
 \label{nov21-19-1}
 \ee
   The
unbroken symmetry $SU(3)_V$ rotates left and right quarks
together (this is the well known flavor $SU(3)$);
$U(1)_B$ also remains unbroken.

Spontaneous breaking of global symmetry always leads to the presence of
Nambu--Goldstone bosons. Naively, one expects that there are 9
Nambu--Goldstone bosons: 8 of them come from symmetry breaking
$SU(3)_L\times SU(3)_R \to SU(3)_V$, and one from $U(1)_B \times U(1)_A
\to U(1)_B$ (since the original symmetry is explicitly broken by
quark masses, these should be pseudo-Nambu--Goldstone bosons with
non-zero mass). However, there are only 8 light pseudoscalar
particles whose properties
are well described by Nambu--Goldstone theory: these are
 $\pi^{\pm}$, $\pi^0$,
$K^{\pm}$, $K^0$, $\bar{K}^0$, $\eta$. Indeed, their masses squared
are proportional to quark masses, e.g.,
$m_\pi^2 = (m_{u} + m_d)  \langle \bar{q} q \rangle/f_{\pi}^2$.
Importantly,  yet another pseudoscalar
$ \eta^\prime$ is heavy and does not behave like
pseudo-Nambu--Goldstone boson.

The reason for this mismatch (absence of the 9th pseudo-Nambu--Goldstone boson)
is that $U(1)_A$ is not, in fact,
a symmetry of QCD even in the chiral limit. The corresponding axial
current suffers, at the quantum level, Adler--Bell--Jackiw (triangle, or
axial) anomaly,
\[
\d_\mu J_A^\mu \neq 0 \; .
\]
This means that the axial charge is not conserved, and thus the $U(1)_A$
is explicitly broken. We discuss this phenomenon in little more detail
in Sec.~\ref{sec:EW-B-violation} in the conext of electroweak baryon number
non-conservation.

The strong 
CP-problem~\cite{'tHooft:1976up,Callan:1976je,Jackiw:1976pf}
emerges in the following way.
One considers quark mass terms in the Standard Model
Lagrangian, which are obtained from the Yukawa interaction terms
with non-zero Higgs expectation value. The common lore is that
one can perform unitary rotations of quark fields to make
quark mass terms real (and in this way generate CKM matrix
in quark interactions with $W$-bosons). This is not quite true,
precisely because one cannot freely use $U(1)_A$-rotation.
In fact, by performing $SU(3)_L \times SU(3)_R \times U(1)_B$-rotation,
one casts the mass term of light quarks into the form
\[
L_m = \mbox{e}^{i\theta} \cdot m_{ij}^{CKM} \bar{q}_{L,i} q_{R, j} + h.c.\; ,
\]
where $ m_{ij}^{CKM}= \mbox{diag} (m_u, m_d, m_s)$
is real diagonal matrix, and $\theta$ is some phase.
Naively, this phase can be rotated away by axial rotation of
all three light
quark fields, $q_i \to \mbox{e}^{-i \theta \gamma_5/2} q_i$,
but, as we discussed, this is not an innocent field redefinition.
What happens instead is that this transformation generates
an extra term in the QCD Lagrangian
\begin{equation}
\label{chap8-true-axion-1+}
\Delta L=\frac{\alpha_s}{8\pi}\cdot\theta \cdot G_{\mu\nu}^a\tilde
G^{\mu\nu\;a}\;, 
\end{equation}
where $\alpha_s$ is the $SU(3)_c$ gauge coupling,
$G_{\mu\nu}^a$ is the gluon field strength,
$\tilde
G^{\mu\nu\;a} = \frac{1}{2} \epsilon^{\mu\nu\lambda\rho}G^a_{\lambda\rho}$
is the dual tensor.
The term
\eqref{chap8-true-axion-1+} is invariant under gauge
symmetries of the Standard Model, but it violates
P and CP. Similar term, but with another parameter $\theta_0$
instead of $\theta$, can already exist in the initial QCD Lagrangian.
The combined parameter
\[
\bar{\theta} = \theta + \theta_0
\]
is a ``coupling constant'' that cannot be removed by
field redefinition, and QCD with non-zero $\bar{\theta}$
violates CP.

 Let us show explicitly that
 the parameter $\bar{\theta}$ is physical, i.e., some physical quantities depend
 on $\bar{\theta}$.
To this end, we perform chiral rotation of
light quark fields  $q_i \to \mbox{e}^{+i \bar{\theta} \gamma_5/2} q_i$
to get rid of the term
\eqref{chap8-true-axion-1+}
and generate the phase in the quark mass terms
\[
L_m = \sum_i \mbox{e}^{i \bar{\theta}} m_i  \bar{q}_{L,i} q_{R, i} + h.c. 
\]
Let us consider for simplicity two light quark flavors $u$ and $d$
with equal masses $m_u = m_d \equiv m_q \sim 4$~MeV and calculate the
vacuum energy density in such a theory. We use perturbation theory
in quark masses, and work to the leading order.
Then the $\bar{\theta}$-dependent part of the
vacuum energy density is $V(\bar{\theta}) = - \langle L_m \rangle$.
We recall that $\langle \bar{q} q \rangle$ is non-zero in the chiral
limit, see
\eqref{nov21-19-1}, and observe that it is real,
provided that the term \eqref{chap8-true-axion-1+} is absent
(no spontaneous CP-violation in the chiral limit). Importantly,
$\langle \bar{q} q \rangle$ does not have an arbitrary phase, since
the arbitrariness of this phase would mean that $U(1)_A$ is a
(spontaneously broken) symmetry, which is not the case, as we
discussed above. Thus, we obtain
\be
V(\bar{\theta}) =  - \langle L_m \rangle = - 2m_q \langle \bar{q} q
\rangle \cos \bar{\theta}
= - \frac{m_\pi^2 f_\pi^2}{4} \cos \bar{\theta} \; .
\label{nov21-19-5}
\ee
This shows explicitly that $\bar{\theta}$ is a physically relevant parameter.
We note in passing that the expression for $V(\bar{\theta})$ is,
in fact, more complicated, especially
for $m_u\neq m_d$ and also for
three quark flavors, but the main property ---
minimum at $\bar{\theta}=0$ --- is intact.

Thus, $\bar{\theta}$ is a new coupling constant that can take any value
in the interval $(-\pi, \pi)$.
There is no reason to think that 
$\bar{\theta} = 0$.
The term \eqref{chap8-true-axion-1+} has
a dramatic phenomenological consequence:
it generates
 electric dipole moment (EDM) of neutron
$d_n$, which is estimated as~\cite{Kim:2008hd}
\begin{equation}
\label{chap8-true-axion-4++}
d_n\sim \bar{\theta}\cdot 10^{-16}\cdot e\cdot \mbox{cm}\;.
\end{equation}
Neutron EDM is strongly constrained  experimentally, 
\begin{equation}
\label{chap8-true-axion-4*}
d_n\lesssim 3\cdot 10^{-26}\cdot e\cdot \mbox{cm}\; .
\end{equation}
This leads to the bound on the parameter
 $\bar{\theta}$, 
\[
|\bar{\theta}|<0.3 \cdot 10^{-9}\;.
\]
The problem to explain so small value of 
$\bar{\theta}$ is precisely the strong 
CP-problem.

A solution to this  problem  does not exist
within the Standard Model.
The solution is offered by models with axion.
The idea of these models is to promote $\bar{\theta}$-parameter to
a field, which is precisely the axion field.
This can be done in various ways. Two well-known ones
are 
Dine--Fischler--Srednicki--Zhitnitsky~\cite{Dine:1981rt,Zhitnitsky:1980tq}
(DFSZ)
and Kim--Shifman--Vainshtein--Zakharov~\cite{Kim:1979if,Shifman:1979if} 
(KSVZ) mechanisms\footnote{Earlier and
  even simpler is Weinberg--Wilczek
  model~\cite{Weinberg:1977ma,Wilczek:1977pj}, but it is ruled out
  experimentally.}. In either case, one introduces a complex scalar
field $\Phi$ and makes sure that without QCD effects, the theory is
invariant under global Peccei--Quinn $U(1)_{PQ}$ symmetry.
Under this symmetry, the field $\Phi$ transforms as
$\Phi \to \mbox{e}^{i\alpha} \Phi$. One also arranges that the QCD effects
make this symmetry anomalous, very much like $U(1)_A$, so that
under the $U(1)_{PQ}$-transformation, the Lagrangian obtains an additional
contribution
\be
\Delta L=C \frac{\alpha_s}{8\pi}\cdot\alpha \cdot G_{\mu\nu}^a\tilde
G^{\mu\nu\;a} \; ,
\label{nov21-19-2}
\ee
where $C$ is a model-dependent constant of order 1.
A simple example 
is KSVZ model: one adds a new quark $\psi$ which interacts with $\Phi$
as follows:
\be
L_{int} = h\Phi \bar{\psi}_L \psi_R + h.c.
\label{nov21-19-3}
\ee
where $h$ is Yukawa coupling.
Then the Peccei--Quinn transformation is
\[
\Phi \to \mbox{e}^{i\alpha} \Phi \; , \;\;\;\;\;
\psi \to \mbox{e}^{i\alpha \gamma^5/2} \psi \; ,
\]
while ``our'' quark fields are $U(1)_{PQ}$-singlets.
In the same way as above, this transformation induces the
term \eqref{nov21-19-2}, as required.

Now, one arranges the scalar potential for $\Phi$ in such a way that
the Peccei--Quinn symmetry is spontaneously broken at very high energy.
If not for QCD effects, the phase of $\Phi$ would be a massless
Nambu--Goldstone boson, the axion. At low energies one writes
$\Phi = f_{PQ} \cdot \mbox{e}^{i\theta (x)}$, where $f_{PQ}$ is the
Peccei--Quinn vacuum expectation value. In the absence of
QCD, the field $\theta$ is rotated away
from the non-derivative part of the action by the Peccei--Quinn rotation,
while it reappears  in the form \eqref{nov21-19-2}
when QCD is switched on. 
We see that the parameter $\bar{\theta}$ is indeed promoted to
a field, and this parameter disappeas upon
shifting $\theta (x) \to \theta(x) - \bar{\theta}$; we are free to set $\bar{\theta}=0$.
Now, there is a potential for the field $\theta$; it is given precisely
by eq.~\eqref{nov21-19-5} with $\bar{\theta}$ replaced by $\theta$.
Hence, the low energy axion Lagrangian reads
\[
L_a = \frac{f^2_{PQ}}{2} \d_\mu \theta \d^\mu \theta - V(\theta) \; .
\]
As usual,
the first term here comes from the kinetic term for the field $\Phi$.
We recall that the minimum of $V(\theta)$ is at $\theta=0$;
at this
value CP is not violated, the strong CP problem is solved!
We now make field redefinition, $\theta = a/f_{PQ}$ and find from
\eqref{nov21-19-5} that the quadratic axion Lagrangian is
\[
L_a =  \frac{1}{2} \d_\mu a \d^\mu a - \frac{m_a^2}{2} a^2 \; ,
\]
where
\be
m_a = \frac{m_\pi f_\pi}{2 f_{PQ}} \; .
\label{nov21-19-10}
\ee
The axion is {\it pseudo-}Nambu-Goldstone boson.

To summarize, for large Peccei--Quinn scale
$f_{PQ} \gg M_W$,
axion is a light particle whose interactions with the
Standard Model fields are very weak.
Like for any Nambu--Goldstone field, the tree-level
interactions of axion
with quarks and leptons
are described by
the generalized Goldberger--Treiman formula 
\begin{equation}
\label{chap8-true-axion-add-1+RR+}
L_{af}=\frac{1}{f_{PQ}}\cdot\d_\mu a\cdot J^\mu_{PQ}\; .
\end{equation}
Here
\begin{equation}
\label{chap8-true-axion-add-1+}
J^\mu_{PQ}=\sum_f e^{(PQ)}_f\cdot\bar f\gamma^\mu\gamma^5 f\;.
\end{equation}
The contributions of fermions to the current
$J^\mu_{PQ}$ are proportional to their PQ charges
$e^{(PQ)}_f$;
these charges are model-dependent.
There is necessarily interaction of axions with gluons, see
\eqref{nov21-19-2},
\begin{equation}
\label{chap8-true-axion-add-1a}
L_{ag}=C_g\frac{\alpha_s}{8\pi}\cdot\frac{a}{f_{PQ}}\cdot
G_{\mu\nu}^a \tilde G^{\mu\nu\;a}
\ee
Finally, there is axion-photon coupling
\be
\label{chap8-true-axion-add-1*}
L_{a\gamma}= g_{a\gamma\gamma} \cdot a
F_{\mu\nu} \tilde F^{\mu\nu} \; , \;\;\;\;\;
g_{a\gamma\gamma}=C_\gamma\frac{\alpha}{8\pi f_{PQ}}
\;,
\end{equation}
The dimensionless constants
$C_g$ and $C_\gamma$ are  model-dependent
and, generally speaking,
not very much different from 1.
The main free parameter is $f_{PQ}$, while
the axion mass is related to it via eq.~\eqref{nov21-19-10}; numerically,
\begin{equation}
\label{chap8-true-axion-add-2+}
m_a =  6~\mu\mbox{eV}~\cdot
\l
\frac{10^{12}~\mbox{GeV}}{f_{PQ}}
\r
\;.
\end{equation}
There are astrophysical bounds on the strength of axion interactions
$f_{PQ}^{-1}$ and hence on the axion mass. Axions in theories
with
$f_{PQ}\lesssim 10^{9}$~GeV, which are heavier than
about $10^{-2}$~eV, would be intensely produced in stars and supernovae
explosions. This would lead to contradictions with 
observations. So, we are left with very light axions,
$m_a \lesssim 10^{-2}$~eV. These very light and very
weakly interacting axions are interesting dark matter
candidates\footnote{We note in passing that axions may be heavy
  instead~\cite{Rubakov:1997vp}. This case
  is irrelevant for dark matter.}.

\subsection{Axions in cosmology}

Axions can serve as dark matter if they
do not decay in the lifetime of the Universe.
The main decay channel of the light axion is decay into two
photons. The axion width is calculated as
\[
\Gamma_{a \to \gamma \gamma} = \frac{m_a^3}{4\pi} \l C_\gamma
\frac{\theta}{8\pi f_{PQ}} \r^2 \; ,
\]
where the quantity in parenthesis is the axion-photon coupling,
see \eqref{chap8-true-axion-add-1*}.
We recall the relation \eqref{nov21-19-10}
 and obtain axion lifetime 
\[
\tau_a=\frac{1}{\Gamma_{a\to\gamma\gamma}}
=\frac{64\pi^3 m_\pi^2f_\pi^2}{C_\gamma^2 \alpha^2m_a^5}
\sim
10^{24}~\mbox{s}~\cdot\l\frac{\mbox{eV}}{m_a}\r^5\;.
\]
By requiring that this lifetime exceeds the age of the Universe,
$\tau_a>t_0\approx 14$~billion years, we find a very weak bound on the
mass of axion as dark matter candidate, 
$m_a<25~\mbox{eV}$.

Thermal production of axions in the early Universe not very relevant,
since even if they were in thermal equilibrium at high temperatures,
their thermally produced
present number density is substantially smaller than that of
photons and neutrinos, and with their tiny mass they do not
contribute much into the energy density\footnote{If axions were in
  thermal equilibrium, they contrubute to the effective number
  of ``neutrino'' species
  $N_{eff}$. This contribution, however, is smaller than the
  current precision~\cite{Aghanim:2018eyx} of the determination of $N_{eff}$,
  which is equal to is 0.17.}.
This is a welcome property, since thermally produced axions, if they
composed substantial part of dark matter,
would be {\it hot} dark matter, which is ruled out.

There are at least two
mechanisms of axion production in the early Universe
that can provide not only right axion abundance but
also  small initial velocities of axions. The latter
property makes axion a {\it cold} dark matter
candidate, despite its
very small  mass.


One
mechanism~\cite{Preskill:1982cy,Abbott:1982af,Dine:1982ah} is called
misalignment scenario. It
assumes that the Peccei--Quinn symmetry is spontaneously
broken before the beginning of the hot epoch, $\langle \Phi \rangle \neq 0$.
This is indeed the case in inflationary framework, if $f_{PQ}$ is
higher than both inflationary Hubble parameter (towards the inflation end)
and the reheat temperature of the Universe. 
In this case the axion field (the phase of the field $\Phi$)
is homogeneous over the entire visible universe, and initially it can
take any value $\theta_0$ between $-\pi$ and $\pi$. 
As we have seen in \eqref{nov21-19-5},
the axion potential is
proportional to the quark condensate
$\la \bar{q} q \ra$. This condensate vanishes 
at high temperatures, $T \gg
\Lambda_{QCD}$, and
the 
axion potential is negligibly small.
As the temperature decreases, the axion potential builds up.
Accordingly, the axion mass
increases from zero to $m_a$; hereafter
$m_a$ denotes the zero-temperature axion mass.
The axion field practically does not
evolve when
$m_a (T) \ll H(T)$ and at the time when
$m_a (T) \sim H(T)$ it starts to roll down from the initial value
$\theta_0$ to the minimum $\theta=0$ and then it oscillates.
During all these stages of evolution, the axion field
is homogeneous in space. The homogeneous oscillating field
can be interpreted as a collection of scalar quanta
with zero spatial momenta, the axion condensate.
This is indeed cold dark matter.

Let us estimate the present energy density
of axion field in this picture.
The oscillations start at the time
$t_{osc}$ when $  m_a (t_{osc}) \sim H(t_{osc})$.
At this time, the energy density of the axion field is estimated as
\[
\rho_a (t_{osc}) \sim  m_a^2 (t_{osc}) a_0^2=
m_a^2 (t_{osc}) f_{PQ}^2 \theta_0^2 \; .
\]
The number density of axions at rest at 
the beginning of oscillations
is estimated as
\be
 n_a (t_{osc}) \sim \frac{\rho_a (t_{osc})}{m_a (t_{osc})}
\sim  m_a (t_{osc}) f_{PQ}^2 \theta_0^2
\sim  H (t_{osc}) f_{PQ}^2 \theta_0^2 \; .
\nonumber 
\ee
This number density, as any number density of non-relativistic
particles, then decreases as
$a^{-3}$.
Axion-to-entropy ratio at time
$t_{osc}$ is
\[
\frac{n_a}{s} \sim   \frac{H (t_{osc}) f_{PQ}^2}{\frac{2\pi^2}{45} g_* T_{osc}^3}
\cdot  \theta_0^2 \simeq 
\frac{f_{PQ}^2}{\sqrt{g_*} T_{osc} M_{Pl}}\cdot  \theta_0^2 \; ,
\]
where we use the usual relation $H = 1.66 \sqrt{g_*} T^2 /
M_{Pl}$.  
The axion-to-entropy ratio remains constant after the beginning of
oscillations, so the present mass density of  axions is
\be 
\rho_{a, 0} = \frac{n_a}{s} m_a s_0 \simeq \frac{m_a
f_{PQ}^2}{\sqrt{g_*} T_{osc} M_{Pl}} s_0 \cdot \theta_0^2 \; .
\label{ch8-addit2}
\ee
%
To obtain a simple estimate, let us set
$T_{osc} \sim \Lambda_{QCD}
\simeq 200~\mbox{MeV}$ and make use of  \eqref{chap8-true-axion-add-2+}.
We find
\be
 \Omega_a \equiv \frac{\rho_{a, 0}}{\rho_c} \simeq
\l \frac{10^{-6}~\mbox{eV}}{m_a} \r \theta_0^2 \; .
\label{ch8-addit3}
\ee
The natural assumption about the initial phase is 
$\theta_0\sim \pi/2$. Hence, axion of mass
$m_a=(\mbox{a~few}) \cdot 10^{-6}$~eV is a good dark matter 
candidate. Note that axion of lower mass
  $m_a<10^{-6}$~eV
may also serve as dark matter particle, if for some reason
the initial phase
$\theta_0$ is much smaller than $\pi/2$.

More precise estimate is obtained by taking into account
the fact that
that the axion mass smoothly depends on 
temperature:
\begin{equation}
\Omega_a\simeq0.2\cdot\theta_0^2\cdot
\l\frac{4\cdot10^{-6}~\mbox{eV}}{m_a}\r^{1.2} 
\nonumber
\end{equation}
We see that our crude estimate 
\eqref{ch8-addit3} is fairly accurate.

We note that in the misalignment scenario,
and in the inflationary framework, the initial phase
$\theta_0$ is not quite homogeneous in space.
At inflationary stage, vacuum fluctuations of all
massless or light scalar fields get enhanced.
As a result, scalar fields become inhomogeneous
on scales exceeding the inflationary Hubble scale
$H_{infl}^{-1}$. The amplitudes of these inhomogeneities
(for canonically normalized fields) are equal to
$H_{infl}/(2\pi)$. Phase perturbations give rise to perturbations of
axion dark matter energy density, which are uncorrelated with
perturbations of conventional matter. These uncorrelated dark matter
perturbations are called isocurvature (or entropy) modes.
Cosmological observations show that their contribution cannot exceed
a few per cent of the dominant adiabatic mode. This leads to
a constraint~\cite{Visinelli:2009zm}
on inflationary Hubble parameter $H_{infl}$ or, equivalently,
on the energy scale of inflation (energy density of the inflaton field)
\[
V_{infl}^{1/4} \lesssim 10^{13}~\mbox{GeV} \; .
\]
This makes the misalignment mechanism somewhat contrived.
Reversing the argument, detection of the dark matter entropy mode
would be an interesting hint towards the nature of dark matter.

Another mechanism of axion production in the early
Universe works under the assumption which is
opposite to the main assumption of the misalignment scenario.
Namely, one assumes that the Peccei--Quinn symmetry is restored
at the beginning of hot epoch, and gets spontaneously broken
at temperature of order $T\sim f_{PQ}$ at hot stage.
Then the phase of the field $\Phi$ is uncorrelated
at distances exceeding the size of the  horizon at that time.
In principle, one should be able to predict the value
of $f_{PQ}$ and hence $m_a$ in this scenario, since there is
no uncertainty in the initial conditions. However,
the dynamics in this case is quite complicated.
Indeed, the uncorrelated phase gives rise to the production of
global cosmic strings~\cite{Vilenkin:1982ks}
 --- topological  defects that
exist in theories with spontaneously broken global
$U(1)$ symmetry ($U(1)_{PQ}$ in our case;
for a discussion see, e.g.,
Ref.~\cite{Battye:1999bd}). At the QCD transition epoch,
defects of another type, axion domain walls, are created.
Then all these defects get destructed, giving rise to
the production of axions.   The analysis of this dynamics
has been made by various authors, see, e.g.,
Refs.~\cite{Klaer:2017ond,Kawasaki:2018bzv}, but it is fair
to say that there is no compelling prediction for $m_a$ yet.
A reasonable estimate of the axion mass is
(Ref.~\cite{Klaer:2017ond} claims $m_a = 2.6\cdot 10^{-5}$~eV)
\[
m_a =(\mbox{a~few})\cdot 10^{-5}~\mbox{eV} \; .
\]

To end up with cosmological aspects of axion dark matter, we note that
it has interesting phenomenology in the present Universe.
Axions tend to form mini-clusters~\cite{Kolb:1993zz}
which can be disrupted
and form streams of dark matter~\cite{Tinyakov:2015cgg}.
Axions also form
bose-stars~\cite{Levkov:2018kau}. All this exotica is
relevant to both astrophysics and axion search.

\subsection{Axion search}

\begin{figure}[htb!]
  \begin{center}
   \vspace{-1cm}
  \includegraphics[angle=0,width=0.5\textwidth]{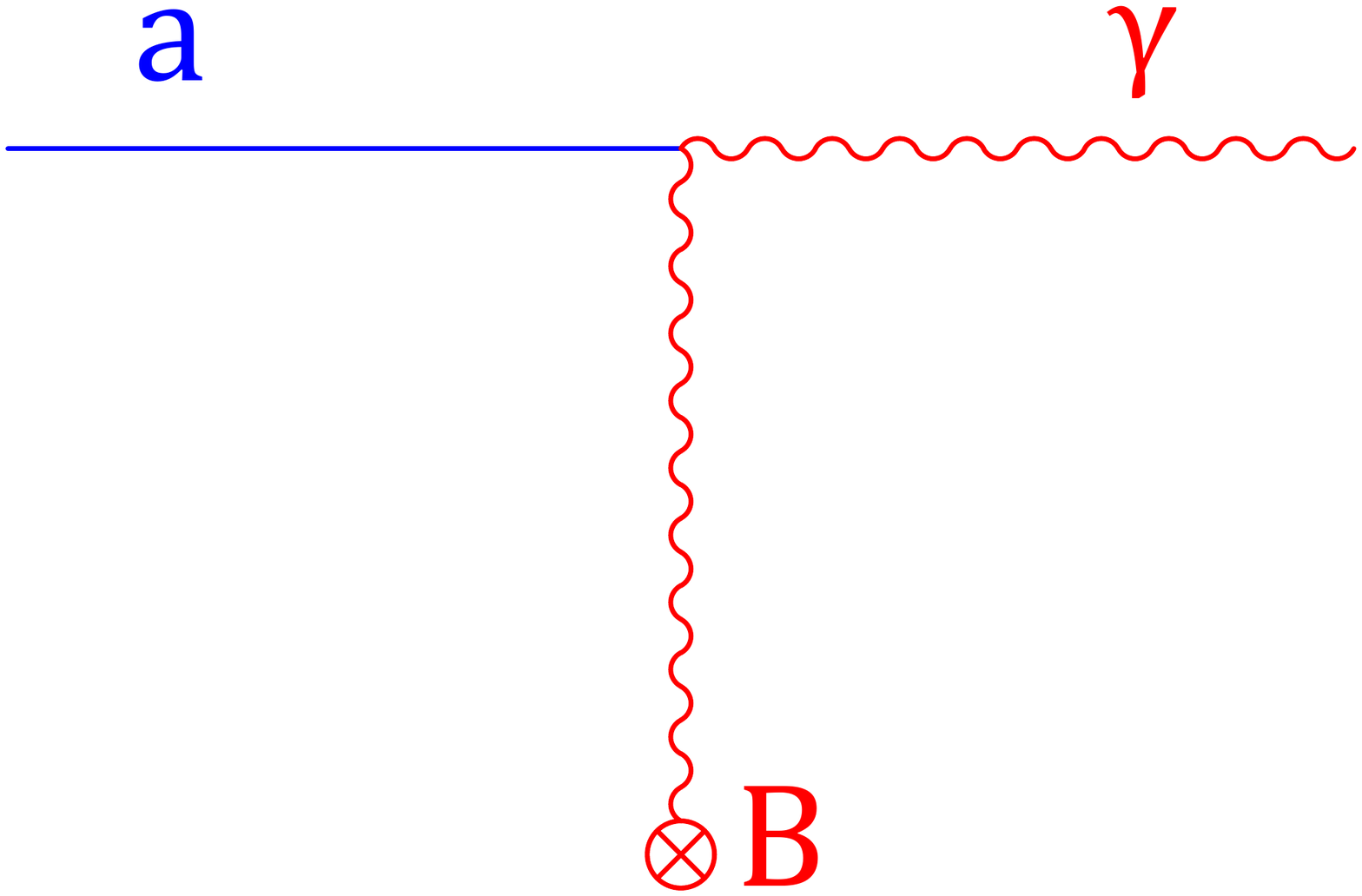}
  \end{center}
  \vspace{-1cm}
\caption{Axion-photon conversion in magnetic field.
\label{a-gamma}
}
\end{figure}

\begin{figure}[htb!]
    \vspace{-3cm}
    \begin{center}
    \hspace{2cm}  
\includegraphics[angle=90,width=0.75\textwidth]{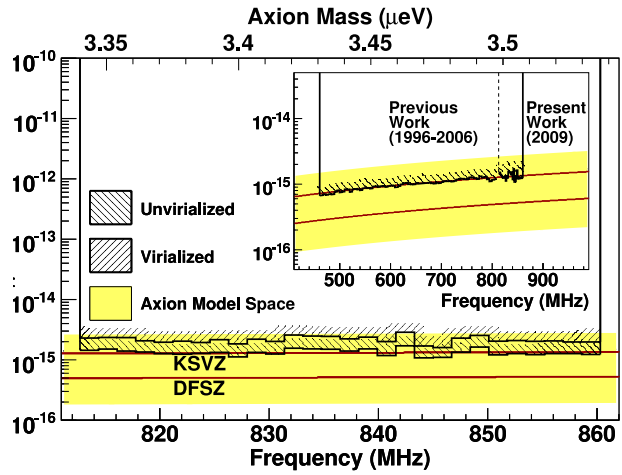}
\end{center}
\begin{center}
  \hspace{-4.5cm}
  \includegraphics[angle=0,width=1.0\textwidth]{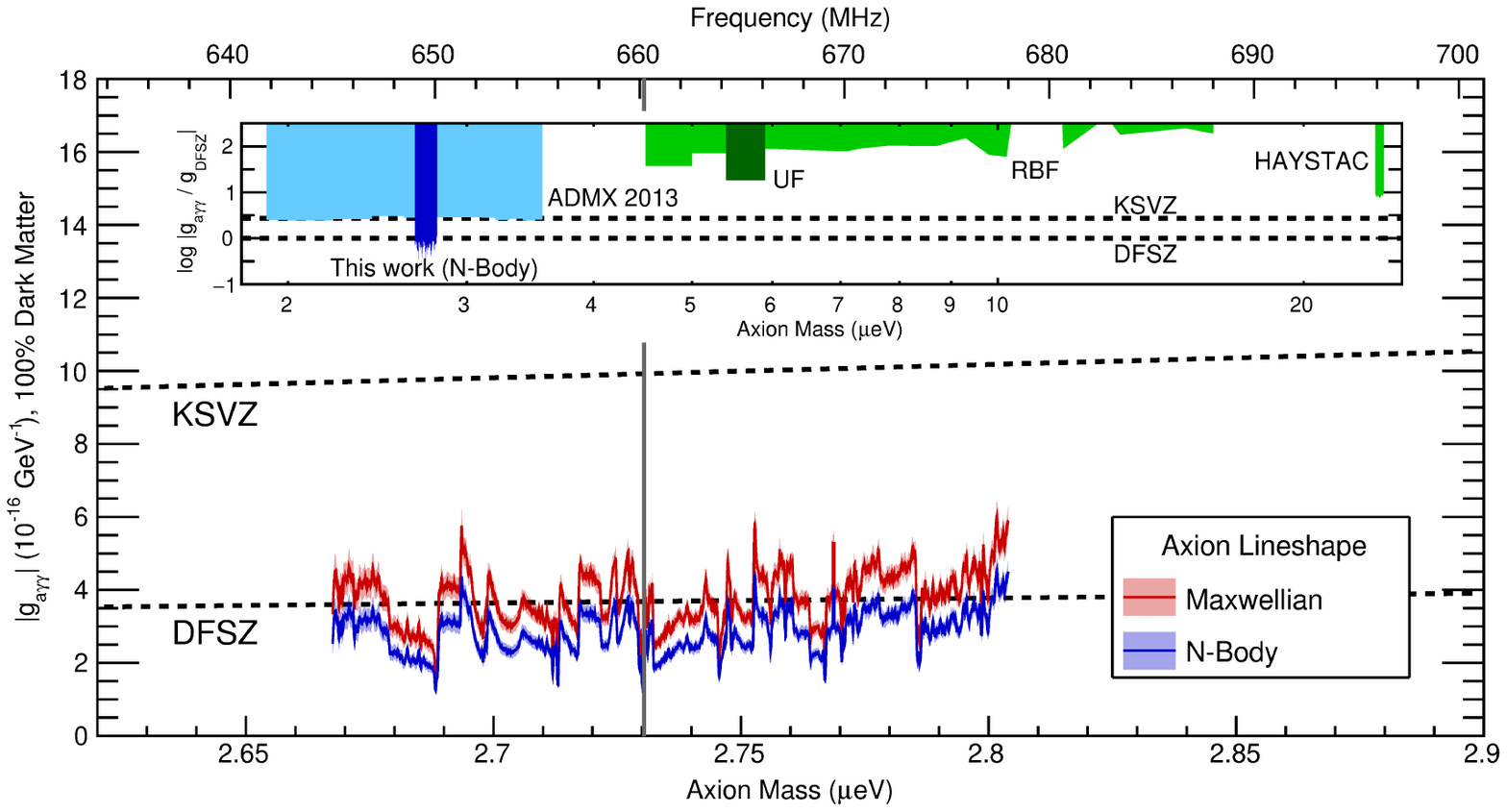}
\end{center}
\vspace{-6.5cm}
\caption{Limits on the axion-photon coupling for various  axion masses.
  Lines labeled KSVZ and DSVZ refer to predictions of the two  axion
  models
  under the assumption that axions make the whole of dark matter. 
  Shown are limits published by ADMX collaboration
  in 2010~\cite{Asztalos:2009yp} (upper panel) and
  in 2018~\cite{Du:2018uak}. Note the limited ranges of masses
  spanned durin the long period of time. Note also that the
  recent limits (lower panel) reach almost entire range of
  axion-photon couplings predicted by varios axion models.
\label{admx-long}
}
  \end{figure}

Search for dark matter axions with mass
$m_a\sim10^{-5} - 10^{-6}$~eV is difficult, but not impossible.
One way is to search for axion-photon conversion
in a resonant cavity filled with strong magnetic field.
Indeed, in the background magnetic field, the axion-photon
interaction~\eqref{chap8-true-axion-add-1*} leads to the conversion
$a \to \gamma$, see Fig.~\ref{a-gamma}. 
Axions of mass $10^{-5} - 10^{-6}$~eV are converted
to photons of frequency $\nu = m/(2\pi) = 2 - 0.2$~GHz (radiowaves;
$m=10^{-6}~\mbox{eV} \longleftrightarrow \nu= 240$~MHz). To collect
  reasonable number of conversion photons, one needs cavities of high quality
  factor $Q$, which have small bandwidths. This means that one goes in
  small steps in $m_a$, and the whole search takes long time.
  This is illustrated in Fig.~\ref{admx-long}.

\begin{figure}[!htb]
\begin{center}
\includegraphics[width=0.6\textwidth]{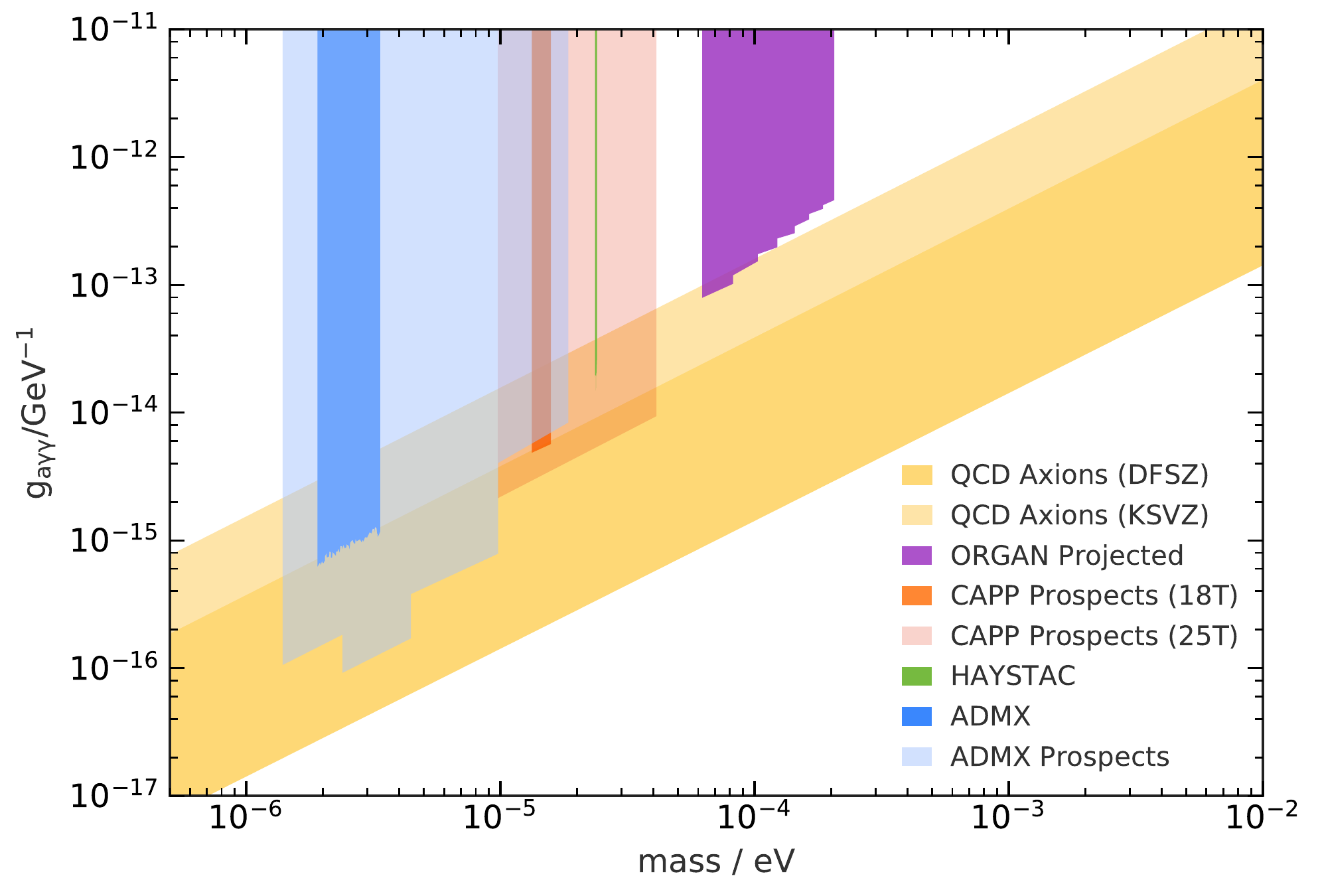}
\end{center}
\caption{
\label{capp}
Future prospects of dark matter axion search with resonant
cavities~\cite{Battesti:2018bgc}.
}
\end{figure}

  The hunt for dark matter axions has been intensified recently.
  A new set of resonant cavity experiments, CAPP, is under preparation,
  see Fig.~\ref{capp}. A new approach to search for heavier dark matter
  axions
  with $m_a \gtrsim 4\cdot 10^{-5}$~eV has been suggested by
  MADMAX interest group~\cite{Majorovits:2017ppy}. Other axion
  search experiments are reviewed in
  Refs.~\cite{Battesti:2018bgc,Graham:2015ouw}.

  \subsection{Axion-like particles (ALPs)}

  There may exist light, weakly interacting scalar or pseudoscalar
  particles other than axions. They are called axion-like particles,
  ALPs, and they may emerge as pseudo-Nambu--Goldstone
  bosons of some new approximate global symmetry. We have discussed one
  example,  fuzzy dark matter,
  in Sec.~\ref{subs:hints}.
  Unlike the axion case, where axion-photon coupling is related,
  albeit in somewhat model-dependent way, to its mass via
  eqs.~\eqref{chap8-true-axion-add-1*} and \eqref{chap8-true-axion-add-2+},
  ALP mass and coupling to photons are both arbitrary parameters.
  Also, ALPs may interact with the Standard Model fermions, and that
  coupling is again a free parameter.
  ALPs may or may not be dark matter candidates; search for them is of
  interest independently of dark matter problem.

  If ALP is a dark matter candidate, instruments described in
  previous subsection --- ``haloscopes'' --- are capable for searching
  for dark matter ALPs, and it makes sense to extend the search to
  as wide mass range as possible. In this regard, it is worth mentioning
  that CASPEr experiment~\cite{Budker:2013hfa} is going to be sensitive
  to very light ALPs, $m\lesssim 10^{-9}$~eV, and very small
  ALP-fermion couplings.

\begin{figure}[htb!]
    \vspace{-1cm}
\begin{center}
  \includegraphics[angle=0,width=0.5\textwidth]{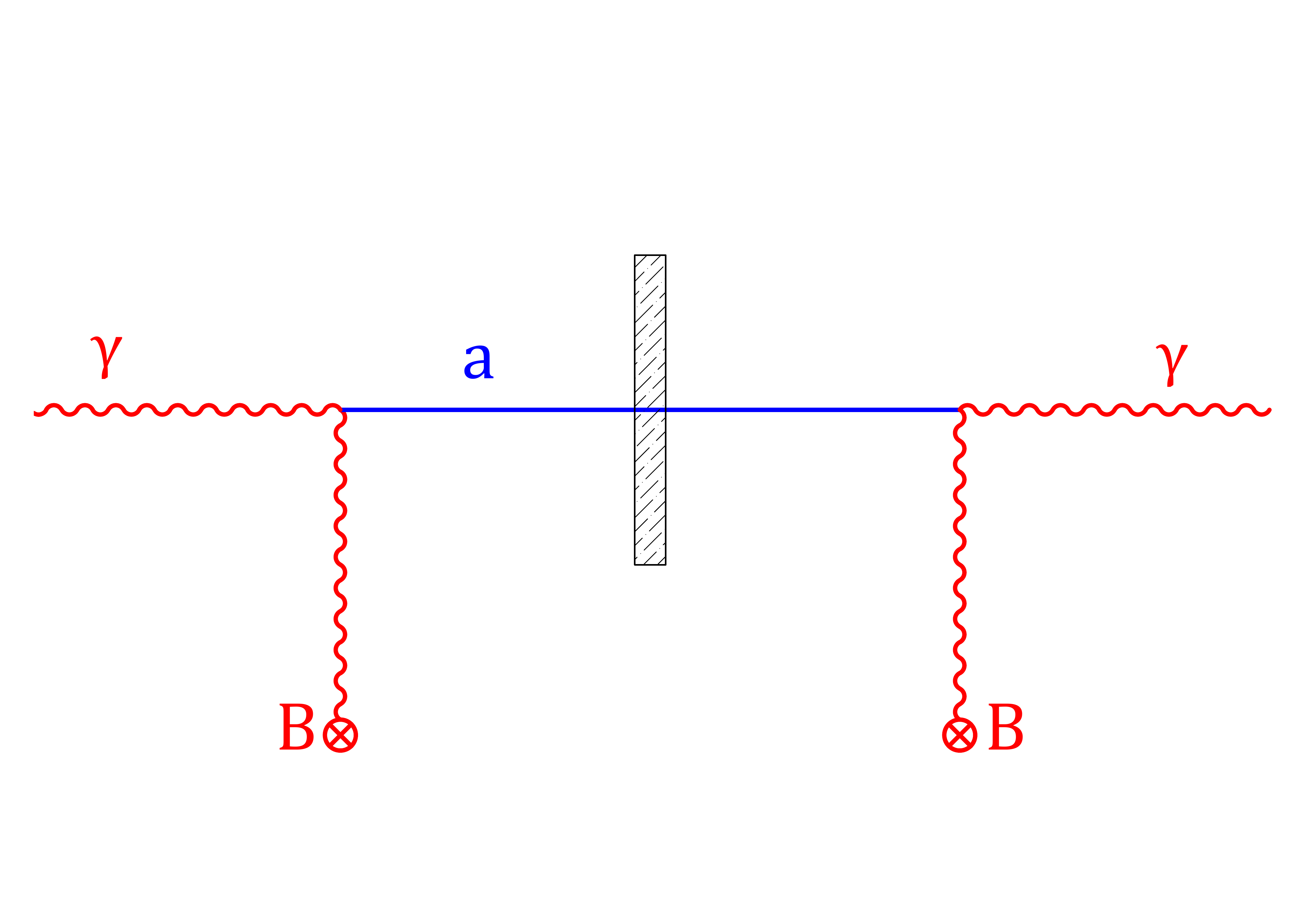}
\end{center}
\vspace{-1cm}
\caption{``Light shining through a wall'': laser light shining from the
  left is converted into axions in magnetic field of a magnet placed
  before the wall,
  axions pass through the wall and are converted into photons
  by a magnet behind the wall; the latter photons are detected by
  highly sensitive photon detector.
\label{light-through-wall}
}
\end{figure}

  ALPs may be produced in the Sun, and their flux may be detectable
  by ``helioscopes'', instruments
  searching for the axion-photon conversion in the magnetic
  field of a magnet looking at the Sun. One such instrument,
  CAST, has been operating for long time, whereas other experiments,
  IAXO and TASTE, are planned. Another way to search for ALPs
  makes use of the idea of ``light shining through a wall'',
  see Fig.~\ref{light-through-wall}; this idea is implemented in
  ALPS-I, ALPS-II expertiments. For a review of these approaches see,
  e.g., Ref.~\cite{Graham:2015ouw}. Finally, ALPs can be searched in
  beam-dump experiments and in decays
  of $K$- and $B$-mesons. Interesting limits are obtained by  CHARM and BaBar
  experiments, and a promising planned experiment is SHiP at
  CERN~\cite{Alekhin:2015byh}.

\begin{figure}[!htb]
     \vspace{-2cm}
\begin{center}
\includegraphics[width=0.8\textwidth]{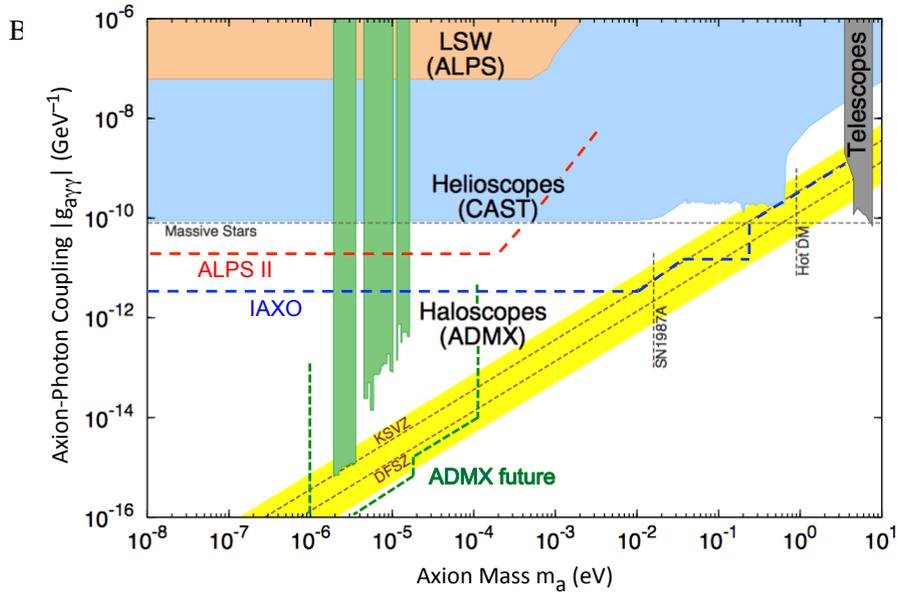}
\end{center}
\vspace{-1cm}
\caption{
\label{axion-summary}
Bounds on ALPs: ALP-photon coupling vs  ALP
mass~\cite{Graham:2015ouw}. Inclined straight strip with lines labeled
``KSVZ'' and ``DFSZ'' is the range of predictions of
axion models. Shaded regions are limits from existing experiments,
dashed lines show sensitivities of future searches.
}
\end{figure}

  Bounds and propects for search for
   light ALPs are summarized  Fig.~\ref{axion-summary}.

\section{Warm dark matter: sterile neutrinos}
\label{sterilnu}

As we discussed in Section~\ref{subsec:missing}, there
are arguments, albeit not yet conclusive, which
 favor warm, rather than cold, dark matter.
If WDM particles 
were in kinetic equilbrium at some epoch in the early Universe,
then their mass should be in the range $3 - 10$~keV.
Reasonably well motivated particles of this mass are
sterile neutrinos.

Sterile neutrinos --- massive
leptons $N$ which do not participate
in the Standard Model gauge interactions ---
are most probably required for giving masses
to ordinary, ``active'' neutrinos. The masses of sterile neutrinos
cannot be predicted theoretically.
Although sterile neutrinos
of WDM mass $m_{N} = 3 -10$~keV are not particularly plausible
from particle physics prospective, they are not pathological either.
In the simplest 
case
the creation of sterile neutrino
states
$|N\ra$ in the early Universe occurs due to their mixing with
active neutrinos 
$|\nu_\alpha\ra$, $\alpha=e,\mu,\tau$. In the approximation of 
mixing between two states only, we have
\be
|\nu_\alpha\ra=\cos\theta |\nu_1\ra + \sin\theta
 |\nu_2\ra\;, \;\;\;\;\;
|N\ra=-\sin\theta |\nu_1\ra + \cos\theta
 |\nu_2\ra\;,
\nonumber
\ee 
where
$|\nu_\alpha\ra$ and $|N\ra$ are active and sterile neutrino states,
$|\nu_1\ra$ and  $|\nu_2\ra$ are mass eigenstates of masses
$m_1$ and $m_2$, where we order $m_1 < m_2$,
and $\theta$ is the vacuum mixing angle between
sterile and active neutrino. This  mixing should be weak,
$\theta \ll 1$, otherwise sterile neutrinos would decay too
rapidly, see below.
The heavy state is mostly sterile neutrino
$|\nu_2\ra \approx
|N\ra$, and 
$m_2\equiv m_N$ is the sterile neutrino mass.

\begin{figure}[htb!]
\vspace{-1cm}
   \begin{center}
  \includegraphics[angle=0,width=0.5\textwidth]{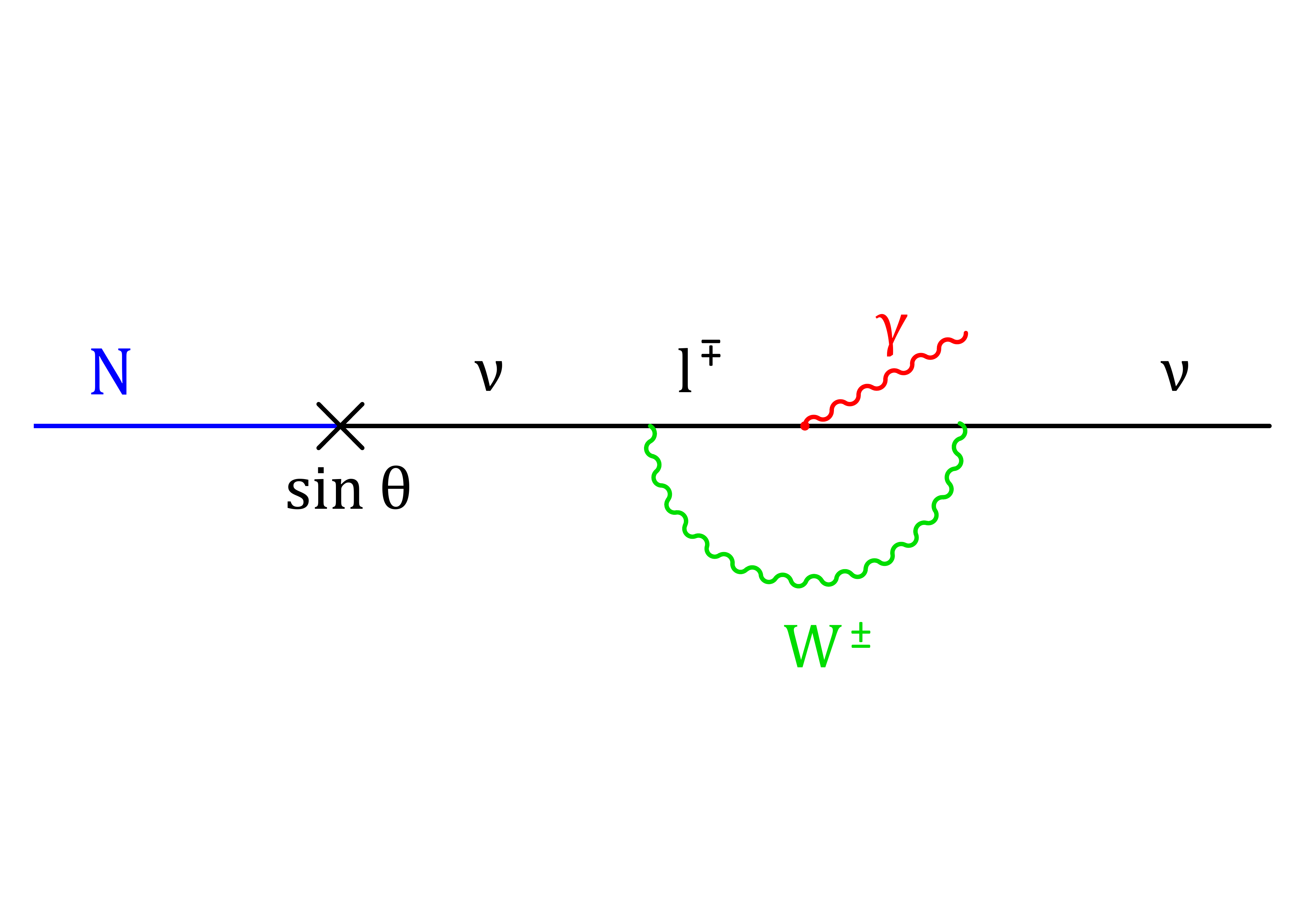}
   \end{center}
   \vspace{-1.5cm}
\caption{Sterile meutrino decay $N \to \nu_\alpha + \gamma$.
\label{snu-gamma}
}
 \end{figure}

\begin{figure}[!htb]
\hskip 0.1\textwidth
\includegraphics[width=0.7\textwidth]{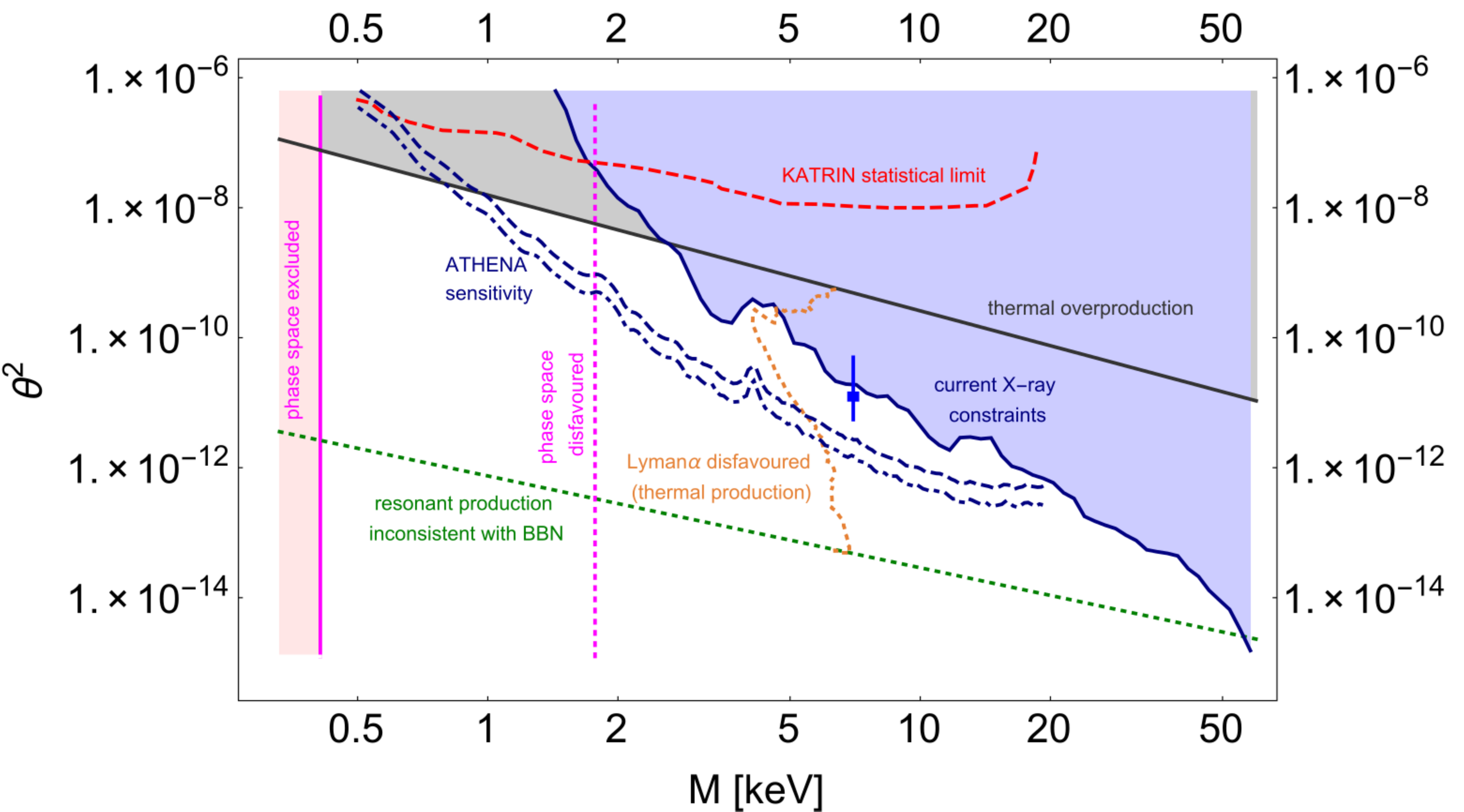}
\caption{Limits on sterile neutrino parameters
(mass $M$, mixing angle squared $\theta^2$) obtained from 
X-ray telescopes~\cite{Boyarsky:2018tvu}. Straight solid line refers to
sterile neutrino dark matter produced in non-resonant
oscillations, eq.~\eqref{chap7-sterile-4++}.
Region between this line and dotted line corresponds to
resonant mechanism that works in the Universe with
fairly large lepton asymmetry. Vertical lines show
very conservative limits coming from phase space and Lyman-$\alpha$
considerations, see Sec.~\ref{subsec:missing}. Regions left of these
lines are disfavored. In fact, for
non-resonant mechanism, the phase space constraint is $M \gtrsim 6$~keV.
Bullet with vertical interval shows the point corresponding
to putative 3.5~keV
line.
\label{snu-photon}
}
\end{figure}

The calculation of sterile neutrino abundance is fairly
complicated, and we do not reproduce it here. If there is no
sizeable lepton asymmetry in the Universe,
 the estimate is
\begin{equation}
\label{chap7-sterile-4++}
\Omega_{N}\simeq 0.2 \cdot \l\frac{\sin \theta}
{10^{-4}}\r^2 
\cdot \l\frac{m_N}{1~\mbox{keV}}\r^2\;.
\end{equation}
The energy spectrum of sterile neutrinos is nearly thermal.
Thus, sterile neutrino of mass
$m_\nu\gtrsim 1$~keV and small mixing angle
$\theta_\alpha\lesssim 10^{-4}$ would serve as dark matter candidate.
However, this range of masses and mixing angles is ruled out.
The point is that due to its mixing with
active neutrino,
sterile neutrino can decay into active neutrino and photon,
see Fig.~\ref{snu-gamma}.
\[
N \to \nu_\alpha + \gamma \; .
\]
The sterile neutrino decay 
width is proportional to $\sin^2 \theta$.
If sterile neutrinos are dark matter particles, their decays
would produce a narrow line in X-ray flux from cosmos
(orbiting velocity of dark matter particles in galaxies 
is small, $v \lesssim 10^{-3}$, hence the photons produced in their
two-body decays are nearly monochromatic). Leaving aside
a hint towards 3.5~keV line advocated in
Refs.~\cite{Bulbul:2014sua,Boyarsky:2014jta}
(see the discussion of its status in Ref.~\cite{Boyarsky:2018tvu}),
one makes use of
strong limits on such a line and translates them into 
limits on $\sin^2 \theta$.  These limits as function
of sterile neutrino mass, are shown in 
Fig.~\ref{snu-photon};
they rule out the range of masses giving the right mass density
of dark matter, eq.~\eqref{chap7-sterile-4++}.

A (rather baroque) 
way out~\cite{Shi:1998km} is to assume that there is failry large
lepton asymmetry in the Universe. Then the oscillations
of active neutrino into sterlie neutrino may be enhanced
due to the MSW effect, as at some temperature they occur in the
Mikheev--Smirnov resonance regime. In that case the right
abundance of sterile neutrinos is obtained at smaller $\theta$,
and may  be consistent with X-ray bounds. This is also shown in
Fig.~\ref{snu-photon}.

\begin{figure}[!htb]
\hskip 0.1\textwidth
\includegraphics[width=0.7\textwidth]{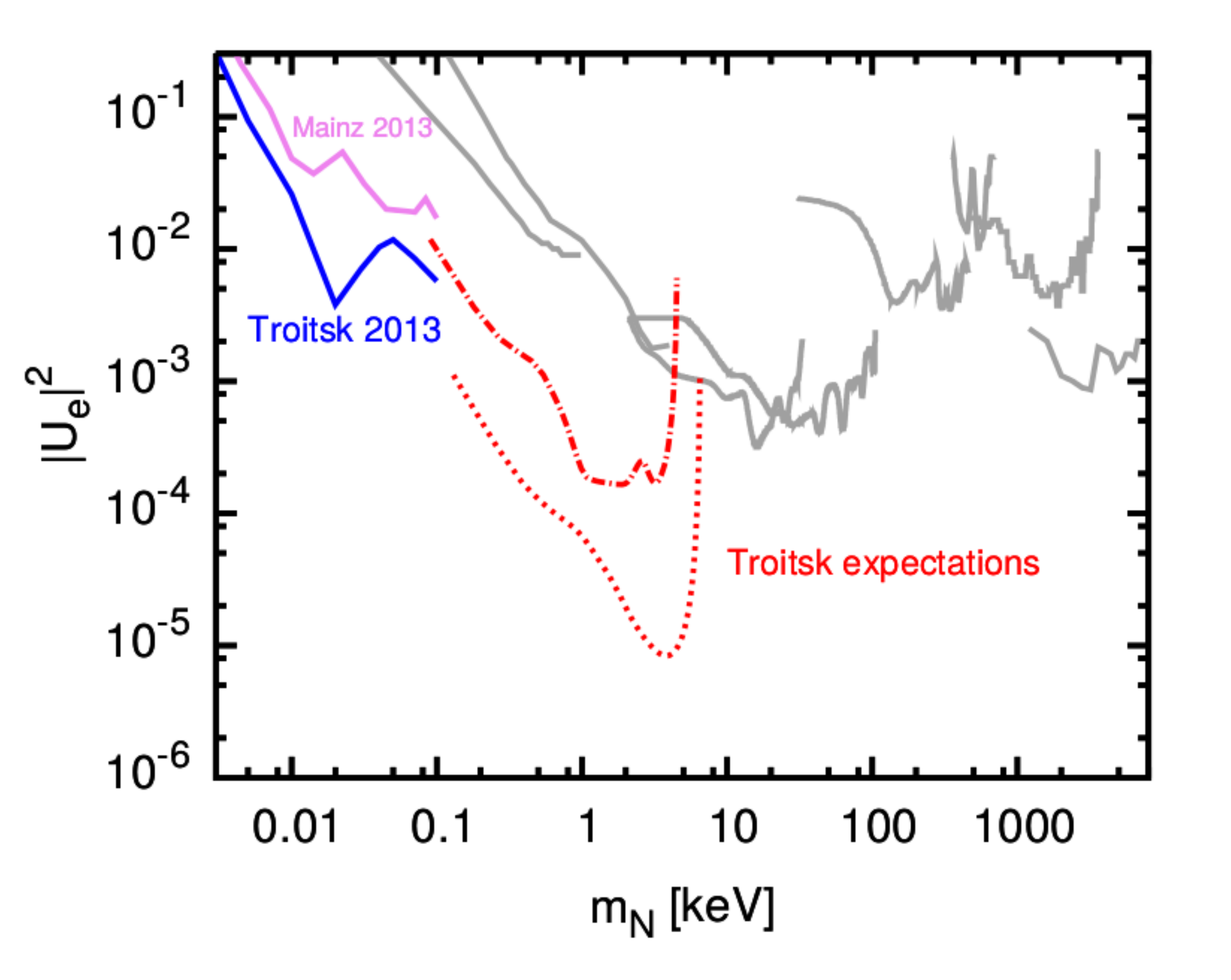}
\hskip 0.1\textwidth
\caption{
  Existing laboratory limits on sterile neutrino mixing with
  electron neutrino, $|U_e|^2 = \theta_{N \, \nu_e}^2$, and projected sensitivty
  of Troitsk nu-mass experiment~\cite{Abdurashitov:2015jha}.
\label{troitsk}
}
\end{figure}

Direct laboratory searches for strile neutrino are currently
sensitive to substantially larger sterile-active mixing angles.
This is shown in Fig.~\ref{troitsk} and also in Fig.~\ref{snu-photon},
projected KATRIN limit, dashed line.

\vspace{0.3cm}

\section{Dark matter summary}

In the first place, the mechanisms discussed here are
by no means the only ones
capable of producing dark matter, and particles we discussed
are by no means 
the only dark matter candidates. Other dark matter
candidates include gravitinos,
axinos, Q-balls, 
very heavy relics produced 
towards the end of inflation (wimpzillas), primordial black holes, 
etc. Hence, even though there are grounds to hope
that the dark matter problem will be solved soon, there is no guarantee
at all. Indeed, some of the candidates, like gravitino or sterile
wimpzilla,
interact
with the Standard Model particles so weakly that their direct
discovery is hopeless. Concerning the candidates we have
presented, we make a few comments.

\begin{itemize}

\item With the exception of axions/ALPs, the plausible
  candidates are strongly constrained already. However, as we pointed out,
  this does not mean much, since the actual values of parameters
  may still be in the unexplored region of the parameter space.

\item The null results obtained so far suggest that
  it makes sense to look for less motivated candidates, and
  employ diverse search strategies.
  This happens already: we note in this regard existing and proposed
  experiments like
  NA64, SHiP, Troitsk nu-mass, Katrin, etc.

\item Astrophysics and cosmology may well provide hints towards
  the nature of dark matter (CDM vs WDM vs SIMP vs fuzzy DM, etc.)
  
\item WIMPs are attacked from different directions.
    If dark matter particles are indeed WIMPs, and the relevant
energy scale is of order 1~TeV, then the Hot Big Bang theory will be
probed experimentally up to temperature of $(\mbox{a~few})\cdot
(10 - 100)~$GeV
and down to age $10^{-9} - 10^{-11}~$s (compare
to 1~MeV and 1~s accessible today through Big Bang Nucleosynthesis). 
With microscopic physics to be
known from collider experiments, the WIMP abundance will be reliably
calculated and checked against the data from observational cosmology.
Thus, WIMP scenario offers a window to
a very early stage of the evolution of the Universe.

\item Searches for dark matter axions and ALPs
  and signal from light sterile
neutrino make use of completely different methods. Yet there is
good chance for discovery, if either of these particles make dark
matter.

\end{itemize}

All this shows that the situation with dark matter is
controversial but extremely  interesting.

\section{Baryon asymmetry of the Universe}
\label{sec:bau}

As we discussed in Section~\ref{sub:RD}, the baryon asymmetry of the 
Universe is characterized by the baryon-to-entropy ratio, which at high
temperatures is defined as follows, 
\[
\Delta_B = \frac{n_B - n_{\bar{B}}}{s}=
\frac{1}{3} \frac{n_q - n_{\bar{q}}}{s} \; ,
\]
where  $n_q$ and $n_{\bar{q}}$ are the number densities
of quarks and antiquarks, respectively
(baryon number of a quark equals $1/3$), and $s$ is
the entropy density. If the baryon number is conserved and the
Universe expands adiabatically (which is the case at least after the
electroweak epoch, $T \lesssim 100$~GeV), $\Delta_B$ is time-independent
and equal to its present value 
$  \Delta_B \approx 0.86 \cdot 10^{-10}$, see eq.~\eqref{mar28-15-1}.
At early times, at temperatures well above 100~MeV, cosmic plasma
contained many quark-antiquark pairs, whose number density
was of the order of the entropy density,
$n_q + n_{\bar{q}} \sim s$.
Hence, in terms of quantities characterizing the very early epoch, the
baryon asymmetry may be expressed as
\[
  \Delta_B \sim \frac{n_q - n_{\bar{q}}}{n_q + n_{\bar{q}}} \; .
\]
We see that there was one extra quark per about 10 billion
quark-antiquark pairs! It is this tiny excess that is responsible for
the entire baryonic matter in the present Universe: as the Universe
expanded and cooled down, antiquarks annihilated with quarks,
and only the excessive quarks remained and formed baryons.

There is no logical contradiction to suppose that the tiny excess of
quarks over antiquarks was built in as an initial condition. This would be
very contrived, however. Furthermore,
inflationary scenario
predicts that the Universe
was baryon-symmetric at inflation (no quarks, no antiquarks).
Hence, the baryon asymmetry must be explained 
dynamically~\cite{Sakharov:1967dj,Kuzmin:1970nx}, by some
mechanism of its generation in the early Universe.

\subsection{Sakharov conditions}

There are  
three necessary conditions for the generation of the baryon asymmetry
from initially baryon-symmetric state. These are  Sakharov conditions:

(i) baryon number non-conservation;

(ii) C- and CP-violation;

(iii) deviation from thermal equilibrium.

All three conditions are easily understood.
(i) If baryon number were conserved, and initial net baryon number in
the Universe  vanishes, the Universe today would still be baryon-symmetric.
(ii) If C or CP were conserved, then the rate of
reactions with particles would be the same as the rate of reactions
with antiparticles, and no asymmetry would be generated. 
(iii) Thermal equilibrium means that the system is stationary (no 
time-dependence at all). Hence, if the initial baryon number is zero, it is
zero forever, unless there are deviations from thermal equilibrium.
Furthermore, if there are processes that violate baryon number,
and the system approaches thermal equilibrium, then the baryon number
tends to be washed out rather than generated (with qualification, see below).

At the epoch of the baryon asymmetry generation, all three Sakharov 
conditions have to be met simultaneously. There is a qualification, however.
These conditions would be literally correct if there were no other
relevant quantum numbers that characterize the cosmic medium.
In reality, however, lepton numbers also play a role. As we will see shortly,
baryon and lepton numbers are rapidly violated by anomalous
electroweak processes at temperatures above,
roughly, 100~GeV. What is conserved in the Standard Model
is the combination $(B-L)$, where $L$ is the total lepton
number\footnote{Masses of neutrinos, if Majorana, violate
  lepton number. This effect, however, is by itself neglgible.}.
So, there are two options. One is to generate the baryon asymmetry
at or below the electroweak epoch, $T \lesssim 100$~GeV, and make sure
that the electroweak processes do not wash out the baryon asymmetry
after its generation. This leads to the idea of electroweak baryogenesis
(another possibility is Affleck--Dine baryogenesis~\cite{Affleck:1984fy}).
Another is
to generate  $(B-L)$-asymmetry before the electroweak epoch,
i.e., at $T \gg 100$~GeV:
if the Universe is $(B-L)$-asymmetric above 100~GeV, the electroweak
physics reprocesses $(B-L)$ partially into baryon number and partially into
lepton number, so that in thermal equilibrium with conserved $(B-L)$ one has
\be
B = C\cdot (B-L) \; , \;\;\;\;\;\;\; L  = (C-1) \cdot (B-L) \; ,
\label{nov25-19-1}
\ee
where $C$ is a constant of order 1 ($C=28/79$ in the Standard
Model at $T \gtrsim 100$~GeV). In the second scenario, 
the first Sakharov condition applies to
$(B-L)$ rather than baryon number itself.

There are two most commonly discussed mechanisms of baryon number
non-conservation. 
One         emerges in Grand Unified Theories and is
due to the exchange of super-massive particles.  The scale of these
new, baryon number violating interactions is the Grand Unification
scale, presumably of order $M_{GUT} \simeq
10^{16}~\mbox{GeV}$. It is not very likely, however,
that the baryon asymmetry was generated due to this mechanism:
the relevant temperature would have to be of order $M_{GUT}$, 
and so high reheat
temperature after inflation is difficult to obtain.

Another mechanism is non-perturbative~\cite{'tHooft:1976up} 
and is related to the triangle
anomaly in the baryonic current (a keyword here is 
``sphaleron''~\cite{Klinkhamer-Manton,Kuzmin:1985mm}).
It exists already in the Standard
Model, and, possibly with mild modifications, operates in all its
extensions. The two main features of this mechanism, as applied to the
early Universe, is that it is effective over a wide range of
temperatures, 
$100~\mbox{GeV} < T < 10^{11}~\mbox{GeV}$, and, as we pointed out above,
that it conserves
$(B-L)$. A detailed analysis can be found in
the book~\cite{Rubakov:2002fi} and in references therein, as well
as in lecture notes of similar School~\cite{Rubakov:2017zvc}, and here we
only sketch its main ingredients.

\subsection{Elecroweak baryon number non-conservation}
\label{sec:EW-B-violation}

Let us consider the
baryonic current,
\[
B^\mu = \frac{1}{3} \cdot \sum_{i} \bar{q}_{i} \gamma^\mu q_{i} \; ,
\]
where the sum runs over all quark flavors. Naively, it is
conserved, but at the quantum level its divergence is non-zero
because of the triangle anomaly (we discussed similar effect in the
QCD context in Sec.~\ref{sec:strongCP}; there,  the axial
current $J_A^\mu$ is not conserved even in the chiral limit),
 \[
    \partial_\mu B^\mu =  
\frac{1}{3} \cdot 3_{colors}
\cdot 3_{generations} \cdot  \frac{g^2}{16 \pi^2}
F^a_{\mu \nu} \tilde{F}^{a \, \mu \nu} \; ,
\]
where $F^a_{\mu \nu}$ and $g$ are the field strength of the $SU(2)_W$
gauge field and the $SU(2)_W$ gauge coupling, respectively, and
$\tilde{F}^{a \, \mu \nu} =\frac{1}{2} \epsilon^{\mu \nu \lambda \rho} F^a_{\lambda \rho}$
is the dual tensor, cf. eq.~\eqref{chap8-true-axion-1+}.
Likewise, each leptonic current ($\alpha = e, \mu, \tau$) is anomalous
in the Standard Model (we disregard here neutrino masses and mixings,
which violate lepton numbers too),
\be
    \partial_\mu L^\mu_\alpha = \frac{g^2}{16 \pi^2} 
 F^a_{\mu \nu} \tilde{F}^{a \, \mu \nu}.
\label{mar31-15-1}
\ee
A non-trivial fact is that there exist 
large field fluctuations,  $F^a_{\mu \nu} ({\bf x}, t) \propto g^{-1}$,
such that
\be
Q \equiv \int~d^3x dt~   \frac{g^2}{16 \pi^2} 
\cdot 
F^a_{\mu \nu} \tilde{F}^{a \, \mu \nu} \neq 0 \; .
\label{mar31-15-2}
\ee
Furthermore, for any physically relevant
fluctuation, the value of $Q$ is integer (``physically relevant'' means that
the gauge field strength vanishes at infinity in space-time).
 In four space-time dimensions such fluctuations exist
only in
{\it non-Abelian} gauge theories.

Suppose now that a fluctuation with non-vanishing $Q$ has occured.
Then the baryon numbers in the end and beginning of the process are different,
\be
 B_{fin} - B_{in} =
 \int~d^3x dt~  \partial_\mu B^\mu = 3 Q \; .
\label{may7-2}
\ee
Likewise
\be
   L_{\alpha,~fin} - L_{\alpha,~in} = Q \; .
\label{may7-3}
\ee
This explains the selection rule mentioned above:
 $B$ is violated, $(B-L) \equiv (B-\sum_\alpha L_\alpha)$ is not.

 At zero temprature, the  field fluctuations that induce baryon
and lepton number violation are vacuum fluctuations, called 
instantons~\cite{Belavin:1975fg}.
Since these  are {\it large} field fluctuations, their probability
is exponentially suppressed. The suppression factor in the
Standard Model is\footnote{Similar fluctuations of gluon field in QCD
  are not suppressed, since QCD is strongly coupled at low energies.
  This explains why the axial current $J_A^\mu$ is not conserved
  even approximately.}
\[
\mbox{e}^{- { \frac{16\pi^2}{g^2}}} \sim 10^{-165} \; .
\]
Therefore, 
the rate of baryon number violating processes at zero temperature
is totally negligible. 
On the other hand, at high temperatures
there are large {\it thermal} fluctuations
(``sphalerons'') whose rate is not necessarily small.
And, indeed, 
$B$-violation in the early Universe
is rapid as compared to the cosmological expansion at
sufficiently high temperatures, provided that
(see Ref.~\cite{ew-rev} for details)
\be
\langle\phi \rangle_T < T \; ,
\label{may7-4}
\ee
where
$ \langle\phi \rangle_T$ is the
 Englert--Brout--Higgs expectation value at temperature $T$.

 \subsection{Electroweak baryogenesis: what can make it work}

 Rapid electroweak baryon number non-conservation at high temperatures
 appears to open up an intriguing possibility that
 the baryon asymmetry was generated just by these electroweak processes.
 This should occur at electroweak temperatures, $T_{EW} \sim 100$~GeV,
 since whatever baryon asymmetry is generated by electroweak processes
 at higher temperatures, it would be washed out by the same processes
 as the Universe cools down to $T_{EW}$.
 There are two obstacles, however:
 \begin{itemize}

 \item CP-violation (2nd Sakharov condition)
   is too weak  in the Standard Model:  the CKM mechanism
alone is insufficient to generate the realistic value of the
baryon asymmetry.

\item  Departure from thermal equilibrium (3d Sakharov condition)
  is problematic as well.  At temperatures of order
$T_{EW}\sim 100$~GeV, the Universe expands very slowly:
the cosmological time scale at these temperatures,
\be
H^{-1} (T_{EW})= \frac{M_{Pl}^*}{T_{EW}^2}
\sim 10^{-10}
~\mbox{s} \; ,
\label{may8-1}
\ee
 is very large by the electroweak physics standards.

 \end{itemize}

 Let us discuss what can make the electroweak mechanism work.
 We begin with the second obstacle. It appears that the only
 way to have strong departure from thermal equilibrium at
 $T_{EW}\sim 100$~GeV is the first order phase transition.
Indeed,
at temperatures well above 100~GeV electroweak symmetry is restored,
and the expectation value of the Englert--Brout--Higgs field 
$\phi$ is zero, while it is non-zero in vacuo.
\begin{figure}[htb!]
  \vspace{-6cm}
\begin{center}
\includegraphics[width=0.6\textwidth]{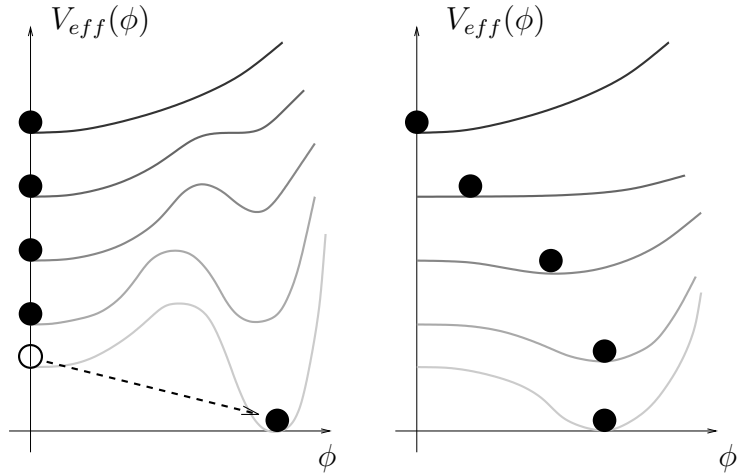}
\begin{picture}(10,10)(0,0)
{\large 
\put(-260,160){$V_{eff}(\phi)$}
\put(-112,160){$V_{eff}(\phi)$}
\put(-160,-5){$\phi$}
\put(-10,-5){$\phi$}
}
\end{picture}
\end{center}
\caption{Effective potential as function of $\phi$ at different temperatures.
Left: first order phase transition. Right: second order phase transition.
Upper curves correspond to higher temperatures. Black blobs show the
expectation value of $\phi$ in thermal equilibrium. The arrow
in the left panel illustrates the transition from the metastable,
supercooled state to the ground state.
 \label{1-2order}}
\end{figure}
This
suggests that there may be a phase transition
from the phase with $\langle \phi \rangle = 0$ to the phase
 with $\langle \phi \rangle \neq 0$. In fact, the situation is 
subtle here, as $\phi$ is not gauge invariant, and hence cannot serve
as an order parameter, so the notion of phases 
with $\langle \phi \rangle = 0$ and
$\langle \phi \rangle \neq 0$
is vague. 
This  is similar to liquid-vapor
system, which does not have an order parameter and,
depending on  pressure, may or may
not undergo vapor-liquid phase transition as temperature decreases.

Continuing to use somewhat sloppy terminology, we recall that
in thermal equilibrium any system is at the global minimum of 
its {\it free
energy}. To figure out the expectation value of $\phi$ at a given 
temperature, one introduces the temperature-dependent effective potential
$V_{eff}(\phi;T)$, which is equal to the free energy density
in the system under the constraint that the average field is equal to
a prescribed
value $\phi$, but otherwise there is thermal equilibrium.
Then the global minimum of $V_{eff}$ at given temperature
is at the equilibrium value of $\phi$, while local minima
correspond to metastable states.

The interesting
case for us is the first order phase transition. In this case,
the system evolves as follows.
At high temperatures, there exists one minimum of $V_{eff}$ at
$\phi =0$, and the expectation value of the Englert--Brout--Higgs field is zero.
As the temperature decreases, another minimum appears at finite
$\phi$, and then becomes lower than the minimum at $\phi =0$, see 
left panel of Fig.~\ref{1-2order}.
However, the minima with $\phi =0$ and $\phi \neq 0$ are
separated by a barrier of $V_{eff}$,
the probability of the transition from the phase
$\phi=0$ to the phase $\phi \neq 0$ is very small for some time,
and the system gets overcooled. The transition occurs when the
temperature becomes sufficiently low, and the transition
probability sufficiently high.
This is to be contrasted
to the case, e.g., of the second order phase transition,
right panel of Fig.~\ref{1-2order}.
In the latter case, the field slowly evolves,
as the temperature decreases, from zero to non-zero vacuum
value, and the system remains very close to thermal equilibrium
at all times.


The dynamics of the first order phase transition is
highly inequilibrium. Thermal fluctuations
spontaneously create
bubbles of the new phase inside the old phase. These bubbles
then grow, their walls eventually collide, and the new phase finally
occupies entire space. The Universe boils.
In the cosmological context, this process
happens when the bubble nucleation rate  per Hubble time
per Hubble volume is roughly of order 1,
 i.e., when a few bubbles are created
in Hubble volume in Hubble time.
The velocity of the bubble wall in the relativistic
cosmic plasma is roughly of the order of the speed of light
(in fact, it is somewhat smaller, from $0.1$ to $0.01$).
Hence, the bubbles grow large before their
walls collide: their size at collision is roughly of order of the
Hubble size (in fact, one or two orders of magnitude smaller).
In other words, the biblles are born microscopic, their initial
sizes are
determined by the electroweak scale and are roughly of order
\[
R_{init} \sim (100~\mbox{GeV})^{-1} \sim 10^{-16}~\mbox{cm} \; .
\]
Their final sizes at the time the bubble walls collide are of order 
\[
R_{fin} \sim 10^{-2} - 10^{-3}~\mbox{cm} \; ,
\]
as follows from 
(\ref{may8-1}). One may hope that the baryon asymmetry
may be generated during this inequilibrium process.

Does this really happen in the Standard Model? Unfortunately, no:
with the Higgs boson mass
$m_H = 125$~GeV, there is no phase transition in the Standard Model
at all; there is smooth crossover instead~\cite{Kajantie:1995kf}.

Nevertheless, the first order phase transition
may be characteristic of some  extensions of the Standard Model.
Generally speaking, 
one needs 
the existence of new bosonic fields that have
large enough couplings to the
Englert--Brout--Higgs field(s).
To have an effect on the dynamics of
the transition, the new bosons must be present in the cosmic plasma
at the transition temperature, $T_{EW} \sim 100$~GeV, so their masses
should not be very much higher than $T_{EW}$. 

Let us turn to the first obstacle, CP-violation.
In the course of the first order phase transition, 
the baryon asymmetry is generated
in the interactions of quarks and leptons 
with the bubble walls. Therefore, CP-violation
must occur at the walls. Now, the walls are made of the
scalar field(s), and this 
points towards the necessity of CP-violation
in the scalar sector, which may only be the case in a theory
containing scalar fields other than the Standard
Model  Englert--Brout--Higgs field.

\begin{figure}[!htb]
\hskip 0.1\textwidth
\includegraphics[width=0.7\textwidth]{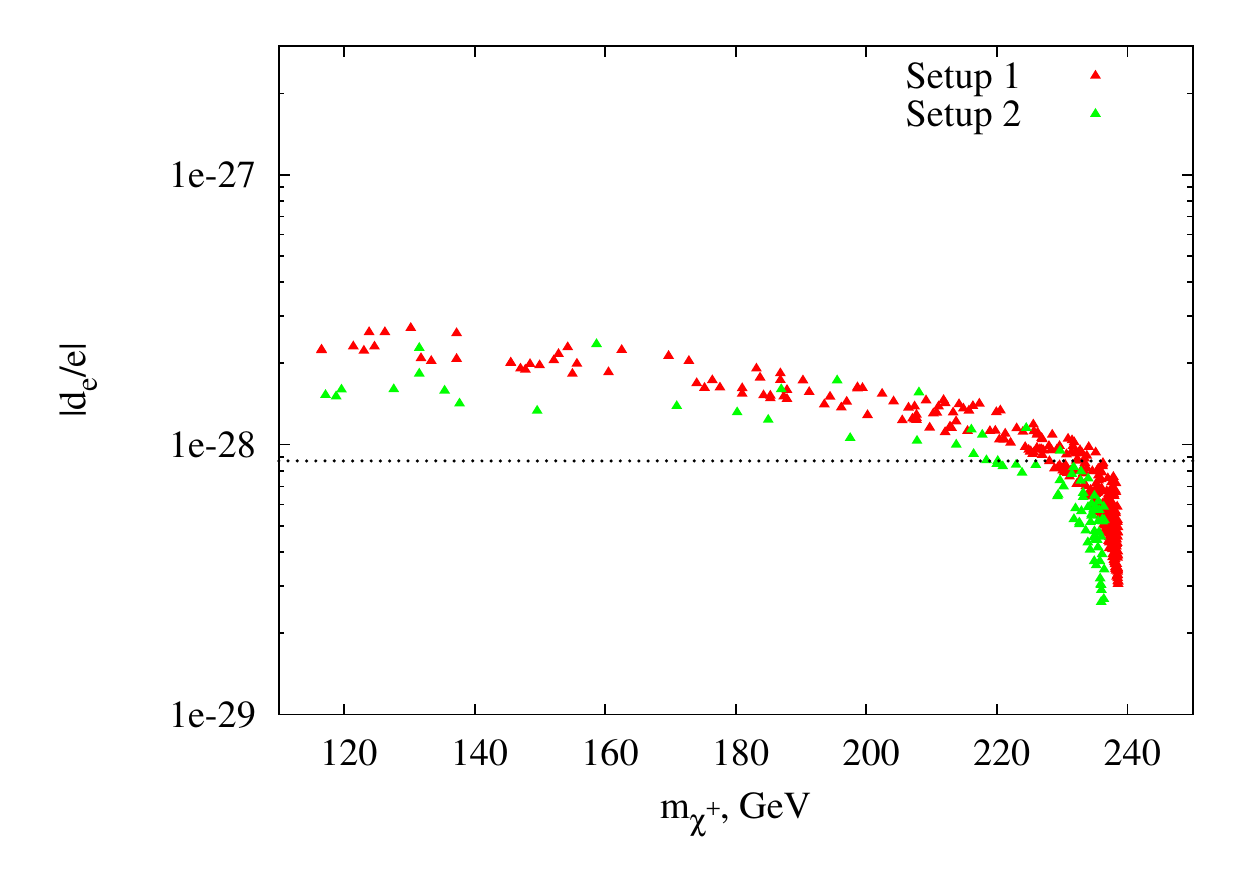}
\caption{Electron EDM predicted by
  Non-Minimal Split Supersymmetric Standard
  Model with parameters suitable for
   electroweak baryogenesis.
  Current limit  $d_e < 1.1 \cdot 10^{-29} e\cdot \mbox{cm}$ rules
  out all these models.
\label{Demidov}
}
\end{figure}

In concrete models with successful electroweak baryogenesis,
CP-violation responsible for the baryon asymmetry often
leads to sizeable electric dipole moments (EDMs) of neutron and electron.
The limits on EDMs are so strong that many such models are actually
ruled out. An example is the Non-Minimal Split Supersymmetric Standard
Model, which only a few years ago had successfull 
 electroweak baryogenesis~\cite{Demidov:2016wcv}. The predictions
of this model for electron EDM are shown in Fig.~\ref{Demidov}.
In 2016, when Ref.~\cite{Demidov:2016wcv} was written,
part of the parameter space
was still allowed, but the recent ACME limit~\cite{Andreev:2018ayy}
\[
d_e < 1.1 \cdot 10^{-29} e\cdot \mbox{cm}
\]
rules out the entire parameter space with efficient electroweak
baryogenesis.

To summarize, electroweak baryogenesis requires  considerable
extension of the Standard Model, often with masses of new particles in
the TeV
range or lower.  Hence, this mechanism will most likely
be ruled out or confirmed by the LHC or its successors.
Moreover, limits on electron
and neutron EDMs make the design of such an extension very difficult.
Still, the issue is not decided yet, and the effort to construct
the models with successfull electroweak baryogenesis
continues~\cite{EWBG}.


\subsection{Baryogenesis in sterile neutrino oscillations}

Let us mention another baryogenesis mechanism interesting from the viewpoint of
terrestrial experiments, namely, leptogenesis in oscillations of
sterile neutrinos~\cite{Akhmedov:1998qx,Asaka:2005pn}.
The general idea of leptogenesis~\cite{Fukugita:1986hr}
is that one or another mechanism generates {\it lepton} asymmetry
in the Universe before the electroweak transition, and electroweak
sphalerons automatically
reprocess part of the lepton asymmetry into baryon asymmetry, see
eq.~\eqref{nov25-19-1}. The particular version of leptogenesis
that we briefly discuss here assumes that there are at least two
heavy Majorana neutrinos in the mass range $1 - 10$~GeV, and that there is
strong enough CP-violation in the sterile neutrino sector. Then asymmetries
in sterile neutrino sector may be generated and transmitted to
active neutrino sector via Yukawa interactions responsible for
see-saw masses of acive neutrinos.
In the case when there is effectively two sterile neutrino species
participating in leptogenesis, correct value of the baryon asymmetry is
obtained when the two sterile neutrnos are nearly degenerate,
\[
\frac{|M_1^2 - M_2^2|}{M_{1,2}^2} \lesssim 10^{-6} \; ,
\]
which makes the model rather contrived. However, with three sterile neutrino
species, the degeneracy is no longer required~\cite{Drewes:2012ma}.
The sterile neutrinos of masses in
GeV range and parameters suitable for leptogenesis in their oscillations
are typically
accessible through rare decays of $B$-mesons, $Z$-bosons, as well as in
future beam dump experiments such as SHiP.

An important point concerning this and virtually all other leptogenesis
 mechanisms is that CP-violation in the sector of active neutrinos,
 which will hopefully be discovered in oscillation experiments,
 does not have direct relevance to leptogenesis: the value of lepton,
 and hence baryon asymmetry is determined by CP-violating parameters
 {\it in the sterile neutrino sector.}

 \subsection{Baryogenesis summary}

 We briefly considered here two mechanisms of baryogenesis
 which may be directly
 tested, at least in principle, in particle physics experiments.
 These are certainly not the only mechanisms proposed, and, arguably,
 not
 the most plausible mechanisms. One particularly strong competitor is
 thermal leptogenesis~\cite{Fukugita:1986hr}, for reviews see, e.g.,
 Ref.~\cite{leptogenesis}. Its idea is that the lepton asymmetry
 is generated in decays of heavy Majorana sterile neutrinos.
 The masses of these new particles are well above the experimentally
 accessible energies. On the one hand, this is in line with the
 see-saw idea; on the other,  direct  proof of this mechanism does not
 appear possible.  Interestingly,
 thermal leptogenesis  works only
 with light  active
 neutrinos: the neutrino masses inferred from cosmology
 and oscillation experiments are just
 in the right ballpark.

 There are numerous alternative mechanisms of baryogenesis.
 To name a few, 
 we have already mentioned Affleck--Dine baryogenesis~\cite{Affleck:1984fy};
 early discussions concentrated mostly on GUT
 baryogenesis~\cite{Kolb:1983ni};  there is even a possibility to
 generate the baryon asymmetry at inflationary epoch~\cite{Babichev:2018afx}.
 Unfortunately,
 most of these proposals will be very difficult, if at all 
 possible, to test. So, there is no guarantee at all that
 we will understand in foreseeable
 future the origin of matter in the Universe.

\section{Before the hot epoch}
\label{sec:pertu}

With Big Bang Nucleosynthesis theory and observations, and due to
evidence, albeit indirect, for relic neutrinos,
 we are confident of the theory of the early Universe
at temperatures up to 
$T\simeq 1$~MeV, which correspond to age of $t\simeq 1$~s.
With the LHC, we are learning the Universe
up to temperatures $T\sim 100$~GeV and down to age $t \sim 10^{-10}$~s.
Are we going to have a handle on even earlier epoch?

Let us summarize the current status of this issue.
\begin{itemize}

  \item On the one hand,
we are confident that the hot cosmological epoch was not
the first one; it was preceded by some other, entirely
different stage.

\item On the other hand, we do not know for sure what was that
  earlier epoch; an excellent guess is inflation, but alternative
  scenarios are not ruled out.

\item It is conceivable (although not guaranteed) that
  future cosmological observations will enable us to understand the
  nature of the pre-hot epoch.

\end{itemize}

All this makes the situation very interesting. It is fascinating
that by studying the Universe at large we may be able to learn about
the earliest cosmological epoch which happened at extremely high energy
density and expansion rate of our Universe.

\subsection{Cosmological perturbations}

The key players in this Section are cosmological perturbations.
These are inhomogeneities in the energy density and assoiated
gravitational potentials, in the first place.
It is these inhomogeneites that, among other things, serve as seeds
for structures -- galaxies, clusters of galaxies, etc.
This type of inhomogeneities is called scalar perturbations,
as they are described by 3-scalars. There may exist perturbations
of another type, called tensor; these are primordial gravity
waves. Tensor modes have not been obesrved (yet), so
we mostly concentrate on scalar perturbations. While 
perturbations of the present size of order 10~Mpc and smaller
have large amplitudes today and are non-linear, 
amplitudes of all known perturbations
were small in the past, and  the linearized theory
is applicable. Indeed, CMB temperature anisotropy
tells us that the perturbations at recombination epoch
were roughly at the level 
\be
\delta \equiv \frac{\delta \rho}{\rho} = 10^{-4} - 10^{-5} \; .
\label{mar29-15-1}
\ee
 We are sloppy here in characterizing the
scalar perturbations by the density contrast
$\delta \rho/\rho$; we are going to skip technicalities
and use this notation in what follows.

Linearized perturbations are most easily studied in
momentum space,
since the background FLRW metric~\eqref{FRW} does not explicitly depend 
on ${\bf x}$.
The spatial Fourier
 transformation reads 
\[
\delta ({\bf x},t)=\int~e^{i{\bf k}{\bf x}}\delta({\bf k},t)~d^3k \; .
\]
Each Fourier mode $\delta({\bf k},t)$ obeys its own linearized equation and
hence
can be treated separately. Note that
the physical distance between neighboring
points is $a(t) d {\bf x}$. Thus, ${\bf k}$ is 
{\it not} the physical momentum (wavenumber); the physical momentum
is ${\bf k}/a(t)$. While for a given mode the comoving (or coordinate)
momentum ${\bf k}$ remains constant in time,
the physical momentum gets redshifted as the Universe expands, see
also Section~\ref{sub:FLRW}. In what follows we set the present value of the
scale factor equal to 1,
$ a_0 \equiv a(t_0) = 1$;
then ${\bf k}$ is the {\it present} physical momentum and
$2\pi/k$ is the present physical wavelength, which is also called comoving
wavelength.

Properties of scalar perturbations are mesured in various ways.
Perturbations of fairly large spatial scales (fairly low ${\bf k}$)
give rise to
CMB temperature anisotropy and polarization, 
so we have very detailed knowledge of them.
Somewhat shorter wavelengths are studied by analysing
distributions of galaxies and quasars at present and in
relatively near past. There are several other methods, some of which
can probe even shorter wavelengths. There is good overall consistency
of the results obtained by different methods, so we have reasonably good 
understanding of many aspects of the scalar perturbations.

Cosmic medium in our Universe has several components that
interact only gravitationally: baryons, photons, neutrinos,
dark matter. Hence, there may be and, in fact, there are
perturbations in each of these components. As we pointed out
in 
Section~\ref{sec:dm}, electromagnetic interactions between
baryons, electrons and photons were strong 
before recombination,
so to reasonable approximation
these species made single fluid, and it is appropriate to talk
about perturbations in this fluid. After recombination, baryons and 
photons evolved independently.

\subsection{Subhorizon and superhorizon regimes.}

It is instructive to compare the wavelength of a perturbation with
the horizon size. To this end, recall (see Section~\ref{sub:RD})
that the horizon size $l_H(t)$ is the size of the largest region which
is causally connected by the time $t$, and that 
\[
l_H (t) \sim 
H^{-1}(t) \sim t
\]
at radiation domination and later, see 
eq.~\eqref{mar23-15-7}. The latter relation, however, holds 
{\it under assumption that the hot epoch was the first one in cosmology},
i.e., that the radiation domination started right after the
Big Bang. This assumption is at the heart of what can be called
hot Big Bang theory. We will find that this assumption in fact
is {\it not valid} for our Universe; we are going to see this ad absurdum,
so let us stick to the hot Big Bang theory for the time being.

The physical wavelength of a perturbation
grows slower than the horizon size. As an example, at radiation domination
\[
\lambda (t) = \frac{2 \pi a(t)}{k} \propto \sqrt{t} \; ,
\]
while at matter domination $\lambda(t) \propto t^{2/3}$.
For obvious reason, the modes with $\lambda(t) \ll H^{-1}(t)$
and  $\lambda(t) \gg H^{-1}(t)$ are called subhorizon and superhorizon
at time $t$, respectively. We are interested in the modes which are
subhorizon {\it today}; longer modes are homogeneous throughout the
visible Universe and are not observed. 
However, {\it the wavelengths which are subhorizon today
were superhorizon at some earlier epoch}. In other words, the
physical momentum $k/a(t)$ was smaller than $H(t)$ at early times;
at time $t_\times$ such that 
\[
q(t_\times) \equiv \frac{k}{a(t_\times)} = H(t_\times) \; ,
\]
the mode entered the horizon, and after that evloved in the
subhorizon regime $k/a(t) \gg H(t)$.
It is straightforward to see that for all cosmlogically interesting wavelengths,
horizon crossing occurs at temperatures below 1~MeV, i.e.,
at the time we are confident about (repeating the calculation of
Sec.~\ref{subsec:missing} we find that the present
wavelength of order 100~kpc entered the horizon
at $T\sim 4$~keV). So, there is no guesswork at
this point.

\begin{figure}[b!]
\begin{center}
\includegraphics[width=0.7\textwidth]{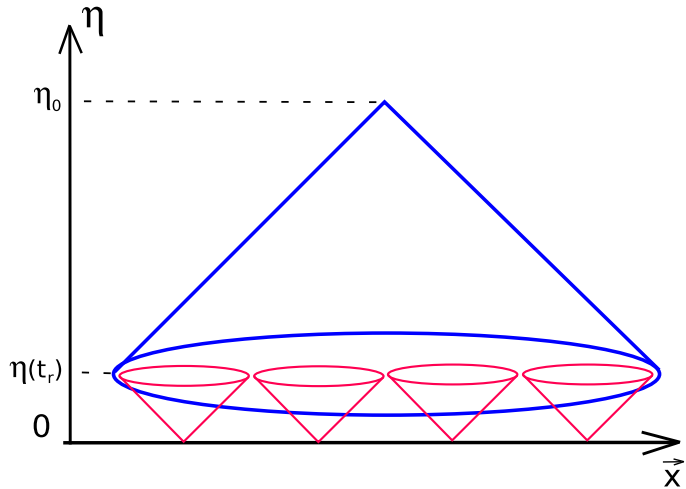}
\end{center}
\caption{ Causal structure of space-time in the hot Big Bang theory.
  $\eta_r$ and $\eta_0$ are conformal times at recombination and today,
  respectively.
 \label{noinfl-horizon}
 }
\end{figure}

Another way to look at the superhorizon--subhorizon behaviour
of perturbations is to introduce a new time
coordinate (cf. eq.~\eqref{sep13-11-6}),
\be
\eta = \int_0^t \frac{dt'}{a(t')} \; .
\label{mar29-15-10}
\ee
Note that this integral converges at lower limit in the hot 
Big Bang theory. In terms of this time coordinate, the FLRW metric
\eqref{FRW} reads
\[
ds^2 = a^2 (\eta) (d\eta^2 - d{\bf x}^2) \; .
\]
In  coordinates $(\eta, {\bf x})$, the light cones
$ds=0$ are the same as in Minkowski space, and $\eta$ is the
coordinate size of the horizon, see Fig.~\ref{noinfl-horizon}. 
Every mode of perturbation has 
time-independent coordinate wavelength $2\pi/k$, and
at small $\eta$ it is in superhorizon regime, $2\pi/k \gg \eta$.

\subsection{Hot epoch was not the first}

This picture falsifies the 
hot Big Bang theory. Indeed, within this theory,
we see the horizon
at recombination $l_H(t_{rec})$ at an angle 
$\Delta \theta \approx 2^\circ$, as schematically shown in 
Fig.~\ref{noinfl-horizon}. By causality,  at recombination
there should be 
no perturbations of larger wavelengths, as 
any perturbation can be generated within the causal light cone
only. In other words, CMB temperature must be isotropic when
averaged over angular scales exceeding $2^\circ$; there should be no
cold or warm regions of angular size larger than  $2^\circ$.

We now take a look at  the CMB 
photographic picture shown in Fig.~\ref{Planck-sky}.
It is seen by naked eye that there are cold and warm regions
whose angular size much exceeds $2^\circ$; in fact, there are
perturbations of all angular sizes up to those comparable
to the entire sky. We come to an important conclusion:
the scalar perurbations were built in at the very beginning
of the hot epoch, i.e., 
the cosmological perturbations were
generated before the hot epoch.

Another manifestation of the fact that the scalar perturbations were
there already at the beginning of the hot epoch is the
existence of peaks in the angular spectrum of CMB temperature, as seen in
Fig.~\ref{Planck-Dl}.
In general, perturbations in the baryon-photon medium before recombination
are acoustic waves (cf. Sec.~\ref{sec:imprint}),
\be
\delta_B ({\bf k}, t)= A ({\bf k}) 
\e^{i{\bf kx}}\cos\left[\int_0^t~v_s\frac{k}{a(t')}dt'+\psi_{\bf k}
\right] \; ,
\label{mar29-15-6}
\ee
where $v_s$ is sound speed, $A ({\bf k})$ is time-independent amplitude and
$\psi_{\bf k}$ is a time-independent phase. 
This expression is valid, however, in the subhorizon regime only, i.e.,
at late times.
The two solutions in superhorizon regime at radiation domination
are
\begin{subequations}
\label{mar29-15-5}
\begin{align}
\delta_B (t) &= \mbox{const} \; ,
\label{mar29-15-5a} \\
\delta(t)_B &= \frac{\mbox{const}}{t^{3/2}} \; .
\label{mar29-15-5b}
\end{align}
\end{subequations}
%
If the perturbations existed at the very beginning
of the hot epoch, they were superhorizon at sufficienly early times,
and were described by the solutions~\eqref{mar29-15-5}. The consistency
of the whole cosmology requires that the amplitude of perturbations was
small at the beginning of the hot stage. The solution~\eqref{mar29-15-5b}
rapidly decays away, and towards the horizon entry the perturbation
is in constant mode~\eqref{mar29-15-5a}. So, the initial condition
for the further evolution is unique modulo amplitude $A ({\bf k})$,
and hence the phase $\psi({\bf k})$ is uniquely determined:
we have $\psi({\bf k}) = 0$ for modes entering horizon at radiation
domination. As discussed in
Sec.~\ref{sec:imprint}, this leads to oscillatory behavior
of baryon-photon perturbations at recombination
{\it as function of $k$}, and
translates into oscillations of CMB temperature multipole
$C_l$ as function of multipole number $l$.

Were the perturbations generated in a causal way at radiation domination,
they would be always subhorizon. In that case the solutions~\eqref{mar29-15-5}
would be irrelevant, and there would be no reason for a particular
choice of phase $\psi_{\bf k}$ in eq.~\eqref{mar29-15-6}. One would rather
expect that $\psi_{\bf k}$ is a random function of ${\bf k}$,
so $\delta_B ({\bf k}, t_r)$ would not oscillate as function of ${\bf k}$,
and oscillations of $C_l$ would not exist.
This is indeed the case for specific mechanisms of the generation
of density perturbations at hot epoch~\cite{Strings-CMB-plots}.

We conclude that
the facts that the CMB angular spectrum has oscillatory behavior
and that there are sizeable temperature fluctuations at $l<50$
(angular scale greater than the angular size  $2^\circ$ of the horizon
at recombination) 
unambiguously tell us that the density perturbations were indeed
superhorizon at hot cosmological stage. The hot epoch was
preceded by some other epoch --- the epoch of the generation of
perturbations.

\subsection{Inflation or not?}

\begin{figure}[htb!]
\begin{center}
\includegraphics[width=0.7\textwidth]{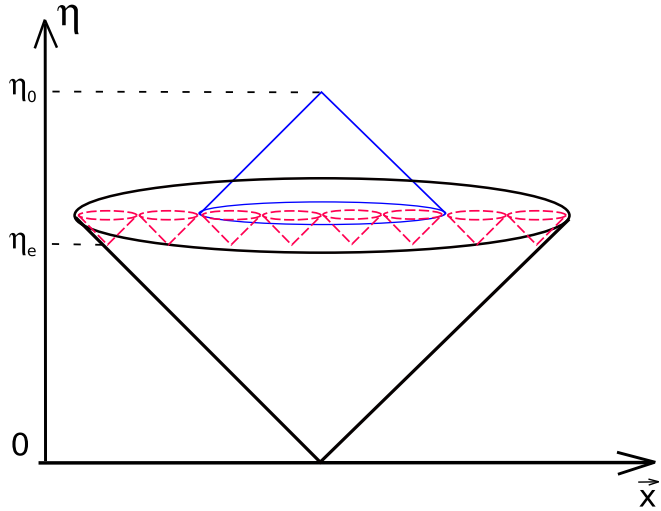}
\end{center}
\caption{ Causal structure of space-time in the real Universe
 \label{infl-horizon}
 }
\end{figure}

The pre-hot epoch must be long in terms of the  
time variable $\eta$ introduced in eq.~\eqref{mar29-15-10}.
What we would like to have is that the large part of the Universe
be causally connected towards
the end of that epoch, see Fig.~\ref{infl-horizon}.
Long duration in $\eta$ does not necessarity mean long duration in
physical time $t$; in fact, the  the
pre-hot epoch may be very short in physical time.

An excellent hypothesis on the pre-hot stage is inflation,
the epoch of nearly exponential expansion~\cite{inflation},
\[
  a(t) = \mbox{e}^{\int H dt} \; , \;\;\;\;\;\;\;\;\;
H \approx \mbox{const} \; .
\]
If this epoch lasts many Hubble times,
the whole visible Universe, and likely much greater
region of space, is causally connected already at very early times.

From the viewpoint of perturbations,
the physical momentum $q(t) = k/a(t)$ decreases
(gets redshifted) at inflation, while the Hubble parameter
stays almost constant. So, every mode is first subhorizon
($q(t) \gg H(t)$), and
later superhorizon ($q(t) \ll H(t)$). This situation is
opposite to what happens at radiation and matter domination;
this is precisely the pre-requisite for generating the density
perturbations. Indeed,  inflation does generate primordial
density 
perturbations~\cite{infl-perturbations}, whose properties are consistent
with everything we know about them.

Inflation is not the only hypothesis proposed so far.
One alternative
option is the bouncing Universe scenario, which assumes
that the cosmological evolution begins from contraction,
then the contracting stage terminates
at some moment of time (bounce) and is followed by expansion.
A version is the cycling Universe scenario with many cycles
of contraction--bounce--expansion, see Ref.~\cite{lehners}
for reviews.
Another scenario is that the Universe starts out from
nearly flat and static state with nearly vanishing energy density.
Then the energy density increases (!), and according to the Friedmann
equation, the expansion
speeds up.
This goes under the name of Genesis scenario~\cite{Creminelli:2010ba}.
Theoretical realizations of these scenarios are surprizingly difficult,
but not impossible, as became clear recently.

\section{Towards understanding the earliest epoch }

Since cosmological perturbations originate from the earliest
epoch that occured before the hot stage, properties of these perturbations
will hopefully give us a clue on that epoch. Presently,
we know only very basic things about the cosmological perturbations.
Let us discuss this point, and at the same time consider promsing
directions
where further study may lead to breakthrough.

Of course, since the properties we know of
are established by observations, thery are valid
within certain error bars. Conversely, deviations from the results
listed below, if observed, would be extremely interesting.

\subsection{Adiabaticity of scalar perturbations}

Primordial scalar perturbations are {\bf adiabatic}.
This means that there are perturbations in the energy density,
but {\it not in composition}. More precisely, the baryon to entropy
ratio and dark matter to entropy ratio are constant in space,
\be
\delta \left(\frac{n_B}{s} \right)= \mbox{const}
\; , \;\;\;\;\;\; \delta \left(\frac{n_{DM}}{s}\right) = \mbox{const} \; .
\label{sep22-11-10}
\ee
This is consistent with the generation of the baryon asymmetry
and dark matter at the hot cosmological epoch: in that case,
all partciles were in thermal equilibrium early at the hot
epoch,  and as long as physics
behind the baryon asymmetry and dark matter generation is the
same everywhere in the Universe, the baryon and dark matter
abundances (relative to the entropy density) are necessarily
the same everywhere. In principle,
there may exist {\it entropy} (or isocurvature) perturbations
that violate (one of) the relations \eqref{sep22-11-10}.
No admixture of the
entropy perturbations have been detected so far, but it is worth
emphasizing that even small admixture will show that many popular
mechanisms for generating dark matter and/or baryon asymmetry
have nothing to do with reality. One will have to think, instead,
that the baryon asymmetry and/or dark matter were generated
before the beginning of the hot stage. A notable example is the
axion misalignment mechanism discussed in Section~\ref{subs:axions}.

\subsection{Gaussianity}

The primordial scalar perturbations are {\bf Gaussian random field}.
Gaussianity means that the three-point and all odd correlation functions
vanish, while the four-point  and  higher order
even correlation functions are expressed through the two-point
function via  Wick's theorem:
\begin{eqnarray}
\langle \delta({\bf k}_1)  \delta({\bf k}_2)  \delta({\bf k}_3) \rangle &=& 0
\nonumber \\
\langle \delta({\bf k}_1)  \delta({\bf k}_2)  \delta({\bf k}_3) 
\delta ({\bf k}_4)\rangle & =
& \langle \delta({\bf k}_1)  \delta({\bf k}_2)
\rangle \cdot \langle \delta({\bf k}_3)  \delta({\bf k}_4)   \rangle 
\nonumber\\
&~&{\ +}~
{ \mbox{permutations~of~momenta}} \; .
\nonumber
\end{eqnarray}
We note that this
 property is characteristic of {\it vacuum fluctuations of
non-interacting (linear) quantum fields.}
Free quantum field has the general form
\[
\phi ({\bf x}, t) =
\int d^3k e^{-i{\bf kx}} \left(f^{(+)}_{\bf k}(t) a^\dagger_{\bf k}
+  e^{i{\bf kx}} f^{(-)}_{\bf k}(t) a_{\bf k} \right) \; ,
\]
where $a_{\bf k}^\dagger$ and $a_{\bf k}$ are creation and annihilation operators. 
For the field in Minkowski space-time one has  
$f^{(\pm)}_{\bf k} (t) = \mbox{e}^{\pm i\omega_k t}$, while enhancement, e.g.
due to the evolution in time-dependent background, means that
$f^{(\pm)}_{\bf k}$ are large. But in any case,
Wick's theorem is valid, provided that the state of the system is vacuum,
$a_{\bf k} |0\rangle = 0$.  Hence, it is quite likely that
the density perturbations originate from the enhanced vacuum
fluctuations of non-interacting or weakly interacting quantum field(s).

Search for {\it non-Gaussianity} is an important topic of current research.
It would show up as a deviation from Wick's theorem.
As an example, the three-point function (bispectrum) may be non-vanishing,
\[\langle
\delta ({\bf k}_1)
\delta ( {\bf k}_2) \delta ({\bf k}_3)
\rangle = \delta( {\bf k}_1 + {\bf k}_2 + {\bf k}_3) ~
G(k_i^2 ;~ {\bf k}_1 {\bf k}_2 ;~
{\bf k}_1  {\bf k}_3) \neq 0 \; .
\]
The functional dependence
 of  $G(k_i^2 ;~ {\bf k}_1 {\bf k}_2 ;~
{\bf k}_1  {\bf k}_3)$ on its arguments
is different in different models of generation of primordial
perturbations, so this shape is a 
potential discriminator.
In some models the bispectrum vanishes, e.g., due to symmetries. 
In that case the trispectrum
(connected 4-point function) may be measurable instead.
For the time being,
non-Gaussianity
has not been detected.

Inflation does the job of producing Gaussian
primordial perturbations very well. At inflationary
epoch, fluctuations of all light fields get enhanced greatly
due to the fast expansion of the Universe.
This is true, in particular, for inflaton,
the field that dominates the energy density
at inflation. Enhanced vacuum fluctuations of the
inflaton are reprocessed into adiabatic
perturbations in the hot medium after
the end of inflation. Inflaton field is very weakly coupled,
so the non-Gaussianity in the primordial scalar perturbations is
very small~\cite{Maldacena:2002vr}. In fact, it is so small that
its detection is problematic even in distant future.
It is worth noting that this refers
to the simplest, single field inflationary models.
In models with more than one relevant field the situation may be different,
and sizeable non-Gaussianity may be generated.

The generation of the density perturbations
is less automatic in scenarios alternative to inflation.
Most models proposed so far can be adjusted in such a way that
non-Gaussianity is not particularly strong, but potentially observable.
In many cases the bispectrum  $G(k_i^2 ;~ {\bf k}_1 {\bf k}_2 ;~
{\bf k}_1  {\bf k}_3)$ and/or trispectrum are different from
inflationary theories.

\subsection{Nearly flat power spectrum}

Another important property is that the
primordial power spectrum of density perturbations
{\bf is nearly, but not exactly flat}.
For 
homogeneous and anisotropic  
Gaussian random field, the power spectrum completely determines its 
only characteristic, the two-point function.
A convenient definition 
is
\be
\langle  \delta  ({\bf k})  
      \delta       ({\bf k}^\prime)  \rangle
= \frac{1}{4 \pi k^3} {\cal P} (k) \delta({\bf k} + {\bf k}^\prime) \; .
\label{sep24-11-1}
\ee
The power spectrum ${\cal P} (k) $ defined in this way
determines the fluctuation in a logarithmic
interval of momenta,
\[
\langle \delta^2 ({\bf x}) \rangle
= \int_0^\infty ~\frac{dk}{k} ~{\cal P}(k) \; .
\]
By definition, the flat, scale-invariant spectrum is such that
${\cal P}$ is independent of $k$. 
The flat spectrum was conjectured by Harrison~\cite{Harrison},
Zeldovich~\cite{Zeldovich} and Peebles and Yu~\cite{Peebles:1970ag}
in the beginning of
1970's, long before realistic mechanisms of the generation
of density perturbations have been proposed. 

In view of the approximate flatness, a natural parametrization is
\be
{\cal P}(k) = A_s \left(\frac{k}{k_*} \right)^{n_s -1} \; ,
\label{sep24-11-6}
\ee
where
$A_s $ is the amplitude, $(n_s-1) $ is the tilt and $k_*$ is a
fiducial momentum, chosen at one's convenience.
The flat spectrum in this parametrization has $n_s=1$.
The 
cosmological data give~\cite{Aghanim:2018eyx}  
\be
n_s = 0.965 \pm 0.004 \; .
\label{nov26-19-1}
\ee
This quantifies what we mean by nearly, bit not exactly flat power spectrum.

The approximate
flatness of the primordial power spectrum 
in inflationary theory  is explained by
the symmetry of the de~Sitter space-time,
which is the space-time of constant Hubble rate,
\[
ds^2 = dt^2 - \mbox{e}^{2Ht} d{\bf x}^{2} \; , \;\;\;\;\;\;\;\;
H = \mbox{const} \; .
\]
This metric is invariant under spatial 
dilatations
supplemented by time translations,
\[
{\bf x} \to \lambda {\bf x} \;, \;\;\;
t \to t - \frac{1}{2H} \log \lambda \; .
\]
Therefore, all spatial scales are alike, as required for
the flat power spectrum. 
At inflation, $H$ and the inflaton field
are almost constant in time, and the
de~Sitter symmetry is an approximate symmetry. For this reason 
inflation automatically generates nearly flat power spectrum.
However, neither $H$ nor inflaton are exactly time-independent.
This naturally leads to the slight tilt in the spectrum.
Overall, this picture is qualitatively
consistent with the result \eqref{nov26-19-1}, though quantitative
prediction depends on concrete inflationary model.

The situation is not so straightforward in alternatives to
inflation: the approximate flatness of
the scalar power spectrum is not at all automatic. So, one has to
work hard to obtain this property. 
Similarly to inflationary theory, 
the flatness of the scalar power spectrum may be
due to some symmetry.
One candidate symmetry is conformal 
invariance~\cite{conf1,conf2}. The point is that
the conformal group includes dilatations,
 $x^\mu \to \lambda x^\mu$. This property indicates that
the theory possesses no
scale, and has good chance
for producing the flat spectrum. This idea is indeed 
realized at least at the toy model level.

\subsection{Statistical isotropy.}

In principle, the power spectrum
of scalar perturbations may depend on the direction of
momentum, e.g., 
\[
{\cal P}({\bf k}) = {\cal P}_0 (k) \left(1 + 
w_{ij} (k) \frac{k_i k_j}{k^2}
+ \dots \right) \; ,
\]
where $w_{ij}$ is a fundamental  tensor in our part
of the Universe (odd powers of $k_i$ would contradict
commutativity of the Gaussian random field $\delta ({\bf k})$).
Such a dependence would 
imply 
that the Universe was anisotropic at the
pre-hot stage, when the primordial
perturbations were generated. This statistical anisotropy
is rather hard to obtain in inflationary models, though it is
possible in inflation with strong vector fields~\cite{soda}.
On the other hand, statistical anisotropy is 
natural in some other scenarios, including conformal 
models~\cite{mlvr+}.
The statistical anisotropy
would show up in correlators~\cite{aniso}
\[
\langle a_{ lm} a_{ l' m'} \rangle \;\;\;\;\;\; \mbox{with}~~{ l'\neq l}
~~
\mbox{and/or}~~{ m' \neq  m} \; .
\]
At the moment, the constraints~\cite{Kim:2013gka} on statistical anisotropy
obtained by analysing the CMB data are getting into the region,
which is interesting from the  viewpoint of some (though not many)
models of the pre-hot epoch.

\subsection{Tensor modes}

The distinguishing property of inflation is  {\it the
generation of tensor modes (primordial gravity waves)}
of sizeable amplitude and nearly flat power spectrum.
The gravity waves are thus smoking gun for inflation (although
there is some debate on this point).  Indeed, there seems to be no way of
generating nearly flat tensor power spectrum in alternatives to 
inflation; in fact, most, if not all, alternative scenarios
predict unobservably small tensor modes.
The reason for their generation at inflation is that
the exponential expansion of the Universe  enhances vacuum fluctuations
of all fields, including the gravitational field itself.
Particularly interesting are gravity waves whose present
wavelengths are huge, 100~Mpc and larger, and periods are
of the order of a billion years and larger. Many inflationary
models predict  their amplitudes to be very large, 
of order $10^{-6}$ or so. Shorter gravity waves are generated
too, but their amplitudes decay after horizon entry at radiation
domination, and today they have much smaller amplitudes making them
inaccessible  to gravity wave detectors like LIGO/VIRGO,
eLISA, etc.
A conventional characteristic of the amplitude of
primordial gravity waves is the tensor-to-scalar ratio
\[
r = \frac{{\cal P}_T}{{\cal P}} \; ,
\]
where ${\cal P}$ is the scalar power
spectrum defined in eq.~\eqref{sep24-11-1} and
${\cal P}_T$ is the tensor power spectrum defined in a similar
way, but for transverse traceless metric perturbations $h_{ij}$.

Until recently, the most sensitive probe of the tensor perturbations
has been the CMB temperature anisotropy~\cite{Rubakov:1982df}.
Nowadays, the best tool 
is the CMB polarization. The point is that a certain class of
polarization
patterns (called B-mode) is generated by tensor perturbations,
while scalar perturbations are unable to create 
it~\cite{Kamionkowski:1996zd}. Hence, 
dedicated experiments aiming at
measuring the CMB polarization may well discover the tensor perturbations,
i.e., relic gravity waves. Needless to say, this would be a
profound discovery. To avoid confusion, let us note that the CMB 
polarization has been already observed, but it belongs to another class of
patterns (so called E-mode) and is consistent with the 
existence of the
scalar perturbations only.

\begin{figure}[htb!]
\begin{center}
\includegraphics[width=0.6\textwidth]{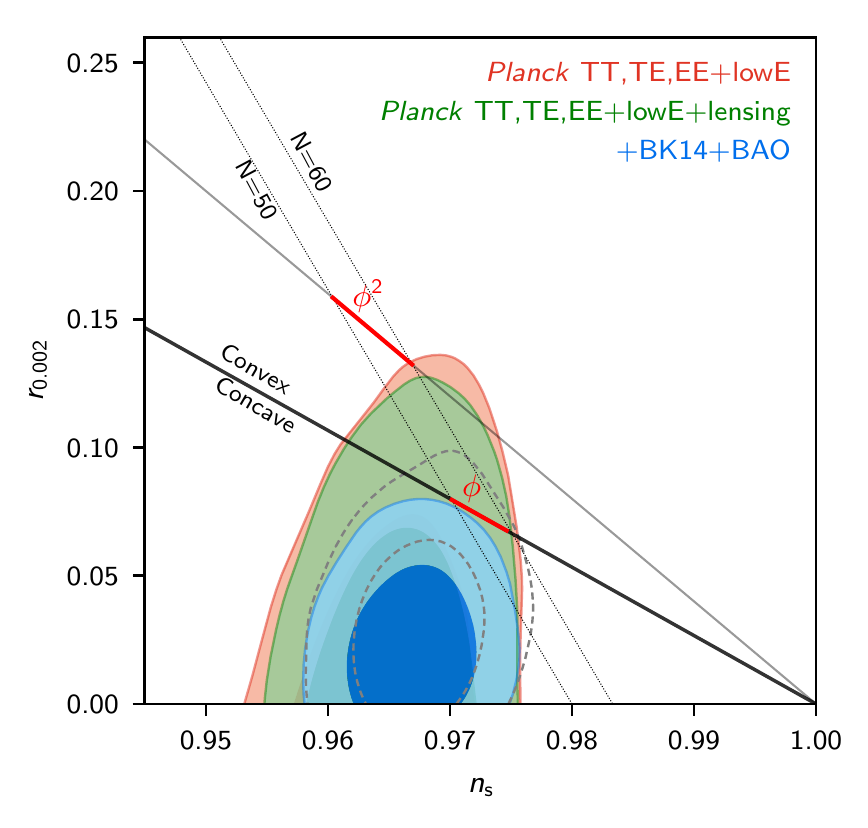}
\caption{Allowed 
regions  (at 68\% and 95\% CL) in the plane $(n_s, r)$, where $n_s$
is the scalar spectral index and $r$ is the tensor-to-scalar
ratio~\cite{Aghanim:2018eyx}, obtained by Planck collaboration alone
and by combining
Planck data with BAO data and CMB polarization data from BICEP2/KEK
experiments. The right corner (the point $(1.0, 0.0)$) is
the Harrison--Zeldovich point (flat scalar spectrum, no
tensor modes). Intervals show predictions of 
inflationary models with quadraic and linear inflaton potentials. 
\label{ns-r}}
\end{center}
\end{figure}

The result of the search for effects of the tensor modes on
CMB temperature anisotropy is shown in Fig.~\ref{ns-r}.
This search has already ruled out some of the popular inflationary models.

\section{Conclusion}

The present situations in particle physics, on one side,
and cosmology, on the other, have much in common.
The Standard Model of particle physics and
Standard Model of cosmology, $\Lambda$CDM, have been shaped.
Both fields enjoyed
fairly unexpected discoveries:  neutrino oscillations and
accelerated expansion of the Universe.

There is strong evidence that the two Standard Models are both incomplete.
Therefore, 
in both fields one hopes for new, revolutionary discoveries.
In the context of these lectures, we hope to 
learn who is dark matter particle; we may learn the origin
of matter-antimatter asymmetry in the Universe;
the discoveries of new properties of cosmological perturbations
will hopefully
reveal the nature of the pre-hot epoch.

However, there is no guarantee of new discoveries 
in particle physics or  cosmology.
Nature may hide its secrets. Whether or not we will be able
to reveal these secrets is
the biggest open question in fundamental physics.

\end{document}